\documentclass[aps,nofootinbib,superscriptaddress,groupedaddress, preprintnumbers,oneside]{revtex4-2}
\pdfoutput=1
\usepackage{amsmath,amsfonts,amssymb,amsthm,amssymb,bbm,bm}
\usepackage{pdfsync}
\usepackage{verbatim}

\usepackage{graphicx,import}
\usepackage[latin1]{inputenc}
\usepackage{hyperref}
\usepackage{empheq}
\usepackage[all]{xy}
\usepackage{stmaryrd}
\usepackage{rotating}
\usepackage{color}  
\usepackage{slashed}
\usepackage{cancel}
\usepackage{array, makecell} %
\usepackage{comment}
\usepackage{tikz-cd}
\usetikzlibrary{arrows.meta,positioning}
\usepackage{subcaption}
\usepackage{hhline}
\usepackage{comment}
\usepackage{cancel}
\makeatletter

\DeclareFontFamily{U}{matha}{\hyphenchar\font45}
\DeclareFontShape{U}{matha}{m}{n}{
    <5> <6> <7> <8> <9> <10> gen * matha
    <10.95> matha10 <12> <14.4> <17.28> <20.74> <24.88> matha12
     }{}
\DeclareSymbolFont{matha}{U}{matha}{m}{n}
\DeclareMathSymbol{\oright}       {2}{matha}{"69}

\newcommand{\doublehat}[1]{%
\begingroup%
  \let\macc@kerna\z@%
  \let\macc@kernb\z@%
  \let\macc@nucleus\@empty%
  \hat{\raisebox{.55ex}{\vphantom{\ensuremath{#1}}}\smash{\hat{#1}}}%
\endgroup%
}

\newcommand{\p}{\partial}

\newcommand{\bit}{\begin{itemize}}
\newcommand{\eit}{\end{itemize}}
\newcommand{\bd}{\begin{description}}
\newcommand{\ed}{\end{description}}

\newcommand{\bc}{\begin{center}}
\newcommand{\ec}{\end{center}}

\newcommand{\C}{{\mathbb C}}

\newcommand{\R}{{\mathbb R}}

\newcommand{\cM}{{\cal M}}

\newcommand{\cJ}{{\cal J}}

\newcommand{\tY}{{\tilde Y}}

\newcommand{\tm}{{\tilde m}}

\newcommand{\cL}{{\mathcal L}}

\newcommand{\cQ}{{\mathcal{Q}}}
\newcommand{\cR}{{\mathcal{R}}}
\newcommand{\tT}{{\tilde T}}

\def\bm{\bar{m}}

\def\be#1\ee{\begin{align}#1\end{align}}
\newcommand{\bea}{\begin{eqnarray}}
\newcommand{\eea}{\end{eqnarray}}
\newcommand{\bs}{\begin{subequations}}
\newcommand{\es}{\end{subequations}}

\newcommand{\la}{\label}

\newcommand{\sgn}{\mathrm{sgn}}
\newcommand{\tr}{{\rm Tr}}
\newcommand{\f}{\frac}

\newcommand{\bz}{{\bar{z}}}

\newcommand{\mS}{\mathscr{S}}

\def\p{\partial}
\def\bQ{\bar{Q}}
\def\bq{\bar{q}}
\def\bR{\bar{R}}

\newcommand{\EM}{ \scriptscriptstyle{\rm {EM}}}

\newcommand{\E}{ \scriptscriptstyle{\rm {E}}}

\newcommand{\M}{ \scriptscriptstyle{\rm {M}}}

\newcommand{\bra}[1]{\langle {#1}|}
\newcommand{\ket}[1]{|{#1}\rangle}

\def\al{\alpha}

\def\d{\delta}

\def\rd{\mathrm{d}}
\def\pa{\partial }

\def\k{{\kappa^2} }

\newcommand{\tcM}{{\widetilde{\mathcal M}}}

\newcommand{\tD}{{\widetilde D }}

\newcommand{\dbms}{\widetilde{\textsf{bms}}}
\newcommand{\ddbms}{\widetilde{\textsf{bms}^*}}

\newcommand{\diff}{\textsf{diff}}

\newcommand{\scri}{\mathscr{I}}
\newcommand{\Ag}{\mathfrak{g}}
\newcommand{\Res}{\text{Res}}

\newcommand{\sds}{\,\tikz[baseline=1]{
		\draw[line width=.6pt] (0,.13) circle (.8ex);
		\draw[line width=.6pt] (-.05ex,.27) -- (-.05ex,0);
		\draw[line width=.6pt] (0,.13) -- (.8ex,.13);}\,}

\usepackage[mathscr]{euscript}

\begin{document}

\title{Mixed-helicity bracket of celestial symmetries}

\author{Daniele Pranzetti}
\email{daniele.pranzetti@uniud.it}
\author{Domenico Giuseppe Salluce}
\email{salluce.domenicogiuseppe@spes.uniud.it}
\affiliation{Universit\`a degli Studi di Udine,
via Palladio 8,  I-33100 Udine, Italy}

\begin{abstract}
Celestial symmetries of gravity and gauge theory can be enhanced to a  $w_{1+\infty}$ algebra and an $S$-algebra respectively, when restricting to a single graviton/gluon helicity sector. Difficulties in combining both sectors in the full theory have been pointed out in the previous literature. 
In this work, we face this problem from the covariant phase space perspective and analyze in detail the structure of the mixed-helicity bracket of the higher-spin charges for both gravity and Yang--Mills theory. We show that, when restricting one of the two helicities to the wedge sector, a closed algebra can be obtained for all spins in terms of a notion of {\it shadow charge} we introduce.
Furthermore, when focusing on the lower spin subalgebra sectors, in the case of gravity, we show that a dual mass extension of the BMS algebra can be consistently constructed; in the case of Maxwell theory, inclusion of magnetic charges allows us to recover a non-vanishing expression for the electromagnetic central charge previously obtained through different methods.
 \end{abstract}

\maketitle

\newpage
\tableofcontents

\section{Introduction}

Study of the symmetry structure of asymptotically flat spacetimes in the last decade or so has revealed a wealth of connections between apparently separate corners and formalisms of both gauge and gravity theories. Since the seminal work of  Bondi--Van der Burg--Metzner--Sachs  (BMS) \cite{Bondi:1960jsa, BMS, Sachs62}, the enhancement of the initially expected Poincar\'e group as the symmetry group of asymptotically null infinity $\scri$ has continued in different steps. This fruitful business was initiated by the original work of Barnich and  Troessaert \cite{Barnich:2009se, Barnich:2011mi, Barnich:2013axa, Barnich:2016lyg} who introduced the so-called extended BMS group where, in addition to super-translations, the Lorentz sector was enlarged, allowing for local conformal transformation on the celestial sphere $S$ foliating $\scri$ and Weyl rescaling preserving a background conformal structure of $S$.  Shortly after, a new extension of the BMS group, called generalized BMS, was proposed by Campiglia and  Laddha \cite{Campiglia:2014yka, Campiglia:2015yka}, and further studied in  \cite{Compere:2018ylh,Campiglia:2020qvc,Campiglia:2024uqq}, where the background structure held fixed is the  scale factor of the celestial sphere metric; in this case
the Lorentz group is enhanced to the full group of smooth diffeomorphisms on $S$. Subsequently, allowing both the scale factor and the conformal structure to vary led to the introduction of the Weyl BMS group \cite{Freidel:2021fxf}. The connection between the conservation of charges associated with these new asymptotic symmetry groups
and graviton soft theorems
\cite{Strominger:2013jfa, He:2014laa,Cachazo:2014fwa, White:2014qia,Kapec:2014opa}, together with the identification of a certain mode of the subleading soft graviton with the stress tensor of a two-dimensional conformal field theory (CFT) \cite{Kapec:2016jld, Cheung:2016iub,Fotopoulos:2019vac}, eventually led to the proposal of a holographic description of gravity in 4D asymptotically-flat spacetimes in terms of a 2D CFT living on the celestial sphere
\cite{ Pasterski:2016qvg, Pasterski:2017kqt,Pasterski:2017ylz,Donnay:2018neh,Stieberger:2018onx,Fotopoulos:2019tpe,Adamo:2019ipt,Donnay:2020guq}.

At the core of this equivalence between soft theorems and charge conservation   is a crucial interplay between 
the
celestial operator product expansion (OPE) of conformal primary operators, in a basis of asymptotic boost eigenstates \cite{Pate:2019lpp, Guevara:2021tvr,Himwich:2021dau}, on one side, and covariant phase space techniques that allow to express the conserved charges as functionals of the radiative phase space variables through an asymptotic expansion of the theory equations of motion, on the other side.
Such interplay is fundamentally characterized by a {\it charge OPE/bracket correspondence} elucidated in \cite{Freidel:2021ytz,Freidel:2023gue}. Such correspondence relates the celestial OPE between a soft charge and a conformal primary operator to the Poisson bracket action of the corresponding hard charge on the same 
primary operator. 
This correspondence is a key entry in the holographic dictionary, which revealed the existence of a 
$w_{1 + \infty}$ structure in gravity  as a further extension of the asymptotic symmetry groups mentioned above, as well as the existence of a
$S$-algebra symmetry in Yang--Mills (YM) theory.
More precisely,
 the infinite tower of conformally soft   symmetries originally manifested by the OPE of conformally soft primary operators \cite{Guevara:2021abz, Strominger:2021lvk,Himwich:2021dau} was later understood as the Poisson bracket algebra of an infinite set of higher-spin-$s$ charges, for integers $s\geq 0$, associated to a tower of recursive differential equations obtained from a large-$r$ expansion of the Einstein and
Yang-Mills equations
\cite{Freidel:2021dfs,Freidel:2021ytz,Freidel:2022skz,Freidel:2023gue}.
These symmetries were
further studied from the covariant phase space perspective  \cite{Hu:2023geb,Geiller:2024bgf,Cresto:2024fhd,Cresto:2024mne,Cresto:2025bfo,Cresto:2025ubl,Agrawal:2024sju}, from the OPE perspective \cite{Mago:2021wje,Ren:2022sws,Banerjee:2023jne,Banerjee:2023bni,Guevara:2025tsm}, and from a
twistor  space perspective  \cite{Adamo:2021lrv,Adamo:2021zpw,Costello:2022wso,Kmec:2024nmu,Mason:2025pbz,Kmec:2025ftx}.

Both in gravity and gauge theory, there are two chiral copies of $w_{1+\infty}$ algebra and $S$-algebra respectively,  associated with the two graviton/gluon helicities. Single copies of the two algebras have been shown to be exact symmetries of
self-dual gravity and self-dual Yang--Mills theory \cite{Adamo:2021lrv,Ball:2021tmb,Bu:2022iak,Kmec:2025ftx,Miller:2025wpq}. However, the fate of these symmetries when combining both helicities in the full theory remains an open question. Difficulties in including both helicities arise at both the level of the charge bracket and the celestial OPE. In the first case, preliminary investigations \cite{Geiller:2024bgf,Pranzetti:2025flv} seem to suggest that the mixed-helicity bracket does not form an algebra. In the second case, the mixed-helicity celestial OPE  is affected by ambiguities when taking a (conformally) double-soft limit \cite{He:2015zea,Fan:2019emx,Ball:2022bgg,Ball:2024oqa,Magnea:2025zut}; such ambiguity, however, could be solved by means of the charge OPE/bracket correspondence, as argued in \cite{Pranzetti:2025flv}.

In this work, we face this problem from the covariant phase space perspective and analyze in detail the structure of the mixed-helicity bracket of the higher-spin charges for both gravity and Yang--Mills theory. We show that difficulties in obtaining a closed algebra can be overcome by introducing a non-local operator acting on the smearing functions for  negative- and positive-helicity, respectively $\tau_s, \bar \tau_s$, entering the charge definition
\be
Q_s(\tau_s)=\int_S \tau_s q_s\,,\quad 
\bQ_s(\bar \tau_s)=\int_S \bar\tau_s \bq_s.
\ee
In the case of gravity, $q_s, \bq_s$ are respectively the positive- and negative-helicity charge aspects,  given by a linear combination of metric components (and their spatial derivatives) at different sub-leading orders in the large-$r$ expansion.
 When restricting to the wedge (global) sector of the smearing function  space, this non-local map $\mS$ can be understood as a generalized version   of shadow transform    \cite{Ferrara:1972uq,Simmons-Duffin:2012juh}, allowing us to introduce the notion of {\it shadow charge}
 \be
  \mS[Q_s]( \tau_s)=Q_s(\mS[\tau_s]).
 \ee
  Our main result is the derivation of a closed-form expression for the linear (in the radiative fields) order bracket of two opposite-helicity charges of arbitrary spin in terms of a shadow charge. Explicitly, arbitrarily choosing, for instance, the 
negative-helicity sector to be global, the mixed-helicity bracket at linear order (denoted by superscript 1) is given by
  \be
\{\bar Q_{s_1}(\bar \tau_{s_1}), Q_{s_2}(\tau_{s_2})\}^1&=\bar \mS  \left[\bQ^1_{s_1+s_2-1}\right]\left(\tau_{s_1,s_2}\right),
\ee
where
\be
\tau_{s_1,s_2}=
(s_2+1)\tau_{s_2}D\bar{ \mS}^{-1} [\bar \tau_{s_1}]
- (s_1-3)
\bar{ \mS}^{-1} [\bar \tau_{s_1}]D\tau_{s_2}.
\ee
An analogous result is derived for Yang--Mills theory as well. These non-localities in the mixed-helicity algebra arise inevitably for spins $s\geq2$ in gravity and $s\geq1$ in Yang--Mills. Of particular interest is the question whether a closed bracket can be obtained without relying on the introduction of shadow charges for the gravitational sector $s=0,1$, where non-localities are a priori not expected. 
More precisely, the set of charges of spin $s=0,1$ comprises {\it complex} super-translations and super-rotations, and these form a closed subalgebra of the $w_{1+\infty}$ algebra, separately in each helicity sector. In particular, the complex super-translation charge aspect involves a real contribution given by the covariant mass and an imaginary contribution given by the {\it covariant dual mass} \cite{Godazgar:2018qpq,Godazgar:2018dvh,Godazgar:2019dkh,Freidel:2021qpz,Freidel:2021dfs}.
One could expect these two subalgebras (one per helicity sector) to provide an extension of the BMS algebra. However, such expectation can only be met if the {\it real} charges close an algebra. This requires the closure of the mixed-helicity bracket as well, providing a BMS extension where the dual mass aspect is included, too. We show that such expectation can indeed be met, upon imposing (anti-)holomorphic conditions on the smearing functions $\tau_0,\tau_1 (\bar \tau_0,\bar \tau_1)$ in addition to the wedge restriction on the spin-1 charges, and a non-trivial extension of the BMS algebra can be obtained for non-compact topologies of the celestial sphere $S$. This last requirement is a reflection of the topological nature of the dual mass charge, as already pointed out in
\cite{Kol:2019nkc,Huang:2019cja,Kol:2020zth,Emond:2020lwi}. This new algebra, which we call {\it dual} BMS and denote $\dbms$, is given by the nested structure
\be
\dbms
=\textsf{so}(3,1) \sds \, (\R_T^S\oplus \R_\tT^S),
\ee
where $T, \tT$ are real functions on $S$ parametrizing respectively super-translations and dual super-translations. We 
show as well that an associated moment map can be constructed for this algebra, confirming that the dual mass can be consistently included in the gravitational phase space. 

In the context of gauge theory, an analogous extension can be performed by including the magnetic charges in Maxwell theory. We show that, by considering boundary conditions compatible with the presence of magnetic monopoles, a central charge arises in the algebra of the electric and magnetic charges due to the non-triviality of the  $U(1)$-bundle for the dual vector potential, in accordance with previous analysis \cite{Freidel:2018fsk,Hosseinzadeh:2018dkh,Geiller:2021gdk}.

The plan of the paper is organized as follows. In Section \ref{sec:prel} we recall background material on the construction of higher-spin charges in gravity and Yang--Mills and their expressions in terms of the discrete basis introduced in \cite{Freidel:2022skz,Freidel:2023gue}, which considerably simplifies bracket computations. In Section \ref{sec:GR} we first obtain the general expression of the mixed-helicity bracket for general spin in gravity; we then focus on the $s=0,1$ sector and derive the dual BMS algebra; finally, we introduce the notion of shadow charge and show how this allows us to obtain a closed mixed-helicity bracket upon imposing the wedge condition. A similar construction is repeated for Yang--Mills theory 
in Section \ref{sec:YM}, including the derivation of the electromagnetic central charge.
Closing remarks and perspectives are presented in
 Section \ref{sec:conc}. Most of the explicit technical derivations are presented in detail in a series of appendices. 

\section{Preliminaries}
\la{sec:prel}

In this section, we review the construction of higher-spin charges for gravity and Yang--Mills theory. We restrict to radiation signals belonging to the Schwartz space, since this provides a phase-space description based on a discrete basis of canonically conjugate fields \cite{Freidel:2022skz,Freidel:2023gue}.

\subsection{Higher-spin charges in gravity}
\subsubsection{Phase space and discrete basis}
The phase space of asymptotically flat spacetimes near $\mathscr{I}^+$ can be described by two canonically conjugate radiative fields: the shear helicity scalar $C(u,z,\bz)$ and the News $N(u,z,\bz)$, being functions of the coordinates of $\mathscr{I}^+$; here, the latter are conveniently chosen to be the Bondi retarded time $u$ and the complex stereographic coordinates $(z,\bar z)$ of the {celestial} 2-sphere $S$ defined at each constant-$u$ cut.\footnote{In the rest of the paper we abbreviate the field dependence on $(z,\bz)$ with simply $(z)$; when (anti-)holomorphic restrictions are imposed, this will be explicitly pointed out.} 
Unless otherwise specified, we assume no puncture to be present on the cut $S$; we analyze the implications of different topologies 
 in Section \ref{sec:dualmass} and \ref{sec:EM}.
 
Positive and negative energy modes of the shear are defined as
\be
\tilde C_+(\omega,z)=\int_{-\infty}^\infty du e^{i\omega u}C(u,z), \quad \tilde C_-(\omega,z)=\int_{-\infty}^\infty du e^{i\omega u}C^*(u,z), \quad \omega>0,
\ee
and their Mellin transforms are given by
\be
\hat C_\pm(\Delta,z)=\int_0^\infty d\omega \omega^{\Delta-1} \tilde C_\pm(\omega,z).
\ee
This leads to the following decomposition of $C(u,z)$ into positive and negative energy sectors
\be\la{sheardec}
C(u,z)=C_+(u,z)+C^*_-(u,z)
\ee
where 
\be
C_{\pm}(u,z)=\f{1}{2\pi}\int_{0}^\infty d\omega e^{-i\omega u}\tilde C_{\pm}(\omega,z).
\ee
Analogous objects can be constructed also for the news
{\be\la{newsdec}
\bar N(u,z)=N_+(u,z)+N^*_-(u,z).
\ee}
The pre-symplectic potential in the radiative phase space of the theory is \cite{Ashtekar:1978zz,Ashtekar:1981bq,Ashtekar:1981sf}
\be \la{eq:ppfull}
\theta_G&:=\f12(\theta+\bar \theta)=\f{1}{\k}\int_{-\infty}^\infty du\left({N}(u,z)\delta {C}(u,z)+\bar {N}(u,z)\delta \bar{C}(u,z)\right),
\ee
where
\be \la{eq:pp}
\theta&=\f{2}{\k}\int_{-\infty}^\infty du{N}(u,z)\delta {C}(u,z)\,,\quad
\bar \theta=\f{2}{\k}\int_{-\infty}^\infty du{\bar N}(u,z)\delta {\bar C}(u,z).
\ee
and $\kappa=\sqrt{32\pi G}$.
{In an unconstrained theory, the Poisson brackets between fundamental fields can be read off directly from the pre-symplectic potential. However, the canonical fields appearing in \eqref{eq:ppfull}
are not independent, but have to satisfy the relations 
\be\la{constr}
\chi_1:=N-\p_u\bar C=0, \quad \chi_2:=\bar N-\p_u C=0.
\ee
Hence, the system is constrained. In particular, the relations \eqref{constr}
define two \emph{second-class} constraints, which means that the matrix of their
Poisson brackets
\be
M_{ij}(u,u',z,z'):=\{\chi_i(u,z),\chi_j(u',z')\}_{PB}, \quad  i,j=1,2
\ee
is invertible on the constraint surface (namely, where \eqref{constr} holds). As a consequence, the Poisson brackets
must be replaced by Dirac brackets according to the standard Dirac procedure \cite{HenneauxBook}.
We denote the resulting bracket simply by $\{\, , \,\}$.
The details of its construction are presented in Appendix \ref{a:DB}.} 
From the presymplectic potential \eqref{eq:ppfull}, one obtains the following Dirac brackets  
\be
 \label{NCbracket}
 \{   N(u,z), C(u', z')\}&=  \{   \bar N(u,z), \bar C(u', z')\}=\f \k 2\delta(u-u') \delta(z,z')\,,
 \ee
 which match the ones used in \cite{He:2014laa} {for a single helicity. The advantage of implementing the constraints \eqref{constr} through the Dirac procedure is that one can obtain brackets for both helicities at the same time, while recovering the factor $1/2$ on the RHS of \eqref{NCbracket}}.
The other non-vanishing brackets among the radiative fields {obtained from the Dirac procedure are}
\be \la{bcc}
\{ \bar C(u,z), C(u', z')\} &= \f {\k}{4}\sgn(u-u') \delta(z,z')\, ,\\ \la{bnn}
  \{   N(u,z),  { \bar N}(u', z')\}&=- \f \k 2\p_{u}\delta(u-u') \delta(z,z')\,,
\ee
where $\sgn(u-u')$ denotes the sign  function. All remaining Dirac brackets among the canonical radiative fields vanish.

Focusing on one helicity, the pre-symplectic potential in \eqref{eq:pp} can also be decomposed into positive and negative energy sectors as $\theta=\theta_++\theta_-$, where
\be
\theta_\pm&=\f{1}{\pi\k}\int_0^\infty d\omega\Tilde{N}_\pm(\omega,z)\delta \Tilde{C}_\pm^*(\omega,z)=\f{2}{\k}\int_{-\infty}^\infty du{N}_\pm(u,z)\delta {C}_\pm^*(u,z).
\la{thetapm}
\ee

If we require $N(u,z)$ to belong to the Schwartz space $\mathcal{S}$, the gravitational signal is completely determined in terms of memory observables $\mathscr{M}_\pm(n,z)$ and Goldstone observables $\mathscr{S}_\pm(n,z)$ \cite{Freidel:2022skz}, defined as
\be
\mathscr{M}_\pm(n,z)=\Res_{\Delta=-n}\hat N_\pm(\Delta,z), \quad \mathscr{S}_\pm(n,z)=\lim_{\Delta\rightarrow n}\hat N_\pm(\Delta,z) \quad \text{where} \quad n\in\mathbb{Z}_+.
\la{MS}
\ee
One can show that 
\be
\mathscr{M}_\pm(n,z)&=\f{i^n}{n!}\int_{-\infty}^{\infty}duu^nN_\pm(u,z),
\la{n-m} \\
{C}_\pm(u,z)&=\f{i}{2\pi}\sum_{n=0}^\infty\,\f{(-iu)^n}{n!}\mathscr{S}_\pm(n,z).\la{c-s}
\ee
In terms of the discrete basis, the pre-symplectic potential \eqref{thetapm} can be re-expressed as
\be
\theta_\pm=\f{1}{i\pi\k}\sum_{n=0}^\infty \mathscr{M}_\pm(n,z)\delta  \mathscr{S}_\pm^*(n,z).
\la{eq3}
\ee
{Consequently, the full pre-symplectic potential \eqref{eq:ppfull} takes the form
\be
\theta_{G}=\f1{2\pi i\k }\sum_{\sigma=\pm}\sum_{n=0}^{\infty}\left(\mathscr{M}_\sigma(n,z)\delta \mathscr{S}^*_{\sigma}(n,z)-\mathscr{M}^*_\sigma(n,z)\delta \mathscr{S}_{\sigma}(n,z)\right).
\ee
In \cite{Freidel:2022skz} it is shown that  $\mathscr{M}$ and $\mathscr{S}$ are linearly dependent, namely we have
\be
\mathscr{S}_\pm(m, z)=\sum_n  {R}_{mn}\mathscr{M}_\pm(n, z),
\quad 
\mathscr{S}^*_\pm(m, z)=\sum_n  {R}_{mn}\mathscr{M}^*_\pm(n, z),
\la{eq6}
\ee
whose inverse relations are
\be
\mathscr{M}_\pm(m, z)=\sum_n  \tilde{R}_{mn}\mathscr{S}_\pm(n, z),
\quad 
\mathscr{S}^*_\pm(m, z)=\sum_n  \tilde{R}_{mn}\mathscr{S}^*_\pm(n, z).
\la{eq6.1}
\ee
Relations \eqref{eq6} can be implemented as constraints on the phase space   applying the Dirac procedure again; this allows us to derive the following Dirac brackets for the canonical variables in the discrete basis (see Appendix \ref{a:DDB})}
\be
\{\mathscr{M}_\pm(n,z),\mathscr{S}_\pm^*(m,z')\}=i\pi\k \delta_{nm}\delta^2(z,z'),
\la{eq4}
\ee
\be
\{\mathscr{M}_\pm(n,z),\mathscr{S}_\mp^*(m,z')\}=
\{\mathscr{M}_\pm(n,z),\mathscr{S}_\pm(m,z')\}=
\{\mathscr{M}_\pm(n,z),\mathscr{S}_\mp(m,z')\}=0.
\la{eq5}
\ee
{One also obtains the following non-trivial brackets where matrices $R$ and $\tilde R$ appear}
\be
\{\mathscr{M}_\pm(n,z),\mathscr{M}_\pm^*(m,z')\}=i\pi\k \Tilde R_{nm}\delta^2(z,z'),
\la{eq9}\\
\{\mathscr{S}_\pm(n,z),\mathscr{S}_\pm^*(m,z')\}=i\pi\k R_{nm}\delta^2(z,z').
\la{eq11}
\ee
{Although the two matrices $R$ and $\Tilde R$ are expected to be mutual inverses, they are found to be divergent,} and therefore need to be regularized. 
Here we report the expressions found in \cite{Freidel:2022skz}, based on the regularization of the quantum vacuum 2-point functions $\bra{0}\mathscr{S}_\pm(n,z)\mathscr{S}_\pm^\dagger(m,z)\ket{0}$ and $\bra{0}\mathscr{M}_\pm(n,z)\mathscr{M}_\pm^\dagger(m,z)\ket{0}$. They read
\be
R_{nm}=\lim_{\epsilon\to0^+}\f{\Gamma(m+n)}{\epsilon^{m+n}},
\la{eq15}
\ee
where $\epsilon$ is an \emph{ultraviolet} regulator, and
\be
\Tilde{R}_{mn}=\lim_{\zeta\to0^+}\bigg[\f{i^{n-m+1}}{2\pi n!(m-1)!}[1-(-)^{n+m}]\f{\Gamma(m+n)}{\zeta^{m+n}}\bigg],
\la{eq19}
\ee
where, $\zeta$ is an \emph{infrared} regulator. 
We stress the fact that matrices $R$ and $\Tilde R$ are expected to be mutual inverses, i.e. they should satisfy
\be
\sum_{m} R_{nm} \Tilde R_{ml}=\sum_{m} \Tilde R_{nm} R_{ml}=\delta_{nl}.
\la{eq7}
\ee

At this stage, it is unclear what kind of regularization one should use to obtain \eqref{eq7} properly within some limit. A possible way out is the Schwartzian regulator used in \cite{Freidel:2022skz}, relying on the approximation of states $\mathscr{M}_\pm^\dagger(n,z)\ket{0}$ and $\mathscr{S}_\pm^\dagger(n,z)\ket{0}$ as Schwartzian states. The nature of this Schwarzian regulator is UV and IR at the same time. Within this regularization scheme, the form of $R$ is the same as \eqref{eq15}. To our scopes, the explicit form of neither $R$   nor $\tilde R$ is needed.

\subsubsection{Higher-spin charges}
In the case of Schwartzian signals, an infinite tower of higher-spin charges \cite{Freidel:2021ytz, Geiller:2024bgf} exists, labelled by conformal dimension and spin $(\Delta, J)=(3,s)$, with $s\in\mathbb{Z}, s\ge-2$. These charges satisfy the following recursive evolution equations
\be
\partial_u\mathcal{Q}_s=D_z\mathcal{Q}_{s-1}+\f{(s+1)}{2}C\mathcal{Q}_{s-2},
\la{eeom}
\ee
which are extracted from a large-$r$ expansion of truncated (for $s>3$) Einstein's equations around $\mathscr{I}^+$. Here $D_z$ denotes a covariant derivative with respect to $z$ on the celestial sphere $S$; it will also be denoted as $D$ in the following, while $D_{\bar z}$ will be similarly shortened as $\bar D$. In this tower, $\mathcal{Q}_{-2}=\partial_u{N}/2$ is the derivative of the News, $\mathcal{Q}_{-1}=D_z{N}/2$ is the energy current, $\cQ_0=\cM_\C=\cM+i\tcM$ is the complex mass aspect, with $\cM$ the covariant mass and $\tcM$ the dual covariant mass, and $\cQ_1=\cJ$ is the covariant angular momentum charge aspect \cite{Freidel:2021qpz,Freidel:2021ytz}.
{If one imposes the final condition
\be\la{final-cond}
\lim_{u\rightarrow\infty}\cQ_s(u,z)=0,
\ee
the evolution equations \eqref{eeom} can be integrated recursively, and charges can be expressed as functionals of the fundamental fields $\cQ_s=\cQ_s[C,N]$. One can then expand $\cQ_s$ in powers of the radiation fields $\cQ_s=\cQ_s^1+\cQ_s^2+\cdots$, where the first two contributions are respectively the soft charge (linear) and the hard charge (quadratic). In this work, we will only consider these two contributions, since they are sufficient to compute linear-order brackets of charges, which is the main focus of this work. 


The physical origin of this perturbative truncation can be traced back to the three-point interactions of Yang--Mills theory and gravity. At tree level, the leading collinear factorization of scattering amplitudes is governed by the corresponding three-point amplitudes, whose splitting functions determine the singular coefficients of celestial OPEs. The conformally soft limits of these OPEs give rise to the celestial current algebras, including the $w_{1+\infty}$ structure in gravity and the corresponding higher-spin current algebra in Yang--Mills theory \cite{Guevara:2021abz,Himwich:2021dau}. From the covariant phase space perspective, through the celestial OPE/charge-bracket correspondence, the same three-point information is encoded in the brackets between the linear  and quadratic  contributions to the charges.
The same perturbative reasoning motivates our investigation of the mixed-helicity bracket.
However, let us stress that this truncation  does not by itself establish the complete nonlinear charge algebra. At higher orders, additional contributions involving cubic and higher-order charge terms must be included. The nonlinear completion of the charges and the extension of the mixed-helicity bracket beyond the order considered here remain distinct questions.


Since the action of $\cQ_s(u,z)$ on the phase space diverges in the limit $u\rightarrow -\infty$, one defines the \emph{renormalized} higher-spin charge aspects $\hat q_s(u,z)$ as the combinations of them that are conserved when the non-radiative conditions $\cQ_{-2}=0=\cQ_{-1}$ are imposed. The limit  $u\rightarrow-\infty$ of these generators is finite and defines the higher-spin charge aspects
\be\la{aspectinf}
q_s(z):=\lim_{u\rightarrow-\infty}\hat{q}_s(u,z).
\ee
While these charge aspects have helicity $s$, their hermitian conjugates $\bar q_s:=q_s^*$ have helicity $-s$.
The higher-spin charges are defined by integrating the charge aspects against some spin-$s$ celestial smearing functions $\tau_s(z)$ over the celestial sphere $S$, i.e.
\be
Q_s(\tau_s):=\f{8}{\k}\int_S\tau_s(z)q_s(z),
\la{Qs}
\ee
where we denote $\int_S:=\int_Sd^2z\sqrt{q}$, $q$ being the determinant of the leading order of the celestial sphere metric.
 At the linear and quadratic orders in the radiative fields, the renormalized aspects are given by
\begin{subequations}\la{ren-p}
 \be
\hat{q}_s^1(u,z):=&\sum_{n=0}^s\f{(-u)^{s-n}}{(s-n)!}D^{s-n}\mathcal{Q}_s^1(u,z),
\la{ren-p1}
\\
\hat{q}_s^2(u,z):=&\sum_{n=0}^s\f{(-u)^{s-n}}{(s-n)!}D^{s-n}\mathcal{Q}_s^2(u,z)+\f12\sum_{\ell=2}^s\sum_{n=0}^{\ell-2}\f{(-)^n(\ell+1)}{(s-\ell)!}D^{s-\ell}\left(\p^{-(1+n)}_{u,0}((-u)^{s-n}C)D^n\cQ_{\ell-2-n}^1\right)(u,z).
\la{ren-p2}
\ee   
\end{subequations}
where
\be\label{iiint}
 \pa_{u,a}^{-n}f(u)  := \int_{a}^{u}\rd u_1 \int_{a}^{u_1} \rd u_2 \cdots \int_{a}^{u_{n-1}} \rd u_n f(u_n)\,.
\ee
The second contribution in the quadratic order \eqref{ren-p2} is a \emph{non-linear correction} for $s\geq 2$ to the initial renormalization prescription originally proposed in \cite{Freidel:2021ytz}. This correction, first obtained for the spin $s=2$  charge in \cite{Freidel:2021dfs} {(see Eq. (86) therein)}, was introduced for a general spin $s$ in \cite{Geiller:2024bgf} {(see Eq. (6.18) therein)}, by demanding charge conservation on a non-radiative cut; the one difference here is that for the lower bound of \eqref{iiint}, we chose $a=0$, as done in \cite{Cresto:2024mne}, instead of $a=\infty$. {While the charge conservation condition does not select a specific base point in \eqref{ren-p2}, our choice is motivated by the systematic approach to obtain the full expressions of the correctly renormalized charges at all orders in the radiative fields  developed in  \cite{Kmec:2024nmu} from a twistor space perspective and in \cite{Cresto:2024mne} from a space-time perspective.} More precisely, the common idea is to introduce a   master charge at each cut of $\scri^+$ as the sum over all spins $s$ of integrals of the aspects $\cQ_s(u,z)$  weighted by Carrollian smearing functions $\xi_s(u,z)$. Demanding the conservation of these charges at non-radiative cuts implies that the functions $\xi_s(u,z)$ satisfy evolution equations dual to \eqref{eeom}. Solving them recursively gives a non-perturbative prescription for the renormalized charge aspects.
{
In \cite{Cresto:2024mne}, the dual equations of motion are solved by imposing the initial conditions $\xi_s(u=0,z)=\tau_s(z)$ at the cut $u=0$. This fixes the integration base point to $a=0$, as reflected in the expressions \eqref{ren-p} for the renormalized charge aspects. Although one could in principle choose a different initial cut $u=a$, we adopt $a=0$ throughout this work. This choice is particularly natural in our framework, as it allows us to express the charges in terms of the discrete basis for the asymptotic phase space in a straightforward way. This stems from the fact that, as defined in \eqref{c-s}, the Goldstone variables $\mS_\pm(n)$ are the coefficients in the Taylor expansion of the shear around $u=0$.
}

We note that the non-linear correction can also be expressed in terms of linear-order renormalized charges, as follows
\be
\hat{q}_s^2(u,z)=\sum_{n=0}^s\f{(-u)^{s-n}}{(s-n)!}D^{s-n}\mathcal{Q}_s^2(u,z)-\f12\sum_{k=0}^{s-2}\left[\left((s+1-k)\hat q^1_{s-2-k}D^k-kD\hat q^1_{s-2-k}D^{k-1}\right)\p_{u,0}^{-1}\left(\f{(-u)^k}{k!}C\right)\right](u,z).
\ee
}
{In terms of the fundamental fields, the renormalized soft and hard charge aspects correspond to the expressions
\begin{subequations}
\la{fund-exp}
 \be
\hat{q}_s^1(u,z)&=\f{1}{2}\sum_{n=0}^s\f{(-u)^{s-n}}{(s-n)!}D^{s-n}(\p^{-1}_{u,\infty} D)^{s+2}\p_u N(u,z), 
\la{q1}\\
\hat{q}_s^2(u,z)&=-\frac14\p_{u,\infty}^{-1}\sum_{\ell=0}^s (-)^{l}(\ell+1) 
  D^{s-\ell}\left[\p_{u,0}^{-\ell+1}\left( \frac{(-u)^{s-\ell}}{(s-\ell)!}C\right)  D^{\ell}   N  \right](u,z)+\f14\f{(-u)^s}{s!}\left[NC\right](u,z),
  \la{q2}
\ee   
\end{subequations}
 The choice of the lower bound of the first anti-derivative in \eqref{q2} is $a=\infty$ in this case, since these expressions are obtained by integrating the time derivatives of expressions \eqref{ren-p} in $u$, using the final condition \eqref{final-cond}. 
}

It can be shown that expressions \eqref{fund-exp}, in the limit $u\rightarrow-\infty$, converge to  
\begin{subequations}
\la{q2full}
\be
q^1_s(z)=&D_z^{s+2}N_s(z), \quad \text{where} \quad N_s(z)=\f{(-)^{s+1}}{2 s!}\int_{-\infty}^\infty du u^s N(u,z),\\
{q}_s^2(z)=&\frac14\sum_{\ell=0}^s (-)^{l}(\ell+1)\int_{-\infty}^\infty du 
  D^{s-\ell}\left[\p_{u,0}^{-\ell+1}\left( \frac{(-u)^{s-\ell}}{(s-\ell)!}C\right)  D^{\ell}   N  \right](u,z) \la{q2i}.
\ee    
\end{subequations}
The exact expressions \eqref{q2full} can be obtained from the master charge approach of \cite{Cresto:2024mne,Kmec:2024nmu} as well, {using the procedure described above, namely, solving the dual  evolution equations for the Carrollian parameters up to linear order in the shear}. 

We can decompose the smeared charge $Q_s(\tau_s)$ into positive and negative energy sectors
\be
Q_s=Q_{s+}+Q_{s-}.
\ee
Charges are completely determined by Memory and Goldstone observables. In the positive-energy sector, the soft charge can be expressed as

\be\la{linQdisc}
    Q^1_{s+}(\tau_s)&=-\f{2}{\kappa^2}i^{s}\int_{S} D^{s+2}\tau_s(z)\mathscr{M}^*_+(s,z).
\ee
Having defined the falling factorial as $(x)_n=x(x-1)\cdots(x-n+1)$, one can write the hard charge as
\be
     Q^2_{s+}(\tau_s)&=\f{2}{\kappa^2}i^{-s}\f{1}{2\pi}\sum_{n=0}^\infty\sum_{l=0}^s\bigg[(-)^{l+s}(l+1)\f{(s+n-l)_{s-l}}{(s-l)!}\int_{S} D^{s-l}\tau_s(z)\mathscr{S}_+(n,z)D^l\mathscr{M}^*_+(s+n-1,z)\bigg].
     \la{eq30}
\ee
In Appendix \ref{A:dbcharge}, we prove that \eqref{eq30} follows from \eqref{q2i}. Therefore, the discrete basis expression \eqref{eq30}, which is used in the rest of this work,  correctly incorporates the non-linear correction. 
However, let us clarify some possible source of confusion that may arise from previous literature. 
The  expression \eqref{eq30}  of the hard charge in terms of the discrete basis was originally derived in \cite{Freidel:2022skz} (see Appendix G.1 there); its YM analogue  \eqref{r2+} was 
originally obtained in \cite{Freidel:2023gue} and both correct expressions were used in that work to compute the loop $w_{1+\infty}$ algebra for gravity and the $S$-algebra for gauge theory up to quadratic order. However, in both \cite{Freidel:2022skz} and \cite{Freidel:2023gue}
 an incomplete renormalization prescription not including the second contribution on the RHS of \eqref{ren-p2}  was used. The reason why the authors still arrived at the   same correct discrete basis expressions is due to the fact that this non-linear correction exactly compensates for boundary terms arising from a series of integrations by parts which were originally discarded in those initial works. 
{Similarly, the action of the quadratic charge on the shear derived in \cite{Freidel:2021ytz} (see Eqs. (67) and (68) therein) is also correct, despite the use of the same incomplete renormalization prescription. As in the previous case, the absence of the non-linear correction is compensated by the neglect of boundary terms in the derivation. The correctness of this expression for the quadratic charge action was confirmed in \cite{Geiller:2024bgf}. 
 }

The hard charges can equivalently be  expressed as
\be
     Q^2_{s+}(\tau_s)&=\f{2}{\kappa^2}i^{s}\f{1}{2\pi}\sum_{n=0}^\infty\sum_{l=0}^s\bigg[(l+1)\f{(3-n)_{s-l}}{(s-l)!}\int_{S} D^{s-l}\tau_s(z)\mathscr{S}^*_+(n,z)D^l\mathscr{M}_+(s+n-1,z)\bigg],
     \la{eq31}
\ee
{which had also been originally} derived in \cite{Freidel:2022skz}. {In Appendix \ref{A:dbcharge}, we explain how \eqref{eq31} can be obtained by rewriting in the discrete basis an alternative expression for the quadratic charge aspect, which can be inferred from the action of \eqref{q2i} on the shear.}  
In the negative energy sector, $\bQ_{s-}(\bar\tau_s)$   is given by the same expressions, up to the substitutions $D\rightarrow\bar D$, $\mathscr{S}_+\rightarrow\mathscr{S}_-$ and $\mathscr{M}_+\rightarrow\mathscr{M}_-$. 

Using the fundamental bracket of the discrete basis \eqref{eq5} (or, equivalently, \eqref{NCbracket}, if one works with the $u$-basis), one can compute the linear order contribution to the charge bracket
\be
\{Q_{s_1}(\tau_{s_1}), Q_{s_2}(\tau_{s_2})\}^1:=\{Q^2_{s_1}(\tau_{s_1}), Q^1_{s_2}(\tau_{s_2})\}+\{Q^1_{s_1}(\tau_{s_1}), Q^2_{s_2}(\tau_{s_2})\},
\ee
thus obtaining the following structure
\be\la{winf}
\{Q_{s_1}(\tau_{s_1}), Q_{s_2}(\tau_{s_2})\}^1&=
Q^1_{s_1+s_2-1} \left((s_2+1) \tau_{s_2}D\tau_{s_1}
-(s_1+1)\tau_{s_1}D \tau_{s_2}
\right),
\ee
which describes an algebra named $w_{1+\infty}$. 

In complete analogy with this procedure, we can construct smeared charges out of the negative-helicity aspects $\bar q_s$, as follows
\be
\bar Q_s(\bar \tau_s)=\int_S \bar\tau_s(z)\bar q_s(z).
\ee
In this manuscript, if not otherwise specified (as in Section \ref{sec:dualmass}), we take the smearing parameter $\bar \tau_s$ to be independent of $\tau_s$, namely $\bar\tau_s\neq \tau_s^*$. This set of charges satisfies the same algebraic structure as \eqref{winf}, i.e.
\be\la{winf-}
\{\bar Q_{s_1}(\bar \tau_{s_1}), \bar Q_{s_2}(\bar\tau_{s_2})\}^1&=
\bar Q^1_{s_1+s_2-1} \left((s_2+1) \bar\tau_{s_2}\bar D\bar\tau_{s_1}
-(s_1+1)\bar\tau_{s_1}\bar D \bar\tau_{s_2}
\right).
\ee
This result is limited to brackets involving charges of the same helicity.
In this work, we will investigate what happens if one considers brackets between charges of opposite helicities.

\subsection{Higher-spin charges in Yang--Mills}
\subsubsection{Phase space and discrete basis}
In close analogy to the case of gravity, we can study the Yang--Mills theory in a 4d Minkowski spacetime. Let us call $G$ the gauge group of the theory, and $\Ag$ the related Lie algebra, spanned by the generators $T_a$, with $a=1,\dots, \text{dim}(\Ag)$. The algebra is defined by the following commutation relations
\be
[T_a,T_b]_\Ag=i{f_{ab}}^cT_c,
\ee
whose structure constants satisfy the normalization condition
\be
\tr(T_aT_b)=\delta_{ab}.
\ee
We introduce the $\Ag$-valued 1-form {gauge field} $A=A_\mu dx^\mu$, with $A_\mu=A_\mu^aT_a$,  and the $\Ag$-valued 2-form {field strength} $F$, satisfying the following Yang--Mills equations
\be
d\star F, \quad F=dA+iA\wedge A,
\la{YM}
\ee
where $\star F$ is the Hodge-dual of $F$.
We use retarded Bondi coordinates $(u,r,z, \bar z)$ in  which the Minkowski metric takes the form 
\be
ds^2=-du^2-2dudr+2r^2\gamma_{z\bar z}dzd\bar z,
\ee
with $\gamma_{z\bar z}=2/(1+z\bar z)^2$ being the leading contribution to the metric of the celestial sphere $S$. 
In the radial gauge $A_r=0$, the large-$r$ expansions of the coefficients of $A$ and $F$ around $\scri^+$ are respectively
\be
A_u&=\sum_{n=0}^\infty \f{A^{(n)}_u}{r^n}, \quad A_z=\sum_{n=0}^\infty \f{A^{(n)}_z}{r^n}, \\
F_{ur}=\f{1}{r^2}\sum_{n=0}^\infty
\f{F_{ur}^{(n)}}{r^n}, \quad
F_{uz}&=\sum_{n=0}^\infty
\f{F_{uz}^{(n)}}{r^n},\quad
F_{rz}=\f{1}{r^2}\sum_{n=0}^\infty
\f{F_{rz}^{(n)}}{r^n},\quad
F_{z\bar z}=\sum_{n=0}^\infty\f{F_{z\bar z}^{(n)}}{r^n}.
\ee
Moreover, we further fix the gauge by imposing $A_u^{(0)}=0$. As a consequence, the phase space of the theory near $\scri^+$ is specified by $A_z^{(0)}$ and its canonically conjugate field $F_{\bz u}^{(0)}=-\p_u A^{(0)*}_z$. 
Positive and negative energy modes of $A_z^{(0)}$ are defined as
\be
\tilde A_+(\omega,z)=\int_{-\infty}^\infty du e^{i\omega u}A_z^{(0)}(u,z), \quad \tilde A_-(\omega,z)=\int_{-\infty}^\infty du e^{i\omega u}{A_z^{(0)*}(u,z)}, \quad \omega>0,
\ee
whose Mellin transforms are
\be
\hat A_\pm(\Delta,z)=\int_0^\infty d\omega \omega^{\Delta-1}\tilde A_\pm(\omega,z).
\ee
This allows us to  decompose $A_z^{(0)}$ into positive and negative energy sectors as
\be
A_z^{(0)}(u,z)=A_+(u,z)+A^*_-(u,z)
\ee
where 
\be
A_{\pm}(u,z)=\f{1}{2\pi}\int_{0}^\infty d\omega e^{-i\omega u}\tilde A_{\pm}(\omega,z).
\ee
Analogous objects can be defined from $F^{(0)}_{\bar z u}$. 

If we require the vector potential to belong to the Schwartz space $\mathcal{S}$, a discrete basis can be used to describe the phase space of the Yang--Mills theory \cite{Freidel:2023gue}. In complete analogy with the case of gravity, we have memory observables $\mathscr{F}_\pm(n,z)$, and Goldstone bosons $\mathscr{A}_\pm(n,z)$, defined as
\be
\mathscr{F}_\pm(n,z):=\Res_{\Delta=-n}\hat F_{\pm}(\Delta,z), \quad 
\mathscr{A}_\pm(n,z):=\lim_{\Delta\rightarrow n} \hat F_{\pm}(\Delta,z), \quad \text{where} \quad n\in\mathbb{Z}_+.
\ee
The symplectic potential of the theory at $\mathscr{I}^+$ reads
\be\la{sf:YM}
\Theta^{YM}=\f{1}{2}(\theta^{YM}+\bar\theta^{YM})=\f{1}{g_{YM}^2}\int_\mathscr{I^+}\tr[F^{(0)}_{\bar z u}(u,z)\delta A^{(0)}_z(u,z)+F^{(0)}_{ z u}(u,z)\delta A^{(0)}_{\bar z}(u,z)],
\ee
where each contribution can be decomposed as $\theta^{YM}=\theta^{YM}_++\theta^{YM}_-$, and similarly for $\bar\theta^{YM}$; in terms of the discrete basis, we have
\be
\theta^{YM}_\pm=\f{1}{2\pi ig^2_{YM}}\sum_{n=0}^\infty\int_S\tr[\mathscr{F}_\pm(n,z)\delta\mathscr{A}^*_\pm(n,z)].
\ee
From this, {in analogy with the case of gravity,} we can {construct the Dirac bracket} of fundamental fields 
\be
\{\mathscr{F}^a_\pm(n,z),\mathscr{A}^{b*}_\pm(m,z')\}=2\pi i g^2_{YM}\delta^{ab}\delta_{nm}\delta^{2}(z,z').
\la{ymcomm}
\ee

\subsubsection{Higher-spin charges}
For a Schwartzian vector potential in Yang--Mills theory, one finds an infinite tower of $\Ag$-valued charge aspects \cite{Freidel:2023gue}  with conformal dimension and spin $(\Delta, J)=(2,s)$ satisfying the recursive differential equations
\be
\partial_u\mathcal{R}_s=D\mathcal{R}_{s-1}+i[A^{(0)}_z,\mathcal{R}_{s-1}]_\Ag, \quad  s\ge0, \quad \mathcal{R}_{-1} =F^{(0)}_{\bar z u},
\la{YM-rec}
\ee
extracted from the large-$r$ expansion of a truncation (for $s>1$) of the Yang--Mills equations \eqref{YM}.
Since such charges diverge as $u\rightarrow-\infty$, one introduces renormalized charge aspects $r_s(z)$ at $\mathscr{I}^+_-$, which are finite by construction. The fields $r_s$ carry spin $s$, while their complex conjugates $\bar r_s$ carry spin $-s$.
The charge aspects admit an expansion in powers of the gauge field as $r_s=r_s^1+r_s^2+\dots$, where the first two terms are known as the soft and hard contributions. The renormalization prescription for them is:
\begin{subequations}\la{cym}
\be
r_s^1(z)=&\lim_{u\rightarrow-\infty}\sum_{l=0}^s\f{(-u)^{s-l}}{(s-l)!}D^{s-l}\mathcal{R}^1_l(u,z), \\
r_s^2(z)=&\lim_{u\rightarrow-\infty}\left[\sum_{l=0}^s\f{(-u)^{s-l}}{(s-l)!}D^{s-l}\mathcal{R}^2_l(u,z)+i\sum_{\ell=1}^s\sum_{n=0}^{\ell-1}\f{(-)^n}{(s-\ell)!}D^{s-\ell}[\p^{-n}_{u,0}((-u)^{s-n}A_z^{(0)}), D^n\cR_{\ell-1-n}^1]_\Ag(u,z)\right] \la{hcym},
\ee    
\end{subequations}
In this case, the second summation in \eqref{hcym} was introduced in \cite{Agrawal:2024sju}, as a correction to the renormalization prescription proposed in \cite{Freidel:2023gue} (see also \cite{Cresto:2025bfo}); from a twistor space approach, this correction was obtained in \cite{Kmec:2025ftx}. The soft components take the explicit form
\be
r_s^1=D^{s+1}F_s, \quad \text{where} \quad F_s:=\frac{(-)^{s+1}}{s!}\int_{-\infty}^\infty duu^sF^{(0)}_{\bar z u}.
\ee
It is convenient to decompose the charge aspects into positive- and negative-frequency parts  
\be
r_s=r_{s+}+r_{s-}.
\ee
In the discrete basis introduced above, the soft and hard parts of $r_{s+}$ take the form
\be
r^1_{{s}+}&=-i^{-{s}}D^{{s}+1}\mathscr{F}^*_{+}({s},z),
\la{r1+}\\
r^2_{{s}+}&=-\f{i^{-s}}{2\pi}\sum_{l=0}^\infty\sum_{n=0}^{s}\binom{l+n}{l}D^n[\mathscr{A}_+(l,z),D^{{s}-n}\mathscr{F}^*_+({s}+l,z)]_\Ag
 \cr
  &=-\f{i^{s}}{2\pi}\sum_{\ell=0}^\infty\sum_{n=0}^{s} 
(-)^{n}
\f{(s +\ell-1)_{s-n}}{(s-n)!} D^{s-n} \left[   \mathscr{F}_+(s+\ell,z,\bz),  \p^{n} \mathscr{A}^*_+(\ell,z,\bz) \right]_\Ag\,.
\la{r2+}
\ee
{These expressions follow from  \eqref{cym}, from analogous manipulations to those used in the gravitational case (see Appendix \ref{A:dbcharge}). The reason why the first row of \eqref{r2+} was already derived in \cite{Freidel:2023gue} with the old renormalization prescription stands in the role of {omitted} boundary terms, as discussed above in the case of gravity.} 
The corresponding expressions for $\bar r_{s-}$ in the negative-energy sector follow from the substitutions $D\rightarrow \bar D$, $\mathscr{F}_+\rightarrow\mathscr{F}_-$ and $\mathscr{A}_+\rightarrow\mathscr{A}_-$.
Higher-spin charges are defined by smearing the charge aspects against $\Ag$-valued spin-$s$ arbitrary functions $\tau_s$ on the celestial sphere $S$
\be
R_{s}(\tau_s)=\int_S \tr(\tau_s(z)r_{s}(z))=\int_S {\tau_s}_a(z)r^a_{s}(z).
\la{expR}
\ee

Using the fundamental Poisson bracket \eqref{ymcomm}, the linear and quadratic part of the bracket of higher-spin charges take the form \cite{Freidel:2023gue}
\be\la{HSalgYM}
\{R_{s_1}(\tau_{s_1}),\,R_{s_2}(\tau_{s_2})\}=-ig_{YM}^2R_{s_1+s_2}([\tau_{s_1},\tau_{s_2}]_\Ag).
\ee

In the quantum theory, this algebra contains as a subalgebra  the wedge (global) $S$-algebra of celestial holography \cite{Guevara:2021abz, Strominger:2021mtt}, obtained upon restricting to smearing functions satisfying 
\be
D^{s+1}\tau_{s}=0.
\la{wedgeYM}
\ee

Also in the case of Yang--Mills, we will study the charge algebra in the mixed-helicity sector (up to linear order), which involves two charges with helicities of opposite signs.

\section{Mixed-helicity bracket in gravity}\la{sec:GR}

We are interested in the linear order contribution to the mixed-helicity Poisson bracket
\be
\{\bar Q_{s_1}(\bar \tau_{s_1}),Q_{s_2}(\tau_{s_2})\}^1=\{\bar Q_{s_1}^1(\bar \tau_{s_1}), Q_{s_2}^2(\tau_{s_2})\}+\{\bar Q_{s_1}^2(\bar \tau_{s_1}), Q_{s_2}^1(\tau_{s_2})\}.
\la{mb}
\ee
 Since positive and negative energy charges always commute, we have
\be
\{\bar Q^1_{s_1}(\bar \tau_{s_1}), Q^2_{s_2}(\tau_{s_2})\}=\{\bar Q^1_{s_1+}(\bar \tau_{s_1}), Q^2_{s_2+}(\tau_{s_2})\}+\{\bar Q^1_{s_1-}(\bar \tau_{s_1}), Q^2_{s_2-}(\tau_{s_2})\}.
\ee

 If we focus on the positive-energy sector, the bracket is given by
\be
\{\bar Q_{s_1+}(\bar \tau_{s_1}), Q_{s_2+}(\tau_{s_2})\}^1&=\{\bar Q_{s_1+}^1(\bar \tau_{s_1}), Q_{s_2+}^2(\tau_{s_2})\}+\{\bar Q_{s_1+}^2(\bar \tau_{s_1}), Q_{s_2+}^1(\tau_{s_2})\}.
\la{linbrack}
\ee

We start with the first term on the RHS. In the computation, we conveniently use expression \eqref{eq31} for the hard charge, since in this case the non-trivial pieces of the bracket involve $\{\mathscr{M}_+(s_1,z),\mathscr{S}^*_+(n,z')\}$, and are straightforward to compute.\footnote{If we used \eqref{eq30} instead, the non-trivial pieces would come from $\{\mathscr{M}_+(s_1,z),\mathscr{M}^*_+(s_2+n-1,z')\}$, thus involving the diverging matrix $\tilde R$, which would require a more careful analysis.}

The final result (details in Appendix \ref{A.1}), is
\be
\{\bar Q_{s_1+}^1(\bar \tau_{s_1}), Q_{s_2+}^2(\tau_{s_2})\}
&=-i^{-(s_1+s_2-1)}\f2{\kappa^2} \sum_{l=0}^{s_2} \int_S \Bigg[(l+1)\f{(s_1+s_2-2)_{s_2-l}}{(s_2-l)!}
\cr
&\times
D^l\bar D^{s_1+2} \bar \tau_{s_1}(z)D^{s_2-l}\tau_{s_2}(z)\mathscr{M}_+  (s_1+s_2-1,z)\Bigg].
\la{eq34}
\ee
The final expression that we get for \eqref{linbrack} is
\be
\{\bar Q_{s_1+}(\bar \tau_{s_1}), Q_{s_2+}(\tau_{s_2})\}^1
&=-i^{-(s_1+s_2-1)}\f2{\kappa^2} \sum_{l=0}^{s_2} \int_S \Bigg[(l+1)\f{(s_1+s_2-2)_{s_2-l}}{(s_2-l)!}
\cr
&\times
D^l\bar D^{s_1+2} \bar \tau_{s_1}(z)D^{s_2-l}\tau_{s_2}(z)\mathscr{M}_+  (s_1+s_2-1,z)\Bigg]
\cr
&+i^{s_1+s_2-1}\f2{\kappa^2} \sum_{l=0}^{s_1} \int_S \Bigg[(l+1)\f{(s_1+s_2-2)_{s_1-l}}{(s_1-l)!}
\cr
&\times
\bar D^l D^{s_2+2}  \tau_{s_2}(z)\bar D^{s_1-l}\bar \tau_{s_1}(z)\mathscr{M}_+^*  (s_1+s_2-1,z)\Bigg].
\la{genexpr}
\ee
We observe that the object on the RHS, under general assumptions, is not a charge. This means that the two copies of the $w_{1+\infty}$ algebra, respectively generated by charges $q_s$ and $\bar q_{s'}$, cannot be joined together to form a wider algebra, because the mixed-helicity bracket is not closed.

By summing up \eqref{eq34} and the negative-energy counterpart, we obtain
\be
\{\bar Q_{s_1}^1(\bar \tau_{s_1}), Q_{s_2}^2(\tau_{s_2})\}
&=-\f8{\kappa^2} \sum_{l=0}^{s_2} \int_S (-)^{s_1+s_2} \Bigg[(l+1)\f{(s_1+s_2-2)_{s_2-l}}{(s_2-l)!}
\cr
&\times
D^l\bar D^{s_1+2} \bar \tau_{s_1}(z)D^{s_2-l}\tau_{s_2}(z)\bar N_{s_1+s_2-1}(z)\Bigg],
\la{eq36}
\ee
where we used 
\be
\bar N_s(z)=\f{(-)^{s+1}}{4}(i^{-s}\mathscr{M}_+(s,z)+i^s\mathscr{M}_-^*(s,z)).
\ee
The expression \eqref{eq36}  matches the result of \cite{Geiller:2024bgf} (see expression (J.4a) there), up to notational differences, as we show in Appendix \ref{A.1}. If we add the second piece of \eqref{mb}, we get
\be
\{\bar Q_{s_1}(\bar \tau_{s_1}), Q_{s_2}(\tau_{s_2})\}^1
&=-\f8{\kappa^2} \sum_{l=0}^{s_2} \int_S (-)^{s_1+s_2} \Bigg[(l+1)\f{(s_1+s_2-2)_{s_2-l}}{(s_2-l)!}
\cr
&\times
D^l\bar D^{s_1+2} \bar \tau_{s_1}(z)D^{s_2-l}\tau_{s_2}(z)\bar N_{s_1+s_2-1}(z)\Bigg]
\cr
&+\f8{\kappa^2} \sum_{l=0}^{s_1} \int_S (-)^{s_1+s_2} \Bigg[(l+1)\f{(s_1+s_2-2)_{s_1-l}}{(s_1-l)!}
\cr
&\times
\bar D^l D^{s_2+2}  \tau_{s_2}(z)\bar D^{s_1-l}\bar \tau_{s_1}(z) N_{s_1+s_2-1}(z)\Bigg]
.
\la{eqcomplete}
\ee
In the case of the mixed-helicity bracket, there is also a $0th$-order contribution playing a role, which is

\be
\{\bar Q_{s_1}(\bar \tau_{s_1}), Q_{s_2}(\tau_{s_2})\}_+^0&=\{\bar Q_{s_1+}^1(\bar \tau_{s_1}), Q_{s_2+}^1(\tau_{s_2})\}
\cr&=\f{4}{\kappa^4}i^{-s_1+s_2}\int_{S\times S'}\bar D^{s_1+2}\bar \tau_{s_1}(z)D^{s_2+2}\tau_{s_2}(z')\{\mathscr{M}_+(s_1,z),\mathscr{M}_+^*(s_2,z')\}
\cr&=\f{4\pi}{\kappa^2}i^{-s_1+s_2+1}\int_{S}\bar D^{s_1+2}\bar \tau_{s_1}(z)D^{s_2+2}\tau_{s_2}(z)\tilde R_{s_1s_2}.
\la{0ord}
\ee
This is a singular contribution, due to the presence of the matrix $\tilde R$.
In the same-helicity bracket, similar contributions are absent as $\{\mathscr{M}_{\pm}(n),\mathscr{M}_{\pm}(m)\}=0$.

\subsection{Dual mass extension of BMS symmetry}\la{sec:dualmass}

In this section, we use the results obtained above on the mixed-helicity bracket for charges of general spin-$s$ to study how the covariant dual mass charge can be included in the gravitational phase space at null infinity to extend the BMS  symmetry group. The goal is to find a closed subalgebra for ${\it real}$ charges of spin $s=0,1$.

In order to investigate whether the sector associated with charges of spin $s=0,1$ forms a closed subalgebra, 
we consider the set $(Q_0, \bar Q_0, Q_1, \bar Q_1)$. In this section, we consider the case where $\bar\tau_s= \tau_s^*$.
We start by studying the bracket $\{\bar Q_0(\bar \tau_0),\,Q_1(\tau_1')\}_+$. Its linear part is obtained by choosing  $s_1=0,\,s_2=1$ in \eqref{genexpr}, which yields

\be
\{\bar Q_{0+}(\bar \tau_0), Q_{1+}(\tau_1')\}^1
&=\f2{\kappa^2} \int_S \Bigg[\bigg(\bar D^2 \bar \tau_0D\tau_1' 
-2D \bar D^2 \bar \tau_0\tau_1' \bigg) \mathscr{M}_+  (0,z) + \bar \tau_0D^3\tau_1 \mathscr{M}^*_+  (0,z)\Bigg].
\la{barQ0Q1lin}
\ee
Since this bracket is not closed, under general assumptions, $\bar Q_0$ and $Q_1$ do not form an algebra. 
When we move to the second order $\{\bar Q_0, Q_1\}^2_+$, we obtain (see Appendix \ref{A.2})

\be
\{\bar Q_0(\bar \tau_0), Q_1(\tau_1')\}^2_+
&= \bar{Q}_{0+}^2(-\bar \tau_0 D\tau_1'+
2\tau_1'D\bar \tau_0)
={Q}_{0+}^2(-\bar \tau_0 D\tau_1'+
2\tau_1'D\bar \tau_0),
\la{quadmixed01}
\ee
whose structure is the same as the positive same-helicity bracket
\be
\{ Q_0( \tau_0), Q_1(\tau_1')\}^2_+
&={Q}_{0+}^2(- \tau_0 D\tau_1'+
2\tau_1'D \tau_0)
\la{quad01}
\ee
found in \cite{Freidel:2023gue}. This fact is not surprising, since the quadratic part of the charge aspect $q_0$ is real due to the boundary conditions on the News.\footnote{Starting with the expression of $q_0$ in the continuous basis (in terms of the radiation fields $N$ and $C$)
\be
q_0(z)&=\lim_{u\to-\infty}q_0(u,z)=\lim_{u\to-\infty}\cM_\C(u,z)
=\lim_{u\to-\infty} \bigg(\frac12 D^2 (\pa_u^{-1}{N})(u,z)  + \frac14 \pa_u^{-1}(C  \pa_u N)(u,z)\bigg)\,,
\la{q0}
\ee
we can write
\be
\bar q_0^2(u,z)&=\frac14 \pa_u^{-1}(\bar C  \pa_u{\bar N})(u,z)
\cr
&=\frac14 \pa_u^{-1}(\pa_u[\bar C {\bar N}]- (\pa_u{C}) N)(u,z)
\cr
&=\frac14 [\bar C {\bar N}- C {N}](u,z)+q_0^2(u,z),
\ee
 from which we can see that $q^2_0(u,z)$ and $\bar q^2_0(u,z)$ are the same thing up to a boundary term which vanishes in the limit $u\to -\infty$. As a consequence, the action of $\bar Q_0^2$ is the same as $Q_0^2$.}
 
As for the commutator $\{\bar Q_1(\bar \tau_1),\,Q_1(\tau_1')\}_+$,  we have
\be
\{\bar Q_{1+}(\bar \tau_{1}), Q_{1+}(\tau_{1}')\}^1
&=\f4{\kappa^2}i \int_S \Bigg[
\bar D^{3}D \bar \tau_{1}(z)\tau_{1}'(z)\mathscr{M}_+  (1,z)+
 D^{3} \bar D \tau_{1}'(z)\bar \tau_{1}(z)\mathscr{M}_+^*  (1,z)\Bigg],
 \la{linearmixed11}
\ee
and (see Appendix \ref{A.2})
\be
\{\bar Q_{1+}(\bar\tau_1), Q_{1+}(\tau'_1)\}^2
&=
\bQ^2_{1+}(2\tau'_1 D\bar\tau_1)
+Q^2_{1+}(2\bar\tau_1 \bar D\tau'_1 )
+
\f{8 i}{\pi k^2}\sum_{n=0}^\infty\int_{S} 
\bar D\tau'_1(z)   D \bar \tau_1(z) \mathscr{S}_+(n,z)\mathscr{M}^*_+(n,z)
.
\la{quadmixed11}
\ee
Even in this case, the bracket does not close.

We should also consider the $0th$-order contributions \eqref{0ord}. We observe that under the hypothesis $\bar D\tau'_1=0$  (to be applied after regularization of $\tilde R$), they both vanish, as
\be
\{\bar Q_{0}(\bar \tau_{0}), Q_{1}(\tau'_{1})\}_+^0&=-\f{4\pi}{\kappa^2}\int_{S}\bar D^{2}\bar \tau_{0}(z)D^{3}\tau'_{1}(z)\tilde R_{01}
=-\f{4\pi}{\kappa^2}\int_{S}\bar D(\bar D\bar \tau_{0}(z)D^{3}\tau'_{1}(z)\tilde R_{01})=0,
\la{zeroord01} \\
\{\bar Q_{1}(\bar \tau_{1}), Q_{1}(\tau'_{1})\}_+^0&=\f{4\pi}{\kappa^2}i\int_{S}\bar D^{3}\bar \tau_{1}(z)D^{3}\tau'_{1}(z)\tilde R_{11}
=\f{4\pi}{\kappa^2}i\int_{S}\bar D(\bar D^2\bar \tau_{1}(z)D^{3}\tau'_{1}(z)\tilde R_{11})=0.
\la{zeroord11}
\ee
Analogously,  imposing $D\bar\tau_1'=0$  kills hermitian conjugate counterparts of such brackets. Another way to get rid of these $0th$-order term is to simply require globality of one of the two sectors, for instance $ D^{s+2}\tau_s=0$. 

Motivated by these observations, in the following, we propose a set of restrictions on the smearing functions to get a closed mixed-helicity bracket for $s=0,1$. 
Let us start with the restriction
\be
\bar D\tau_1=0= D\bar\tau_1,
\la{holotau1}
\ee
which amounts to asking diffeomorphisms on $S$ to be meromorphic superrotations. 
In this case, the $0th$-order contributions of all brackets vanish.
In the sector $\bar 01$ the soft and hard contributions to the bracket, \eqref{barQ0Q1lin} and \eqref{quadmixed01}, reduce to
\be
\{\bar Q_{0}(\bar \tau_0), Q_{1}(\tau'_1)\}_+^1
&= \bar{Q}_{0+}^{1}(-\bar \tau_0 D\tau'_1+
2\tau'_1D\bar \tau_0)+\f2{\kappa^2} \int_S \bar\tau_0D^3\tau'_1 \mathscr{M}^*_+  (0,z)
\cr&={ \bar{Q}_{0+}^{1}(-\bar \tau_0 D\tau'_1+
2\tau'_1D\bar \tau_0)+Q_{0+}^1(-\bar \tau_0D\tau_1'+2\tau_1'D\bar\tau_0)+\f2{\kappa^2} \int_S D^2\bar\tau_0D\tau'_1 D\mathscr{M}^*_+  (0,z)},
\la{cocycle}
\\
\{\bar Q_0(\bar \tau_0), Q_1(\tau_1')\}^2_+
&= \bar{Q}_{0+}^2(-\bar \tau_0 D\tau'_1+
2\tau'_1D\bar \tau_0)
={Q}_{0+}^2(-\bar \tau_0 D\tau_1'+
2\tau_1'D\bar \tau_0)\la{quadb01},
\ee
which do not share the same structure, not only because the linear bracket does not close onto a charge, but also because it includes charge aspect pieces of both helicities. The last term in \eqref{cocycle} is reminiscent of a cocycle (field-dependent central extension). One could point out that, as we said before, at the quadratic order, the spin-0 charge aspect is real, so \eqref{quadb01} can be equivalently expressed as
\be
\{\bar Q_0(\bar \tau_0), Q_1(\tau_1')\}^2_+
= \f12[\bar{Q}_{0+}^2(-\bar \tau_0 D\tau'_1+
2\tau'_1D\bar \tau_0)+{Q}_{0+}^2(-\bar \tau_0 D\tau'_1+
2\tau'_1D\bar \tau_0)],
\ee
but even in this case, there would be a factor 2 mismatch between the charge parts of the linear and quadratic orders.  
The structure of the sector $0\bar1$ is given by the complex conjugates of \eqref{cocycle} and \eqref{quadb01}.

In the sector $\bar11$, we see that expressions \eqref{linearmixed11} and \eqref{quadmixed11} both vanish under this restriction, i.e.
\be
\{\bar Q_{1+}(\bar \tau_{1}), Q_{1+}(\tau_{1}')\}^1&=0,\\
\{\bar Q_{1+}(\bar \tau_{1}), Q_{1+}(\tau_{1}')\}^2&=0.
\ee
Although this seems to be a viable option yielding an algebra representation involving a possible 2-cocycle, we find that the Jacobi identities are violated at the $0th$-order. In fact, we have (see Appendix \ref{a:jac0obstr})
\be
\{\bar Q_1^1(\bar\tau_1),\{\bar Q_0(\bar\tau_0), Q_1(\tau_1)\}^1\}_++&\{Q_1^1(\tau_1),\{\bar Q_1(\bar\tau_1),\bar Q_0(\bar\tau_0) \}^1\}_++\{\bar Q_0^1(\bar\tau_0),\{Q_1(\tau_1),\bar Q_1(\bar\tau_1) \}^1\}_+
\cr&=-\f{4\pi}{\kappa^2}\int_{S}\bar D^{3}\bar\tau_1\bar\tau_0D^3\tau'_1\tilde R_{10}\neq0. \la{jac0obstr}
\ee
This happens because of the alleged cocycle term in \eqref{cocycle}, which violates the $0th$-order cocycle identity, thus implying the obstruction we find in the Jacobi identity \eqref{jac0obstr}. This suggests that the rightmost term in \eqref{cocycle} cannot be understood as a cocycle.
A possible way to remove this undesirable term is to further assume the globality conditions
\be
 D^3\tau_1=0=\bar D^3\bar \tau_1.
\la{dtau1}
\ee
In this case, the obstruction in 
\eqref{cocycle} vanishes, and a closed algebra is consistently built up including both helicity charges.
We stress, however, that on the 2-sphere without punctures the holomorphicity conditions \eqref{holotau1} 
are not independent assumptions, since they follow directly from  \eqref{dtau1} \cite{Cresto:2024fhd}. We will come back to the implications of different celestial sphere topologies at the end of this section.

Since we are looking for an extension of the real BMS algebra including the dual mass, we want to eventually restore the tensor notation where $A$ labels coordinates on $S$. To do this (see \cite{Freidel:2021dfs} for more details), we use the fact that positive and negative helicity scalars $O_s$ and $O_{-s}$ are obtained by contracting spin-$s$ tensors $O_{A_1\dots A_s}$ respectively with frame fields $m^A$ and $\bar m_A$, with $m^A\bm_A=1$, as
\be
O_s=m^{A_1}\cdots m^{A_s}O_{A_1\dots A_s}, \quad O_{-s}=\bar m_{A_1}\cdots \bar m_{A_s}O^{A_1\dots A_s}.
\ee
The frame fields define the sphere metric $q_{AB}$ and the volume form $\epsilon_{AB}$ as\footnote{{For the round sphere metric $\gamma_{AB}$ in complex coordinates, the normalization implies $m^A\p_A=P\p_z$ and $\bar m^A\p_A=P\p_\bz$.
Given $\gamma:=|\det\gamma_{AB}|$, we have
\be
\sqrt\gamma=P^{-2}, \qquad \gamma^{z\bz}=m^z\bar m^\bz=P^2, \qquad\epsilon^{z\bz}=-im^z\bar m^\bz=-iP^2.
\ee
}}
\be
q_{AB}=m_A\bm_B+m_B\bm_A\,,\quad
\epsilon_{AB}=-i(m_A\bm_B-m_B\bm_A).
\ee
We also use the definition of the action of the Cartan derivative $D=m^AD_A$ 
\be
DO_s=m^Am^{A_1}\cdots m^{A_s}D_AO_{A_1\dots A_s}.
\ee
{In the following, we will denote the symmetric, traceless part of a tensor $T_{AB}$ in the indices $A, B$ as
\be
T_{\langle AB\rangle}:=\f12(T_{AB}+T_{BA}-q_{AB}q^{CD}T_{CD}),
\ee}
and for a tensor $T_{ABC}$
\be\la{Tabc}
T_{\langle ABC \rangle}=\f13\left(T_{A\langle BC\rangle}+T_{\langle A|B|C\rangle}+T_{\langle AB\rangle C}-T_{ABC}\right).
\ee

In order to obtain the algebra of real charges in the $s=0,1$ sector, let us   decompose the smearing parameters  into real and imaginary parts as
\be
\tau_0&=T+i\tilde T\\
\tau_1&=\f{1}{2}(Y^A+i\tilde Y^A)\bar m_A.
\ee
The charge aspects $q_0,q_1$ correspond respectively to the  covariant complex mass and angular momentum aspects 
\be
q_0&=m+i\tm,
\la{q0m}\\
q_1&=j_A m^A,
\la{q1ja}\ee
where
\be
m(z)&=\lim_{u\to-\infty}\cM(u,z),
\la{m}\\
\tm(z)&=\lim_{u\to-\infty}\tcM(u,z),
\la{tm}\\
j_A(z)&=\lim_{u\to-\infty}\tfrac12(\cJ_A-u( D_A\cM+\tD_A \tcM))(u,z).
 \la{jA}
 \ee
The real charges read
\be
\Re Q_0(\tau_0)
&=\f12[
 Q_0(\tau_0)+\bar Q_0(\bar\tau_0)]
 =
\int_S (Tm-\tilde T\tilde m), \la{supert}\\
\Re Q_1(\tau_1)
&=\f12[ Q_1(\tau_1)+\bar Q_1(\bar\tau_1)]
=\int_S Y^A j_A.
\la{ReQ1}
\ee
The algebra of these real charges is then given by
\be
\{ \Re Q_0(\tau_0), \Re Q_1(\tau'_1)\}&=
\Re \left( Q_0(\tau'_1 D\tau_0
+\bar \tau'_1 \bar D \tau_0
-\f12\tau_0 (D \tau'_1
+ \bar D \bar\tau'_1))  \right)
\cr
&=\int_S\left[
\left(Y'^AD_A T-\f12TD_AY'^A \right)m
-\left(Y'^AD_A \tT-\f12\tT D_AY'^A \right)\tm
\right],
\la{RQ0}
\ee
and 
\be
\{ \Re Q_1(\tau_1), \Re Q_1(\tau'_1)\}&= \Re (Q_1(\tau'_1D\tau_1 - \tau_1D\tau'_1))
\cr&=-\int_S[Y,Y']^A j_A.
\la{RQ1}
\ee
Defining the total real charges
\be
Q^{\scriptstyle R}(T,\tT,Y)=
\Re Q_0(\tau_0)+\Re Q_1(\tau_1)
=\int_S (Tm-\tilde T\tilde m +Y^A j_A)(z),
\la{QR}
\ee
we  see from \eqref{RQ0}, \eqref{RQ1} that these satisfy the algebra
\be
	\{Q^{\scriptstyle R}(T_1,\tilde{T}_1, Y_1), Q^{\scriptstyle R}(T_2, \tilde{T}_2,Y_2)\}
	=- Q^{\scriptstyle R}(T_{12}, \tilde{T}_{12},Y_{12},),
\la{QRQR-2}
	\ee
where
\begin{subequations}
	\be
	T_{12}&=\f12 T_1 D_AY_2^A
    -Y^A_2D_AT_1-1\leftrightarrow 2
    ,\\
	\tT_{12}&=\f12 \tT_1 D_A Y_2^A
    -Y^A_2D_A\tT_1-1\leftrightarrow 2
    ,\\
    Y_{12}&= [Y_1,Y_2]_S.
    \ee
\la{TY12}
\end{subequations}

The algebra defined by \eqref{TY12} can be understood as the Lie bracket of the (leading order) {\it complex} vector fields
\be
 \xi_{(T,\tT,Y)} := (T+i\tT) \pa_u   + Y^A\pa_A  + \f12 D_AY^A( u\pa_u - r\pa_r)\,,
\la{xiA}
\ee
where $T$ labels the standard real super-translations while $\tT$ imaginary super-translations, and $Y$ conformal transformations on $S$. In fact, we have
\be
\left[\xi_{(T_1,\tT_1,Y_1)}, \xi_{(T_2,\tT_2,Y_2)} \right]_{\mathrm{Lie}} = \xi_{(T_{12},\tT_{12},Y_{12})}.
\ee
A connection between the action of an imaginary BMS supertranslation and the generation of a dual mass charge has previously been pointed out in
\cite{Huang:2019cja,Kol:2019nkc}. More precisely,  BMS super-translation transformation with a complex parameter does not represent a residual diffeomorphism symmetry but rather a map between different spacetime solutions, like the Schwarzschild metric and the Taub--NUT metric \cite{Talbot:1969bpa}, where the NUT charge corresponds to the dual mass charge aspect. For a review of complex coordinate transformations, see, e.g., \cite{Adamo:2014baa,Erbin:2016lzq}.

Standard (real) super-translations are generated by the real part of $Q_0(\tau_0)$ \eqref{supert}.
In fact, if we decompose the shear as
\be
C_{AB}=\bar m_A\bar m_B  C+ m_A m_B  \bar C,
\ee
we have
\be
\{Q_0(\tau_0),C(u,z)\}&=\tau_0\partial_uC(u,z)-2D^2\tau_0, \\
\{Q_0(\tau_0),\bar C(u,z)\}&=\tau_0\partial_u\bar C(u,z).
\ee
In tensor notation, then we obtain that the action of the real  charge \eqref{supert} on the {\it real} shear tensor is given by
\be
\left\{\Re Q_0(\tau_0),C_{AB}(u,z)\right\}&=T\partial_u C_{AB}(u,z)-D_{\langle A} D_{B\rangle}T-D_{\langle A}\tilde D_{B\rangle}\tilde T,
\la{CAB}
\ee
where we use the complex structure $\epsilon_{A}{}^B$ on $S$---defined through the volume form $\epsilon_{AB}$---to define the duality operation\footnote{
Useful relations are
\begin{subequations}
\be
m_A\bar m^B&=\f{1}{2}(\delta_A^B+i{\epsilon_A}^B), \\
 D_A\tilde D_B&=\tilde D_B D_A-q_{AB}q^{CD}\tilde D_C D_D, \\
 \tilde D_A\tilde D_B&=- D_B D_A+q_{AB}q^{CD}D_C D_D,\\
 D_AD_B&=\tilde D_A\tilde D_B+2D_{\langle A} D_{B\rangle},
 \ee
\end{subequations}
where the brackets $\langle \cdot\rangle$ denote the symmetric, trace-free part of a tensor.
}
	\be\la{dualC}
	 {\tilde D}_A=\epsilon_{A}{}^B D_B.
	\ee
The transformation \eqref{CAB} shows how inclusion of the dual mass charge in the gravitational phase space modifies the action of super-translations generated by the sole contribution of the covariant mass in \eqref{supert}. This modification allows us to
 recover the correct action on the shear, once we impose  extra conditions on the super-translation parameters given by  the Cauchy--Riemann equations
\be
\pa_A T=\tilde \pa_A \tT \quad \leftrightarrow \quad \tilde\pa_A T=- \pa_A \tT.
\la{TtT}
\ee
These conditions follow from the requirement of holomorphicity of $\tau_0$ as well, namely
\be
\bar D\tau_0=0= D\bar\tau_0.
\la{holotau0}
\ee
We observe that conditions \eqref{TtT} imply
\be
\triangle T:=D^AD_AT=0, \quad  \triangle \tT:=D^AD_A\tT=0.
\la{LapT}
\ee
In this case, then, we obtain that the Hamiltonian action of the real charge \eqref{supert} on the radiative phase space reproduces the variation of the metric under real super-translations, namely
\be
\left\{\Re Q_0(\tau_0),C_{AB}(u,z)\right\}
&=T\partial_u C_{AB}(u,z)-2D_{\langle A} D_{B\rangle}T
\cr
&=\d_T C_{AB}(u,z).
\la{dTC}
\ee
Without the inclusion of the dual mass term in \eqref{supert}, the rightmost contribution in \eqref{CAB} would not be there, and one would obtain a mismatch in the inhomogeneous (memory) part by a factor of 2.

 This problem was originally remedied in
\cite{He:2014laa} by enlarging the phase space to include the corner soft graviton at $u=\pm\infty$ and introducing a boundary-corner Poisson bracket for its action on the finite-$u$ shear field, implementing the constraint $\tcM|_{u=\pm\infty}=0$. Our analysis shows that there is no need to impose such a constraint at spatial infinity, where a non-vanishing dual mass aspect can be accommodated and used to define a modified real super-translation charge whose Hamiltonian action reproduces the real variation (standard super-translations) of the metric under the (bulk extension of the) vector field \eqref{xiA}.

For completeness,  imaginary super-translations generated by the (bulk extension of the) vector field \eqref{xiA} are instead  represented by the imaginary part of $Q_0(\tau_0)$
 \be
\Im Q_0(\tau_0)&=\f{Q_0(\tau_0)-\bar Q_0(\bar \tau_0)}2
=i\int_S(\tT m+ T\tm),
\la{ImQ0}
 \ee
as 
\be
\left\{\Im Q_0(\tau_0),C_{AB}(u,z)\right\}&=i\tT\partial_u C_{AB}(u,z)-iD_{\langle A} D_{B\rangle}\tT+iD_{\langle A}\tilde D_{B\rangle} T
\cr
&=i\left(\tT\partial_u C_{AB}(u,z)-2D_{\langle A} D_{B\rangle}\tT\right)
\cr
&=\d_{i\tT} C_{AB}.
\ee

In order to understand the structure of the dual mass extension of the BMS algebra, which we call {\it dual} BMS for short and denote $\dbms$, it is important to analyze the implications of the restrictions \eqref{holotau1}, \eqref{dtau1}, \eqref{holotau0} (or equivalently \eqref{LapT}) for different topologies of the cut $S$.
Let us denote $S_n$ the sphere with $n$ points removed. 

\begin{itemize}

\item
On the sphere $S=S_0$, as already pointed out above, the two sets of restrictions \eqref{dtau1} and \eqref{holotau1} are not independent. Solution to \eqref{dtau1} implies that the vector fields $Y^A$ are global conformal Killing vectors, generating the Lorentz subalgebra of $\diff(S)$. The   rotation and boost generators are respectively given by \cite{Barnich:2011mi,Compere:2019gft,Freidel:2024jyf}
\be
\tY^A_{i}&:=\epsilon^{AB}\pa_B n_i, \qquad 
    Y^A_{i}:=  q^{AB}\pa_B n_i,
\ee
where the unit vector $n^i$ represents a point on the sphere.\footnote{
For the round sphere
$
 n^i:=(\sin{\theta}\cos{\varphi},  \sin{\theta}\sin{\varphi}, \cos{\theta})
$. } The real charge \eqref{ReQ1} consists of the set of the six Lorentz generators. 

The further restrictions \eqref{LapT} on the compact sphere admit as continuous solutions only constant functions. This implies that the dual mass contribution to the real charge \eqref{supert} vanishes, as $\tm$ contains only a soft (total derivative) contribution. We are thus left with only four global translations. 

It follows that, on the sphere $S_0$, the real charges \eqref{supert}, \eqref{ReQ1} generate the  Poincar\'e algebra. 
In other words, the dual mass charge cannot be activated on a celestial sphere of compact topology. This confirms the {\it topological} nature of the dual mass charge, as already pointed out in \cite{Kol:2019nkc}.

\item 
In the case where the celestial sphere is punctured, dual super-translations can be non-trivially represented. Let us consider the complex plane, namely $S=S_1\equiv \C$. In this case, tangent diffeomorphisms are still restricted to the global conformal group, with $\tau_1=\sum_{n=0}^2 c_n z^n, c_n\in \C$, but $T,\tT$ no longer need to be constants;  they are harmonic functions conjugate to each other through the 
Cauchy--Riemann equations \eqref{TtT}. 

In this case, then, the total real charges  \eqref{QR} provide a representation of the dual  BMS algebra given by
\be
\dbms
=\textsf{so}(3,1) \sds \, (\R_T^S\oplus \R_\tT^S),
\la{dbms}
\ee
where $T, \tT$ are real functions on $S$ that parametrize, respectively, super-translations and dual super-translations, and commute with each other.
\end{itemize}

\subsection{Non-radiative phase space and $\dbms$ moment map} \la{sec:moment}

We saw that, for a non-trivial topology of the celestial sphere, the dual mass extension of the BMS group gives rise to the Lie algebra $\dbms$ \eqref{dbms}, with its dual $\ddbms$ parametrized by the three aspects $(\mathsf{m},\mathsf{\tm}, \mathsf{j})$ and  canonical pairing   $\langle\cdot|\cdot\rangle:\dbms\times\ddbms\to\R$  given by
\be
    \langle \mathsf{m},\mathsf{\tm}, \mathsf{j} | T,\tT,Y \rangle %
    =\int_{S} (T\mathsf{m}-\tT\mathsf{\tm}+Y^A\mathsf{j}_A)
 .
\ee
This pairing allows us to define the infinitesimal coadjoint action of $(Y,T,\tT)\in \dbms$ on $(\mathsf{m},\mathsf{\tm}, \mathsf{j} ) \in \ddbms$ from
\begin{equation}\la{coadj}
		\langle \delta_{(  T_1,\tT_1,Y_1)} (\mathsf{m},\mathsf{\tm}, \mathsf{j})| T_2,\tT_2,Y_2 \rangle =-\langle \mathsf{m},\mathsf{\tm}, \mathsf{j} | T_{12},\tT_{12}, Y_{12} \rangle.
\end{equation}
From this definition, using \eqref{TY12}, we obtain  the coadjoint action
\begin{subequations}
\be
\delta_{(  T,\tT,Y)} \mathsf{m}&= \left[\cL_Y + \f32 D_AY^A \right] \mathsf{m},\\
\delta_{(  T,\tT,Y)} \mathsf{\tm} &= \left[\cL_Y + \f32 D_AY^A \right] \mathsf{\tm},\\
\la{dTY-1c}
\delta_{(  T,\tT,Y)} \mathsf{j}_A &=\left[ \cL_{Y} +  D_AY^A\right] \mathsf{j}_A + 
\f32 \left(\mathsf{m} \pa_A T - \mathsf{\tm} \pa_A \tilde T\right) 
+ \f 12 \left(T \p_A \mathsf{m} -\tT{\p}_A \mathsf{\tm} \right).
\ee
\la{dTY-1}
\end{subequations}

These algebraic properties of the dual BMS algebra $\dbms$ can be connected to quantities of a gravitational phase space $\Gamma$, carrying the Hamiltonian action of $\dbms$,   through the identification of a moment map $\mu_{\dbms}:\Gamma\to\dbms$.
The first step is thus the identification of such a phase space. 
{Before proceeding, though, an important clarification is in order. The restrictions $D^3\tau_1=0$, $\bar D\tau_1$,
translate into the following constraints on $Y^E$
\be
\bar D\tau_1=0 &\leftrightarrow D_{\langle A}Y_{B\rangle}=0,
\la{cY1}
\\
 D^3\tau_1=0 &\leftrightarrow
 D_{\langle A} D_B D_{C\rangle}Y^C=0.
 \la{cY2}
\ee
Since we have these constraints, the coadjoint action of $\dbms$ on $\mathsf{j}$ is determined up to an equivalence class relation
\be
\delta_{(  T,\tT,Y)} \mathsf{j}_A\sim \delta_{(  T,\tT,Y)} \mathsf{j}'_A=\delta_{(  T,\tT,Y)} \mathsf{j}_A
+D_{\langle A} D_B D_{C\rangle}\boldsymbol{\alpha}^{BC}
+D^C\boldsymbol\beta_{CA},
\la{equiv}
\ee
where $\boldsymbol\beta_{CA}$ is symmetric-traceless.
 This means that the phase space element associated to $\mathsf{j}$ by the moment map is determined by comparing its Hamiltonian transformation with a given element of the equivalence class \eqref{equiv}.
}

As argued in \cite{Freidel:2024jyf}, the natural candidate for $\Gamma$ is the non-radiative phase space $\Gamma_{\rm NR}$ implementing the conditions 
\cite{Freidel:2021qpz}
\be
\cQ_{-2}=0=\cQ_{-1}.
\la{NRcond}
\ee
In $\Gamma_{\rm NR}$, the evolution equations \eqref{eeom} for $s=0,1$ in vectorial form correspond to
\be
\dot \cM = 0=\dot \tcM\,,\qquad 
\dot\cJ_A =  D_A\cM+\tD_A \tcM.
\la{ceom-P}
\ee
From \eqref{ceom-P}, it is immediate to see that we can parametrize
 the non-radiative corner phase space in terms of the $u$-independent  aspects $(m,\tm, j_A) $,
 where
 \be
 m=\cM\,,\quad \tm=\tcM\,,\quad
j_A=\tfrac12(\cJ_A-u( D_A\cM+\tD_A \tcM)).
 \la{hatJ}
 \ee
 The non-radiative phase space $\Gamma_{\rm NR}$  then corresponds to functionals of these $u$-independent aspects defined on the cut $S$ where \eqref{NRcond} holds. 
An example of a non-radiative cut is represented by spatial infinity, namely $u=-\infty$, where  \eqref{NRcond} holds due to our choice of boundary conditions.  This is also the cut where the charges \eqref{Qs} are defined,  providing a non-radiative phase space where their Hamiltonian action can be computed using the discrete basis.

Within this
$\Gamma_{\rm NR}$, the variables \eqref{hatJ} satisfy the following corner symmetry transformations canonically generated by the total real charges \eqref{QR}
(see Appendix \ref{VectBra})
\begin{subequations}
\be
\d_{(T,\tT,Y)} {m} =
\{Q^{\scriptstyle R}(T,\tT,Y), m\}
&=
\left[\cL_Y + \tfrac 32 D_AY^A \right]{m}
, \\
\delta_{(T,\tT,Y)}  {\tm} =
\{Q^{\scriptstyle R}(T,\tT,Y), \tm\}
&=
\left[ {\cL}_Y +\tfrac 32 D_AY^A\right] {\tm} 
, \\
\d_{(T,\tT,Y)} j_A = 
\{Q^{\scriptstyle R}(T,\tT,Y), j_A\}
&=\left[ \cL_{Y} +  D_AY^A\right] j_A + 
\tfrac32 (m  \pa_A T - \tm \pa_A \tilde T)
 + \tfrac 12 \left(T\p_A m -\tT{\p}_A \tm  \right)
 \la{QjA}
 \\
 &+{\rm extra ~terms}
 \nonumber.
\ee
\la{dTY-2}
\end{subequations}
{The extra terms appearing in \eqref{QjA} correspond to total derivative contributions exactly of the form given in the RHS of \eqref{equiv}, whose explicit expressions can be found in  \eqref{RQpA}. }

By comparing these symmetry transformations with the  infinitesimal coadjoint action \eqref{dTY-1}, modulo the equivalence class relation \eqref{equiv}, it is then immediate to find to moment map $\mu_{\dbms}$ to be given by
\be
\mu_{\dbms}(m)=\mathsf{m}\,,\quad 
\mu_{\dbms}(\tm)=\mathsf{\tm}\,,\quad 
\mu_{\dbms}(j_A)=\mathsf{j_A}\,.
\ee

\subsection{Restricting only one helicity sector}\la{sec:asy}

In preparation for the general spin case treated below, let us illustrate an alternative way to get a closed bracket by breaking the symmetry between the two helicity sectors. Going back to the case $\bar\tau_s\neq\tau_s^*$,
 one can consider the following conditions 
\be
 D^2  \tau_0=0\,,
\quad
 D^3 \tau_1=0,
 \la{Dtt01}
\ee
which mean that we assume the smearing functions of positive-helicity charge aspects to be global.

By contrast, the smearing functions of the negative-helicity set remain completely general.

Condition \eqref{Dtt01}, extended to all spins, namely
\be\la{wedge}
D^{s+2}\tau_s=0,
\ee
is also known as the \emph{wedge restriction}, and it is associated with the covariant wedge algebra of gravity \cite{Cresto:2024fhd, Cresto:2024mne}, represented by the bracket of charges of the same helicity.
In this work, this condition will prove to be crucial also in the mixed-helicity sector. In fact, as we will show in Section \ref{shadsec}, if we impose condition \eqref{wedge} on positive-helicity charges, while leaving the negative-helicity ones unconstrained,
a closed mixed-helicity bracket can be obtained for all spins. This will be achieved at the expense of introducing a non-local operator (in the angular variables) connected to the shadow transform. Introducing non-locality is inevitable to close the linear-order of the bracket when $s_1+s_2-1\ge 2$, since one does not have enough orders of angular derivatives $\bar D$ to reproduce $\bar q^1_{s_1+s_2-1}$ from $\bar N_{s_1+s_2-1}$ in the expression \eqref{eqcomplete}. 
Nevertheless, this is not the case of the subsector $s_1,s_2\le 1$. In this case, one can construct a closed mixed-helicity bracket without the introduction of non-localities, by further restricting smearing parameters to be holomorphic, namely, by imposing
\be
 D^2  \tau_0=0=\bar D\tau_0\,,
\quad
 D^3 \tau_1=0
 =\bar D  \tau_1.
 \la{Dtauv2}
\ee
Under these restrictions, {from the linear-order brackets \eqref{barQ0Q1lin} and \eqref{linearmixed11},} we obtain
\begin{subequations}
\be
\{ Q_0(\tau_0),\bar Q_1(\bar\tau_1)\}^1_+&=\bar Q_{0+}^1(-\tau_0\bar D\bar\tau_1),\la{Q0bQ1}
\\
\{ \bar Q_0(\bar\tau_0), Q_1(\tau_1)\}_+^1&=\bar Q_{0+}^1(-\bar\tau_0 D\tau_1+2\tau_1 D\bar\tau_0),\la{bQ0Q1}
\\
\{ \bar Q_1(\bar\tau_1), Q_1(\tau_1)\}_+^1&=\bar Q_{1+}^1(2\tau_1 D\bar\tau_1).
\la{bQ1Q1}
\ee
\la{QQv2}
\end{subequations}

 Clearly, one could have chosen the opposite situation, where it is the positive-helicity sector to remain unconstrained. 

The viability of the set of restrictions \eqref{Dtauv2} is supported by the fact that
the quadratic-order contribution to these charge brackets, following from \eqref{quadmixed01}, \eqref{quadmixed11}, reproduces the same structure \eqref{QQv2}, 
and the Jacobi identity is satisfied up to the linear order. {Moreover, the wedge restriction also conveniently kills the 0th order contributions to the bracket arising in \eqref{zeroord01} and \eqref{zeroord11}.}


\subsection{A closed bracket for all spins} \la{shadsec}

After this long excursion into the $s=0,1$ sector, let us return to the general spin case.
When inspecting the expression \eqref{eqcomplete} for the mixed-helicity bracket, we notice that recovering a charge aspect from the first term would require acting on $\bar N_{s_1+s_2-1}$ with $\bar D^{\,s_1+s_2+1}$, whereas only $s_1+2$ powers of $\bar D$ are available. This seems to indicate that a closed structure cannot be achieved for $s_2\ge2$. By the same argument, the second term cannot be closed into a charge whenever $s_1\ge2$.  
Nevertheless, a closed structure can still be obtained if one introduces a notion of angular anti-derivative, at the cost of allowing non-local smearing functions to appear in the bracket. Below, we propose a procedure that realizes precisely this idea.

We introduce a \emph{propagator} $G_n(z_1;z_2)$, defined by
\be\la{defG} D^n_{2}G_n(z_1;z_2)=\delta^2(z_1,z_2), \ee
together with its complex conjugate
$\bar G_n(z_1;z_2):=(G_n(z_1;z_2))^*$.
This propagator satisfies 
the following identities (see Appendix \ref{Prop}):
\begin{subequations}
\be G_{n}(z_2;z_1)&=(-)^nG_{n}(z_1;z_2);\\ 
D_1^nG_1(z_1;z_2)&=(-)^nD_2^nG_1(z_1;z_2). \ee
\end{subequations}
{Let us denote $V_{(\Delta,J)}$ the $SL(2,\mathbb{C})$ representation acting on $L^2(\mathbb{C)}$, labelled by the pair $(\Delta,J)$, where $\Delta$ is the conformal weight and $J$ is the helicity.}  Associated with $G_n$, we define the multiple anti-derivative operators  
$D^{-n}:V_{(\Delta,J)}\rightarrow V_{(\Delta-n,J-n)}$ and  
$\bar D^{-n}:V_{(\Delta,J)}\rightarrow V_{(\Delta-n,J+n)}$,  
for any $n\in\mathbb{N}$, acting as 
\be D^{-n}f(z)=\int_{S_1}G_{n}(z_1;z)f(z_1), \quad \bar D^{-n}f(z)=\int_{S_1}\bar G_{n}(z_1;z)f(z_1). \ee

Using these definitions,\footnote{
A key identity is
\(
\displaystyle
\int_S g(z)\,D^{-n}f(z)
   =\int_S g(z)\!\int_{S_1}G_{n}(z_1;z)f(z_1)
   =(-)^n\!\int_{S} f(z)\,D^{-n} g(z).
\)}
the mixed-helicity bracket \eqref{eqcomplete} becomes
\be\la{mhbphi} \{\bar Q_{s_1}(\bar \tau_{s_1}), Q_{s_2}(\tau_{s_2})\}^1 &=\bar Q^1_{s_1+s_2-1}\Big(\bar D^{-(s_1+s_2+1)}\bar \varphi[\bar\tau_{s_1},\tau_{s_2}]\Big) -Q^1_{s_1+s_2-1}\Big( D^{-(s_1+s_2+1)} \varphi[\tau_{s_2},\bar\tau_{s_1}]\Big), \ee
where
$\varphi[\tau_{s_2},\bar\tau_{s_1}]\in V_{(s_1+s_2,2)}$ and $\bar\varphi[\bar\tau_{s_1},\tau_{s_2}]\in V_{(s_1+s_2,-2)}$
are defined by
\be \varphi[\tau_{s_2},\bar\tau_{s_1}]&=\sum_{l=0}^{s_1}(l+1)\f{(s_1+s_2-2)_{s_1-l}}{(s_1-l)!}\bar D^l D^{s_2+2}\tau_{s_2}\bar D^{s_1-l}\bar \tau_{s_1}, \\ \bar\varphi[\bar\tau_{s_1},\tau_{s_2}]&=\sum_{l=0}^{s_2}(l+1)\f{(s_1+s_2-2)_{s_2-l}}{(s_2-l)!} D^l \bar D^{s_1+2}\bar\tau_{s_1} D^{s_2-l} \tau_{s_2}. \ee

Expression \eqref{mhbphi} gives the general form of the mixed-helicity bracket; however, no symmetry structure is immediately apparent. As we will see, interesting symmetry features emerge once specific restrictions are imposed on the smearing functions. Before turning to these constraints, it is useful to recall the celestial diamond structure to introduce a few additional definitions.

Charge aspects and soft gravitons fit into finite-dimensional global conformal multiplets in 2d celestial CFT (see \cite{Pasterski:2021fjn, Pasterski:2021dqe, Guevara:2021abz,Pano:2023slc,Freidel:2021ytz} for more details), whose structure is conveniently illustrated by celestial diamonds in the $(\Delta, J)$ plane. Two families of such diamonds can be constructed, depending on the helicity $s$ of the charges.

\begin{figure}[h!]
     \centering
      \begin{subfigure}[t]{0.45\textwidth}
      \centering
\includegraphics[width=\linewidth,height=50mm, keepaspectratio]{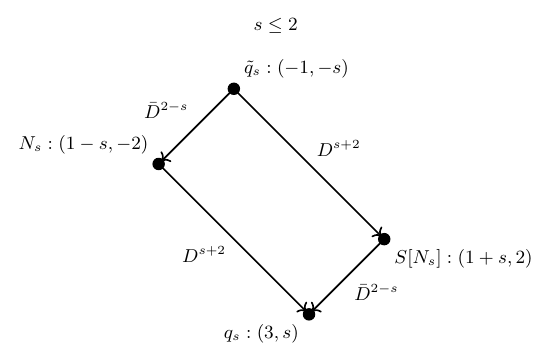}
\caption{}
 \label{fig:CD-GR-2}  
    \end{subfigure}
\quad
 \begin{subfigure}[t]{0.45\textwidth}
      \centering
\includegraphics[width=\linewidth,height=50mm, keepaspectratio]{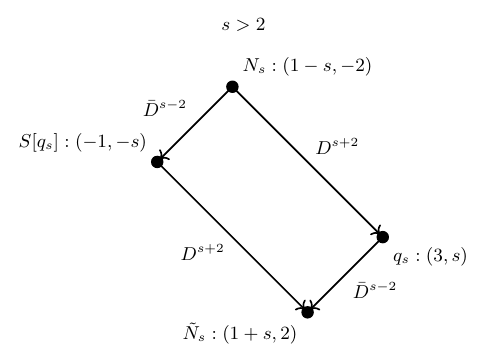}
\caption{}
 \label{fig:CD-GR-1}
    \end{subfigure}
    \caption{ Celestial diamond associated with negative-helicity soft gravitons with (\subref{fig:CD-GR-2}) $s \le2$ and 
    (\subref{fig:CD-GR-1}) $s> 2$.}
    \label{fig:CD-GR}
\end{figure}

For $s\le2$ (see Fig.~\ref{fig:CD-GR-2}), the upper vertex is the \emph{dual charge aspect} $\tilde q_s$, from which all other vertices descend via successive covariant derivatives on the celestial sphere (equipped here with the unit-sphere metric).  
The left and right corners correspond respectively to the sub$^{s}$-leading soft gravitons and to their shadow transforms, while the bottom vertex is the charge aspect $q_s$.\footnote{
One can show that
\(
\bar D^{\,2-s} D^{\,s+2}\tilde q_s = D^{\,s+2}\bar D^{\,2-s}\tilde q_s,
\)
so the two arrows commute and their combined action produces the same bottom vertex $q_s$.
}

In the regime $s\ge2$ ($s=2$ being the limiting case in which the diamond degenerates into a line), the structure is reversed (see Fig.~\ref{fig:CD-GR-1}).  
The top corners are now the sub$^{s}$-leading soft gravitons, the left and right corners the charge aspects and their shadow transforms, and the bottom vertex corresponds to the \emph{dual soft graviton} $\tilde N_s$.\footnote{
Here as well, the order of the two derivative operators does not affect the final result.
}

From the celestial-diamond relations, we obtain
\be
 D^{-(s+2)}\bar D^{\,s-2} q_s =
\begin{cases}
    S[q_s], & s\ge 3, \\
    \tilde q_s, & s\le 2,
\end{cases}
\la{tildeq}
\ee
where the operators $\bar D^{\,s-2}$ and $D^{-(s+2)}$ commute when acting on $q_s$ (this is proven for $s\le2$ in Appendix~\ref{Prop}, while the case $s\ge3$ follows from analogous manipulations). 

This observation motivates the introduction of transformation operators $\mS$ and $\bar\mS$, acting on $SL(2,\C)$ representations $V_{(\Delta,J)}$ according to $\mS:V_{(3,s)}\to V_{(-1,-s)}$, $\bar\mS:V_{(3,-s)}\to V_{(-1,s)}$. Their action on $f_{(3,\pm s)}\in V_{(3,\pm s)}$
is defined by
\be\la{shad} 
\mS[f_{(3,s)}]=D^{-(s+2)}\bar D^{s-2}f_{(3,s)},
\quad 
\bar\mS[f_{(3,-s)}](z)=\bar D^{-(s+2)}D^{s-2}f_{(3,-s)}.
\ee
The inverse maps $\mS^{-1}:V_{(-1,-s)}\rightarrow V_{(3,s)}$, $\bar\mS^{-1}:V_{(-1,s)}\rightarrow V_{(3,-s)}$ acting on $f_{(-1,-s)}\in V_{(-1,-s)}$ and $f_{(-1,s)}\in V_{(-1,s)}$ are given by
\be\la{shadi}
\mS^{-1}[f_{(-1,-s)}]=\bar D^{-(s-2)}D^{s+2}f_{(-1,-s)},
\quad
 \bar\mS^{-1}[f_{(-1,s)}]= D^{-(s-2)}\bar D^{s+2}f_{(-1,s)}.
\ee

When the celestial sphere is flattened to a plane, the propagator takes the form $G_n({z_1};z)=\f{1}{2\pi(n-1)!}\f{(z-{z_1})^{n-1}}{\bar z-\bar {z_1}}$, and for $s\ge3$ all the above expressions reduce to the familiar definition of the \emph{shadow transform} \cite{Ferrara:1972uq,Simmons-Duffin:2012juh}. 
Using the maps \eqref{shad}, the relation \eqref{tildeq} can be recast as
\be
\mS[q_s]&=\begin{cases}
  \tilde q_s \quad &\text{for} \quad s\le2, \\
S[q_s]\quad &\text{for} \quad s\ge3.  
\end{cases} \la{sca}
\ee
We thus see that $\mS$ yields the dual charge aspects for low spin and the standard shadow transform for higher-spin.

In direct analogy with the case of higher-spin charges, we may integrate these transformed charge aspects against new smearing functions $\tilde\tau_s$ with $(\Delta,J)=(3,s)$, thereby defining the corresponding \emph{shadow charges} (with a mild abuse of nomenclature), denoted $\mS[Q_s](\tilde\tau_s)$
\be
\mS[ Q_s](\tilde \tau_s)=\f8\k\int_S\tilde \tau_s\mS[ q_s]. \la{sc}
\ee

From \eqref{sca}, \eqref{shad} and \eqref{shadi}, we obtain the following relations among charges and their shadow
\be\la{shadsmearb}
 \mS[Q_s](\tilde \tau_s)=Q_s(\mS[\tilde\tau_s]), 
 \quad
 Q_s(\tau_s)=
  \mS[Q_s](\mS^{-1}[\tau_s]).
\ee
The corresponding relations involving $\bar\mS$ and $\bar\mS^{-1}$ follow by taking the complex conjugates of \eqref{shadsmearb}.

We now consider the brackets among the charges  $Q_s(\tau_s),
\bar Q_s(\bar \tau_s)
$ in the regime where the positive-helicity sector is global---again, this is just an arbitrary choice, as one could have taken the negative-helicity sector to be global instead---, namely when 
\be\la{glob}
D^{s+2}\tau_s=0\,\quad\forall s.
\ee
In this case, using the relations above, one finds---after a lengthy calculation detailed in Appendix~\ref{A.6})---that the linear-order bracket \eqref{mhbphi} reduces to
\be
\{\bar Q_{s_1}(\bar \tau_{s_1}), Q_{s_2}(\tau_{s_2})\}^1=\bar{ Q}^1_{s_1+s_2-1}\left(\bar \mS[(s_2+1)\tau_{s_2}D\bar{ \mS}^{-1} [\bar \tau_{s_1}]
- (s_1-3)
\bar{ \mS}^{-1} [\bar \tau_{s_1}]D\tau_{s_2}]\right).\la{bqq} 
\ee
This expression is reminiscent of the $w_{1+\infty}$ structure appearing in the same-helicity bracket \eqref{winf}.
The first novelty lies in the appearance of the maps $\bar \mS, \bar \mS^{-1}$ ensuring that the resulting smearing function on the RHS carries the correct conformal weight and spin $(\Delta,J)=(-1,s_1+s_2-1)$. Indeed,
\be
\tau_{s_2}\in V_{(-1,-s_2)}, \quad  \bar\tau_{s_1}\in V_{(-1,s_1)}\implies \bar \mS^{-1}[\bar\tau_{s_1}]\in V_{(3,-s_1)},
\ee
and the Poisson bracket then produces
\be
\tau_{s_2}D \mS^{-1}[\bar\tau_{s_1}],\, \mS^{-1}[\bar\tau_{s_1}] D\tau_{s_2} \in V_{(3,-s_1-s_2+1)},
\ee
which is mapped back to $V_{(-1,s_1+s_2-1)}$ by $\bar\mS$. Another distinction lies in the coefficients. In \eqref{bqq}, the coefficients $(s_2+1)$ and $(s_1-3)$ correspond to the weights  $-2 h=-(\Delta+J)$
associated, respectively, with the smearing parameters $\tau_{s_2}$ and $\bar{\mS}^{-1}[\bar\tau_{s_1}]$. By contrast, the same-helicity bracket expression \eqref{winf} is fully antisymmetric under the exchange $s_1\leftrightarrow s_2$, and its coefficients $(s_2+1)$ and $(s_1+1)$ match the conformal weights of $\tau_{s_2}$ and $\tau_{s_1}$.

For completeness, if, instead of the positive-helicity sector, we had imposed the globality condition on the negative-helicity sector, i.e.
\be
\bar D^{s+2}\bar\tau_s=0\,\quad\forall s,
\ee
we would have
\be
\{\bar Q_{s_1}(\bar \tau_{s_1}), Q_{s_2}(\tau_{s_2})\}^1&=-{ Q}^1_{s_1+s_2-1}\left( \mS[(s_1+1)\bar\tau_{s_1}\bar D{ \mS}^{-1} [ \tau_{s_2}]
- (s_2-3)
{ \mS}^{-1} [ \tau_{s_2}]\bar D\bar \tau_{s_1}]\right).\la{bqqneg} 
\ee

A natural question to ask is whether the structure \eqref{bqq} persists at second order in the bracket. 
In Appendix \ref{Shad2}, we have shown that this occurs for spins $s_1,s_2\le 1$ only when the additional holomorphicity condition $\bar D\tau_{s}=0$ is imposed on the smearing functions.\footnote{
{This is not surprising, since} the algebra 
\eqref{bqq} for $s_1,s_2\le 1$, under the additional assumption $\bar D\tau_{s}=0$, 
is equivalent to 
the local form
\eqref{QQv2}, 
{which survives at the quadratic order.}} We leave the investigation for higher spins to future work, considering also the difficulty to address this question already in the same helicity sector \cite{Freidel:2023gue}.

\subsubsection{Jacobi identities}
The closed charge bracket we derived at linear order is meaningful only if it satisfies the Jacobi identities. Two types of mixed-helicity Jacobi identities need to be checked. The first involves two $Q$-charges and one $\bar Q$-charge:
\be
&\{\bar Q_{s_1}(\bar\tau_{s_1}),\{Q_{s_2}(\tau_{s_2}),Q_{s_3}(\tau_{s_3})\}\}^1
\cr
+&\{Q_{s_2}(\tau_{s_2}),\{Q_{s_3}(\tau_{s_3}),\bar Q_{s_1}(\bar\tau_{s_1})\}\}^1
\cr
+&\{Q_{s_3}(\tau_{s_3}),\{\bar Q_{s_1}(\bar\tau_{s_1}),Q_{s_2}(\tau_{s_2})\}\}^1=0\,,
\la{2q1bq}
\ee
while the second type contains two $\bar Q$-charges and one $Q$-charge:
\be
&\{ Q_{s_1}(\tau_{s_1}),\{\bar Q_{s_2}(\bar\tau_{s_2}),\bar Q_{s_3}(\bar\tau_{s_3})\}\}^1
\cr
+&\{\bar Q_{s_2}(\bar \tau_{s_2}),\{\bar Q_{s_3}(\bar \tau_{s_3}),Q_{s_1}(\tau_{s_1})\}\}^1
\cr
+&\{\bar Q_{s_3}(\bar \tau_{s_3}),\{Q_{s_1}(\tau_{s_1}),\bar Q_{s_2}(\bar \tau_{s_2})\}\}^1=0.
\la{2bq1q}
\ee
Since only the positive-helicity sector of charges is global, these two types of identities are qualitatively different. We have been able to prove \eqref{2q1bq} in Appendix~\ref{JI}, relying only on the assumption (strongly motivated by the evidence provided in \cite{Freidel:2023gue}) that the $w_{1+\infty}$ structure of the same-helicity bracket \eqref{winf} persists at second order---we recall that the condition \eqref{glob} of globality implies that linear charges in the positive-helicity sector vanish.

By contrast, verifying \eqref{2bq1q} requires the explicit form of the mixed-helicity bracket at second order.  

Even if one assumes the validity of the algebra structure \eqref{bqq} at second order as well, computing the Jacobi identities is highly non-trivial, since it involves the action of the maps $\bar \mS$ and $\bar \mS^{-1}$ on products of functions, and disentangling these operations is a challenging task.


\section{Mixed-helicity bracket in Yang--Mills}\la{sec:YM}

The analysis of the mixed-helicity bracket in gravity presented so far can be performed in an analogous fashion for Yang--Mills theory as well. 
In this case, we are interested in computing the linear-order bracket of mixed-helicity charge aspects
\be\la{r1br2}
\{r^a_{s_1}({z_1}),\,\bar r^b_{{s_2}}({z_2})\}^{1}=
\{r_{s_1}^{1a}({z_1}),\,\bar r_{{s_2}}^{2b}({z_2})\}+\{r_{s_1}^{2a}({z_1}),\,\bar r_{{s_2}}^{1b}({z_2})\},
\ee
where
\be
\{r_{s_1}^{2a}({z_1}),\,\bar r_{{s_2}}^{1b}({z_2})\}=\{r^{2a}_{{s_1}+}({z_1}),\,\bar r^{1b}_{{s_2}+}({z_2})\}+\{r^{2a}_{{s_1}-}({z_1}),\,\bar r^{1b}_{{s_2}-}({z_2})\}.
\ee
The final result  takes the form (see Appendix \ref{C})
\be
\{r^{2a}_{{s_1}+}({z_1}),\bar r^{1b}_{{s_2}+}({z_2})\}
&=-i^{{s_1}+{s_2}}g^2_{YM}f_{abc}\sum_{n=0}^{s_1}\binom{{s_1}+{s_2}-n-2}{{s_2}-2}\bar D^{{s_2}+1}_{2}(D^{{s_1}-n}_{1}\delta(z,{z_2})D_{2}^{n}\mathscr{F}^c_+({s_1}+{s_2},{z_2})).
\la{posrcomm}
\ee

Putting the two energy sectors together, one obtains
\be
\{r^{2a}_{{s_1}}({z_1}),\bar r^{1b}_{{s_2}}({z_2})\}
=g^2_{YM}f_{abc}\sum_{n=0}^{s_1}\binom{{s_1}+{s_2}-n-2}{{s_1}-n}\bar D^{{s_2}+1}_{2}(D^{{s_1}-n}_{1}\delta({z_1},{z_2})D_{2}^{n}\bar F^c_{{s_1}+{s_2}}({z_2})),
\la{ymba}
\ee
where one uses that 
\be
F_s(z)=
-i^{-s}\mathscr{F}^*_+(s,z)-i^s\mathscr{F}_-(s,z).
\ee

By summing up the two pieces in \eqref{r1br2} and integrating the charge aspects against the smearing  Lie algebra valued functions on $S$, one obtains the charge bracket

 \be
\{R_{{s_1}}(\tau_{s_1}),\bar R_{{s_2}}(\bar \tau_{{s_2}})\}^1
=-ig^2_{YM}\int_S\biggl(&\sum_{m=0}^{s_1}(-)^{{s_1}+{s_2}}\binom{{s_1}+{s_2}-1}{m}\tr([D^{{s_1}-m}\bar D^{{s_2}+1}\bar \tau_{{s_2}},D^m\tau_{s_1}]_\Ag \bar{F}_{{s_1}+{s_2}}(z))
\cr
+&\sum_{m=0}^{{s_2}}(-)^{{s_1}+{s_2}}\binom{{s_1}+{s_2}-1}{m}\tr([\bar D^{{s_2}-m} D^{{s_1}+1} \tau_{s_1},\bar D^m\bar \tau_{{s_2}}]_\Ag {F}_{{s_1}+{s_2}}(z))\biggr).
\la{smearedmixedfull}
 \ee
We can observe that, analogously to the case of gravity, the bracket does not seem to close. Hence, even in the case of Yang--Mills, the two copies of the higher-spin symmetry algebra, respectively generated by charges $R_{{s_1}}$ and $\bar R_{{s_2}}$, do not form a wider algebra together for general smearing parameters. The problem is again due to the fact that, for general spins $s_1,s_2$, the RHS of \eqref{smearedmixedfull} does not contain enough spatial derivatives to reconstruct a soft charge. Analogously to the gravity case, the introduction of YM shadow charges will allow us to deal with this problem. But before repeating this construction for YM in Section \ref{sec:YMshad}, let us analyze the $s_1=s_2=0$ sector, where the issue with the number of spatial derivatives is not there, and one can try to do away with non-localities. 

\subsection{The lowest-spin sector $s=0$}
In the case ${s_1}={s_2}=0$, 
the charge bracket \eqref{smearedmixedfull} becomes
\be
\{R_0(\tau_0),\bar R_0(\bar \tau_{0})\}^1
&=-ig^2_{YM}\bar R^1_0([\tau_0,\bar \tau_{0}]_\Ag)+ig^2_{YM}\int_S\tr([\bar \tau_{0},\bar D \tau_0]_\Ag\bar F_0(z)+[D \tau_{0},\bar \tau_0,]_\Ag F_0(z))
\cr
&=
-ig^2_{YM}\bar R^1_0([\tau_0,\bar \tau_{0}]_\Ag)
-ig^2_{YM} R^1_0([ \tau_{0},\bar\tau_0]_\Ag)
\cr
&+ig^2_{YM}\int_S\tr\left([\bar \tau_{0},\bar D \tau_0]_\Ag\bar F_0(z)-[ \tau_0,D \bar\tau_{0}]_\Ag F_0(z)\right).
\la{linearYMbis}
\ee
If we go to the second order contribution to the bracket (see Appendix \ref{C} for details), under general assumptions on the smearing functions, we get

\be
\{R_0(\tau_0),\bar R_0(\bar \tau_{0})\}^2=-ig^2_{YM}\bar R^2_0([\tau_0,\bar \tau_{0}]_\Ag)
=-ig^2_{YM} R^2_0([\tau_{0},\bar\tau_0]_\Ag)
,
\la{quadYM}
\ee
{where we used the fact that the aspect $r_0^2$ is real, as follows from \eqref{r2+}.}

Here again, analogously to the $s=0,1$ case of gravity, a consistent algebra for the $s=0$ sector can be obtained if we impose additional restrictions on the smearing parameters. This can be achieved by demanding one of the two helicity sectors to be global and holomorphic. In fact, arbitrarily choosing the positive-helicity parameters to satisfy
\be
D\tau_0=0=\bar D \tau_0,
\la{YM-res}
\ee
the brackets \eqref{linearYMbis},\eqref{quadYM} yield the closed algebra\footnote{Restricting instead the negative-helicity sector, namely $\bar D\bar\tau_0=0= D \bar\tau_0$, would yield the closed algebra
\be
\{R_0(\tau_0),\bar R_0(\bar \tau_{0})\}=-ig^2_{YM} R_0([ \tau_{0},\bar\tau_0]_\Ag).
\ee
}
\be
\{R_0(\tau_0),\bar R_0(\bar \tau_{0})\}=-ig^2_{YM}\bar R_0([\tau_0,\bar \tau_{0}]_\Ag).
\la{YM00}
\ee

\subsection{Electromagnetic duality and central charge}\la{sec:EM}

Continuing the analogy with gravity, where a closed mixed-helicity algebra for the $s=0,1$ sector allowed us to include the dual mass in the gravitational phase space, we can study the role of magnetic charges in electromagnetism. In order to avoid the introduction of the analog singular matrix $\tilde R_{00}$ for electromagnetism, we work in the continuous $u$-basis in this section; we also revert to the vectorial formalism instead of helicity scalars. More importantly, we pay close attention to isolating the radiative modes from the zero modes of the vector potential. This will imply different boundary conditions for the charges at the corner $u=+\infty$, which, as we show, lead to the appearance of an electromagnetic central charge.

The $s=0$ aspect is given by
\be
\cR_{ 0} = \frac{1}{2}\left(\gamma_{z\bz}F_{ru}^{(0)} +  F_{\bz z}^{(0)}\right),
\la{R0} 
\ee
where $F_{ru}^{(0)},  F_{\bz z}^{(0)}$ are respectively the generators of large electric and magnetic transformations.

Under the electromagnetic duality
\be
\tilde F= d\tilde A=\f{2\pi}{e^2} \star F,
\ee
where $\tilde A$ is the dual vector potential and $\star$ denotes the 4D Hodge dual, the aspect \eqref{R0} transforms as
\be
\widetilde \cR_0=\f{2\pi i}{e^2} \cR_0.
\ee
In order to correctly compute the bracket of the electromagnetic (EM) charges, it is crucial to separate properly the radiative ($u$-dependent) part of the vector potential component $A^{(0)}_\bz$ from its zero mode  ($u$-independent) contribution; this can be achieved by writing \cite{Strominger:2017zoo} (see also \cite{Campiglia:2021bap} for an analogous treatment in gravity)
\be
A^{(0)}_\bz(u,z)=\hat A^{(0)}_\bz(u,z)+\alpha_\bz(z),
\ee
where $\hat A^{(0)}_\bz(u,z)$ encodes the radiative component satisfying
\be
\hat A^{(0)}_\bz(+\infty,z)+\hat A^{(0)}_\bz(-\infty,z)=0,
\ee
while
\be
\al_\bz(z)=\f12[A^{(0)}_\bz(+\infty,z)+A^{(0)}_\bz(-\infty,z)]
\ee
is the zero mode given by the corner values of the potential.
Using this parametrization, the EM symplectic form on $\scri^+$ can be split into a radiative (bulk) term and a zero mode (corner) term as 
\be
\Omega_{\scri^+}^{\scriptstyle{\rm EM}}&=\f{1}{e^2}\int du d^2z
\left(\d F^{(0)}_{\bar z u}(u,z)\wedge \delta A^{(0)}_z(u,z)+\d F^{(0)}_{ z u}(u,z)\wedge\delta A^{(0)}_{\bar z}(u,z)
\right)
\cr
&=-\f{1}{e^2}\int du d^2z
\left( \d \p_u \hat A^{(0)}_\bz(u,z)\wedge \delta \hat A^{(0)}_z(u,z)+\d \p_u \hat A^{(0)}_z(u,z)\wedge\delta \hat A^{(0)}_{\bz}(u,z)
\right)
\cr
&-\f{1}{e^2}\int d^2z
\left(\d F_\bz(z)\wedge \d \al_z(z)
+\d F_z(z)\wedge \d \al_\bz(z)
\right).
\la{sf:EM}
\ee
We thus see that the radiative phase space is extended by a corner symplectic form where the vector potential zero modes are paired with the two soft photons
\be
F_\bz:=-\int_{-\infty}^\infty duF^{(0)}_{\bar z u}\,,\quad 
F_z:=-\int_{-\infty}^\infty duF^{(0)}_{ z u}.
\ee
In particular, we have the  bulk bracket
\be
\{\p_u \hat A^{(0)}_\bz(u,z), \hat A^{(0)}_z(u',z')\}=
-e^2\delta(u-u')\delta^2(z,z'),
\la{AuAbra}
\ee
and the corner bracket
\be
\{F_\bz(z), \al_z(z')\}=-e^2\delta^2(z,z').
\la{F0Abra}
\ee

Using the decomposition
\be
F_{\bz z}^{(0)}(u,z)=\hat F_{\bz z}^{(0)}(u,z)+( \p_\bz \al_z-\p_z\al_\bz)(z),
\ee
where
\be
\hat F_{\bz z}^{(0)}(u,z)=\p_\bz \hat A^{(0)}_z(u,z)-\p_z \hat A^{(0)}_\bz(u,z),
\ee
the evolution equation for \eqref{R0} is given by
\be
\p_u\cR_0=\frac{1}{2}\p_u \left(\gamma_{z\bz}F_{ru}^{(0)} + \hat F_{\bz z}^{(0)}\right)=D_zF^{(0)}_{\bar z u},
\ee

Considering the smearing parameter $\tau_0=\epsilon$,  the electromagnetic charges  can be written as
\be
Q_{\EM}(\epsilon)=R_0(\tau_0)
&=\f12 \int d^2z\, \epsilon
\left(\gamma_{z\bz}F_{ru}^{(0)} + \hat F_{\bz z}^{(0)}\right)
+\f12\int d^2z \,\epsilon( \p_\bz \al_z-\p_z\al_\bz)
\cr
&=\int d^2z\, \epsilon D_z F_\bz
+\f12\int d^2z\, \epsilon ( \p_\bz \al_z-\p_z\al_\bz),
\la{QEM}
\ee
where we are demanding the boundary condition
\be
F_{ru}^{(0)}(+\infty,z) + \hat F_{\bz z}^{(0)}(+\infty,z)=0,
\ee
or equivalently
\be
\cR_0(+\infty,z)=\f12 ( \p_\bz \al_z-\p_z\al_\bz)(z).
\la{R0infty}
\ee
Analogously, taking $\bar\tau_0=\bar \epsilon$, we have
\be
\bQ_{\EM}(\bar\epsilon)&=\bR_0( \bar\tau_0 )=\int d^2z \,\bar\epsilon D_\bz F_z
+\f12\int d^2z\, \bar \epsilon ( \p_z \al_\bz- \p_\bz \al_z).
\la{bQEM}
\ee

When acting on the vector potential, then the complex charges \eqref{QEM}, \eqref{bQEM} generate the transformations
\be
\d_\epsilon A^{(0)}_z&=\{Q_{\EM}(\epsilon), A^{(0)}_z\}=e^2\p_z  \epsilon\,,
\quad
\d_\epsilon A^{(0)}_\bz=\{Q_{\EM}(\epsilon), A^{(0)}_\bz\}=0\,,
\la{deA}
\\
\d_{\bar \epsilon} A^{(0)}_z&=\{\bQ_{\EM}(\bar \epsilon), A^{(0)}_z\}=0\,,
\quad
\d_{\bar \epsilon} A^{(0)}_\bz=\{\bQ_{\EM}(\bar \epsilon), A^{(0)}_\bz\}=e^2 \p_\bz \bar \epsilon\,.
\la{bdeA}
\ee
By means of the corner bracket \eqref{F0Abra}, we can compute the algebra
\be
\{Q_{\EM}(\epsilon), \bQ_{\EM}(\bar\epsilon')\}&=0.
\la{EMcc}
\ee

Moving to the same-helicity bracket, we find that it is non-vanishing, as follows
\be
\{Q_{\EM}(\epsilon), Q_{\EM}(\epsilon')\}&=-\f{e^2}{2}\int d^2z \left(\p_z\epsilon\p_\bz\epsilon'-\p_\bz\epsilon \p_z\epsilon'\right),
\ee
where we see the appearance of a central charge.

Let us consider the electric and magnetic parts of this charge. From \eqref{QEM} and \eqref{bQEM}, we obtain
\be
Q_{\E}(\epsilon)&:=Q_{\EM}(\epsilon)+\bar Q_{\EM}(\epsilon)=\int d^2z\, \epsilon(D_z F_\bz+ D_\bz F_z),
\la{QE}
\\
Q_{\M}(\tilde\epsilon)&:=i\left(Q_{\EM}(\tilde\epsilon)-\bar Q_{\EM}(\tilde\epsilon)\right)=i\int d^2z\, \tilde\epsilon(D_zF_\bz- D_\bz  F_z)+i\int d^2z\,\tilde\epsilon( D_\bz\alpha_z- D_z\alpha_\bz).
\la{QM}
\ee
These two charges  yield the algebra
\begin{subequations}
\be
\{Q_{\E}(\epsilon), Q_{\E}(\epsilon')\}&=0,
\\
\{Q_{\M}(\tilde\epsilon), Q_{\M}(\tilde\epsilon')\}&=0,
\\
\{Q_{\E}(\epsilon), Q_{\M}(\tilde\epsilon)\}&=-ie^2\int d^2z\, 
\left(\p_z\epsilon \p_\bz\tilde \epsilon- \p_\bz\epsilon \p_z\tilde \epsilon\right)
=e^2\int d\epsilon\wedge d\tilde\epsilon.
\la{QEQM}
\ee
\end{subequations}

The RHS of \eqref{QEQM} represents an EM central charge. If the celestial sphere has the topology $S=S_0$, this term vanishes; if instead there is a puncture $S=S_1$, namely if the transformation parameters are not everywhere well-defined functions on the celestial sphere $S$ but introduce a singularity,  the central charge can be non-vanishing and given by a contour integral around the pole, namely
\be
\{Q_{\E}(\epsilon), Q_{\M}(\tilde\epsilon)\}&=e^2\oint \epsilon \,d\tilde \epsilon .
\la{cc}
\ee
At this point, the reader may rightfully worry that such contour contributions were not previously taken into account when performing integration by parts to arrive at the bracket \eqref{QEQM}. Indeed, extra care is required when defining the electric and magnetic charges in the presence of poles, as already pointed out in \cite{Freidel:2018fsk}. In this case, in fact, the non-triviality of the $U(1)$-bundle for the dual vector potential implies that a well-defined notion of electric and magnetic charges is given by the expressions
\be
Q_{\E}(\epsilon)&=-\int d^2z\, (D_z\epsilon\, F_\bz+ D_\bz \epsilon\,  F_z),
\la{QE-2}
\\
Q_{\M}(\tilde\epsilon)&=-i\int d^2z\, (D_z\tilde\epsilon\, F_\bz- D_\bz \tilde\epsilon\,  F_z)-i\int d^2z\,( D_\bz\tilde\epsilon\,\alpha_z- D_z\tilde\epsilon\,\alpha_\bz),
\la{QM-2}
\ee
rather than \eqref{QE}, \eqref{QM}; namely, the latter are missing contour contributions around the poles.\footnote{Explicitly, we can write \eqref{QE-2}, \eqref{QM-2} as 
\be
Q_{\E}(\epsilon)&=\int_{S^2{\backslash p}} \!\! \epsilon(D_z F_\bz+ D_\bz F_z)
-i\oint_p  \epsilon({d\bz} F_\bz -d z F_z),
\la{QE-3}
\\
Q_{\M}(\tilde\epsilon)&=i\int _{S^2{\backslash p}}\! \!\tilde\epsilon(D_zF_\bz- D_\bz  F_z)+i\int _{S^2{\backslash p}}\!\!\tilde\epsilon( D_\bz\alpha_z- D_z\alpha_\bz)
+\oint_p\tilde\epsilon({d\bz}(F_\bz-\alpha_\bz) +dz(F_z-\alpha_z)).
\la{QM-3}
\ee
}
The bracket of the charges  \eqref{QE-3}, \eqref{QM-3} requires no integration by parts and the central charge 
\eqref{cc} follows with no ambiguities.

The transformations \eqref{deA},\eqref{bdeA} were previously derived in  \cite{Strominger:2015bla}, where a complexification of the EM large gauge group was proposed; identifying the vector potential components $A^{(0)}_z, A^{(0)}_\bz$ with Kac--Moody currents, such transformation laws suggested the existence of a non-zero level in the Kac--Moody current algebra \cite{Strominger:2013lka,He:2015zea}. We see here that these complex EM charges naturally arise from the asymptotic structure of Maxwell's equations; moreover, for non-trivial topology of $S$, we have shown how the Poisson bracket of the electric and magnetic charges \eqref{QE-2},\eqref{QM-2} indeed gives rise to a non-vanishing EM central charge, confirming the results of \cite{Freidel:2018fsk} where the same EM central charge was obtained by different means through a phase space extension in terms of edge modes. 
The appearance of this central charge is strictly related to the presence of the second term on the RHS of \eqref{QEM}, which arises due to the  boundary condition \eqref{R0infty} instead of 
$\cR_0(+\infty,z)= 0$.

\subsection{A closed bracket for all spins}\la{sec:YMshad}

\begin{figure}[h!]
     \centering
      \begin{subfigure}[t]{0.45\textwidth}
      \centering
\includegraphics[width=\linewidth,height=50mm, keepaspectratio]{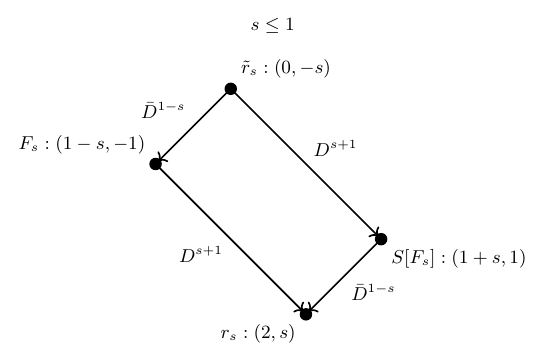}
\caption{}
 \label{fig:CD-YM-2}  
    \end{subfigure}
\quad
 \begin{subfigure}[t]{0.45\textwidth}
      \centering
\includegraphics[width=\linewidth,height=50mm, keepaspectratio]{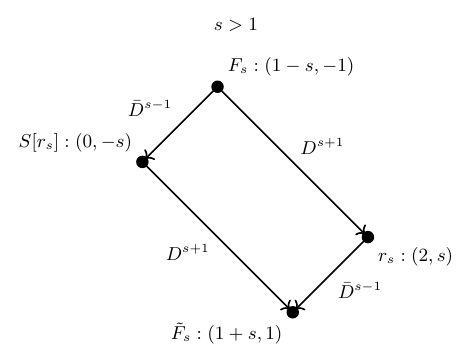}
\caption{}
 \label{fig:CD-YM-1}
    \end{subfigure}
    \caption{ Celestial diamond associated with negative-helicity soft gluon with (\subref{fig:CD-YM-2}) $s \le1$ and 
    (\subref{fig:CD-YM-1}) $s> 1$.}
    \label{fig:CD-YM}
\end{figure}

In order to treat the general spin case, it is again instructive to revisit the celestial diamond structure of Yang--Mills theory.
Also in this case, charge aspects and soft gluons organize themselves into finite-dimensional conformal multiplets in celestial CFT, with $(\Delta,J)$ weights now shifted according to the spin-1 nature of the gluon. As in gravity, the structure rearranges depending on the spin-$s$ of the charges. For $s\ge2$, the celestial multiplet contains the shadow-transformed charge at one corner $S[r_s]$, and multiple covariant derivatives and anti-derivatives along the sphere relate this object to the standard charge aspect (see Fig.~\ref{fig:CD-YM-1}). When $s\le1$, the roles are inverted and the dual charge $\tilde r_{s}$ aspect sits at the top of the multiplet (see Fig.~\ref{fig:CD-YM-2}).
In analogy with the gravitational case, it is convenient to introduce the maps  $\mS:V_{(2,s)}\to V_{(0,-s)}$, $\bar\mS:V_{(2,-s)}\to V_{(0,s)}$ and acting on $f_{(2,s)}\in V_{(2,s)}$ and $f_{(2,-s)}\in V_{(2,-s)}$ 
whose explicit action is given by
\be
\mS[f_{(2,s)}](z)&= D^{-(s+1)}\bar D^{s-1}f_{(2,s)}({z_1}),\la{shadY} \\
\bar\mS[f_{(2,-s)}](z)&=\bar D^{-(s+1)}D^{s-1}f_{(2,-s)}({z_1}),\la{shadYb}
\ee
together with their inverse transformations $\mS^{-1}:V_{(0,-s)}\rightarrow V_{(2,s)}$, $\bar\mS^{-1}:V_{(0,s)}\rightarrow V_{(2,-s)}$ acting on $f_{(0,-s)}\in V_{(0,-s)}$ and $f_{(0,s)}\in V_{(0,s)}$ respectively as
\be
\mS^{-1}[f_{(0,-s)}](z)&=\bar D^{1-s}D^{s+1}f_{(0,-s)}({z_1}), \la{shadYi}\\
 \bar\mS^{-1}[f_{(0,s)}](z)&= D^{1-s}\bar D^{s+1}f_{(0,-s)}(z).\la{shadYib}
\ee

Upon flattening the celestial sphere, these expressions reduce to the usual definition of the shadow transform for all $s\ge2$.
These maps allow us to introduce the \emph{shadow charge aspects}
\be
\mS[r_s]&=\begin{cases}
  \tilde r_s \quad &\text{for} \quad s\le1, \\
S[r_s]\quad &\text{for} \quad s\ge2,  
\end{cases} \la{scaY}
\ee
and the associated \emph{shadow charges}
\be
\mS[ R_s](\tilde \tau_s)=\int_S\tr(\tilde \tau_s\mS[ r_s]). \la{scY}
\ee
Similar identities to \eqref{shadsmearb} are found.

We now examine the mixed-helicity bracket. Imposing the condition 
\be\la{YMglob}
D^{s+1}\tau_s=0,
\ee
which selects the global positive-helicity sector, the linear bracket \eqref{smearedmixedfull} reduces to (see Appendix \ref{D.3} for details)
 \be\la{scbYM}
\{R_{{s_1}}(\tau_{s_1}),\bar R_{{s_2}}(\bar \tau_{{s_2}})\}^1
&=-ig^2_{YM}\bar { R}^1_{{s_1}+{s_2}}\left(\bar \mS[\tau_{s_1},\bar{ \mS}^{-1}[\bar \tau_{{s_2}}]]_\mathfrak g\right)\,.
 \ee
This outcome reflects the higher-spin Yang--Mills algebra in the same-helicity sector, cf. \eqref{HSalgYM}, with the additional feature, as in gravity, that the maps $\mS$ and $\bar\mS$ in the mixed-helicity case naturally adjust the smearing functions so that they acquire the appropriate conformal weight and spin. Here we have
\be
\tau_{s_1}\in V_{(0,-s_1)}, \quad  \bar\tau_{s_2}\in V_{(0,s_2)}\implies \bar \mS^{-1}[\bar\tau_{s_2}]\in V_{(2,-s_2)},
\ee
and the Lie bracket then produces
\be
[\tau_{s_1},\bar \mS^{-1}[\bar\tau_{s_2}]]_\Ag \in V_{(2,-s_1-s_2)}.
\ee
which is mapped back to $V_{(0,s_1+s_2)}$ by $\bar\mS$.

 For completeness, if instead one restricts the negative-helicity sector via $\bar D^{s+1}\bar\tau_s=0$, one obtains the relation

\be
\{R_{{s_1}}(\tau_{s_1}),\bar R_{{s_2}}(\bar \tau_{{s_2}})\}^1
&=-ig^2_{YM} { R}^1_{{s_1}+{s_2}}(\mS[{ \mS^{-1}}[ \tau_{{s_1}}],\bar\tau_{{s_2}}]_\Ag).
\la{RbR}
 \ee
{It is instructive to compare the mixed-helicity bracket derived above with the algebraic structure  discussed in \cite{Kmec:2025ftx} in the context of self-dual Yang--Mills theory. In that work, two towers of charges associated with opposite gluon helicities are constructed from the twistor-space formulation of self-dual Yang--Mills.
However, in \cite{Kmec:2025ftx} only one tower is directly related to the Newman-Penrose scalars, as in \cite{Freidel:2023gue}, and corresponds to our $\bar\cR_s$, with weights $(\Delta,J)=(2,-s)$; the other, denoted $\tilde\cR_s$, is defined  through the Lax formulation and is tied to the the integrability of the theory. 
Here we  work instead in a phase space where the two helicities are treated on the same footing: our two towers of charge aspects, $\cR_s$ and $\bar\cR_s$, are related by complex conjugation, so that $\cR_s$ has weights $(2,s)$, and both are built from the Newman--Penrose scalars. A detailed map between the two constructions is beyond the scope of this work. We observe, however, that $\tilde\cR_s$ carries weights $(\Delta,J)=(1,-2-s)$. These are incompatible with those of the shadow transform $\mS[\cR_s]$, which has $(0,-s)$, but coincide with those of $\partial_u\mS[\cR_{s+2}]$, which suggests a possible nonlocal relation between the two towers.  
Despite this different characterization of one helicity sector, 
the brackets satisfied by the charges in  \cite{Kmec:2025ftx}
exhibits an action of the negative-helicity charges on the positive-helicity ones, with a possible field-dependent cocycle. This structure bears a close resemblance to our mixed-helicity bracket \eqref{RbR}: upon identifying the positive-helicity smearing parameter through the inverse shadow map,  namely $\tilde \tau_s=\mS^{-1}[\tau_s$], the latter takes the form of an adjoint Lie-algebra action
 \be
\{R_{{s_1}}(\mS[\tilde\tau_{s_1}]),\bar R_{{s_2}}(\bar \tau_{{s_2}})\}^1=-ig^2_{YM} { R}^1_{{s_1}+{s_2}}(\mS[[ \tilde\tau_{{s_1}},\bar\tau_{{s_2}}]_\Ag]),
 \ee
followed by the corresponding shadow transformation.
 
There is, however, an important difference between the two constructions. In \cite{Kmec:2025ftx}, the positive-helicity charges, being the generators of the shift symmetry in the BF theory on twistor space, commute among themselves, forming an abelian sector on which the negative-helicity algebra acts. By contrast, the positive-helicity charges entering \eqref{RbR} obey, in general, a non-vanishing same-helicity bracket, with the structure of a  copy of the $S$-algebra. Thus, although the mixed-helicity brackets exhibit related algebraic structures, the complete brackets among the two sets of charges are not identical.
A further distinction concerns the dynamical setting and the conditions under which the mixed-helicity bracket closes. Our result establishes the conditions under which a related mixed-helicity structure arises at linear order in the full Yang--Mills framework. The extension of this result to the complete nonlinear charge algebra, and a more precise correspondence between the respective generators, remain interesting questions for future investigation.
}

 \subsubsection{Jacobi identities}
 
We are left to check the validity of the mixed-helicity Jacobi identities for the closed charge bracket \eqref{scbYM} we found in Yang-Mills. In this case, the two kinds of identities we should have are
 \be\la{sjYM1}
&\{R_{s_1}(\tau_{s_1}),\{R_{s_2}(\tau_{s_2}),\bar R_{s_3}(\bar\tau_{s_3}\}\}^1
\cr+&\{R_{s_2}(\tau_{s_2}),\{\bar R_{s_3}(\bar\tau_{s_3}),R_{s_1}(\tau_{s_1})\}\}^1
\cr+&\{\bar R_{s_3}(\bar\tau_{s_3}),\{R_{s_1}(\tau_{s_1}),R_{s_2}(\tau_{s_2})\}\}^1=0,
 \ee
 and
  \be\la{sjYM2}
&\{\bar R_{s_1}(\bar\tau_{s_1}),\{\bar R_{s_2}(\bar\tau_{s_2}),R_{s_3}(\tau_{s_3}\}\}^1
\cr+&\{\bar R_{s_2}(\bar\tau_{s_2}),\{ R_{s_3}(\tau_{s_3}),\bar R_{s_1}(\bar\tau_{s_1})\}\}^1
\cr+&\{ R_{s_3}(\tau_{s_3}),\{\bar R_{s_1}(\bar\tau_{s_1}),\bar R_{s_2}(\bar\tau_{s_2})\}\}^1=0.
 \ee
We were able to prove only  \eqref{sjYM1} in Appendix \ref{jacYM}, under the fact that the $S$-algebra structure of the same-helicity bracket survives at the second order \cite{Freidel:2023gue}.
On the other hand, verifying \eqref{sjYM2} mimics the case of its gravitational counterpart: the explicit form of the second-order mixed-helicity bracket is required, and the technical issue of disentangling the action of nested maps $\bar \mS$ and $\bar\mS^{-1}$ acting on the smearing functions appears again.

\section{Conclusions}\la{sec:conc}

In this work, we have studied the mixed-helicity brackets of celestial symmetries both in gravity and Yang--Mills theory. We have shown that, in both cases, one can obtain a closed algebra for arbitrary integer values of the charge spins upon introduction of shadow charges yielding a non-local structure of the mixed-helicity algebra \eqref{bqq}, \eqref{scbYM}. These non-localities are a direct consequence of the fact that, when computing the mixed-helicity bracket at linear order, the object one obtains does not, in general, contain enough spatial derivatives to reconstruct a soft charge. This is not the case, though for the $s=0,1$ sector in gravity and $s=0$ in Yang--Mills. We thus analyzed these two cases in more detail and showed that, upon imposing some restrictions on the smearing parameters, one can close the mixed brackets without the appearance of non-localities or cocycle terms.

In particular, in the case of gravity, when the celestial sphere has a non-trivial topology, one can represent the dual mass in the gravitational phase space and derive an extension \eqref{dbms} of the BMS  algebra.  At this stage, this extension does not represent a residual diffeomorphism symmetry algebra in the standard sense, as there is no real-valued vector field corresponding to the action of the dual mass on the asymptotic gravitational phase space. Instead, one can consider complex-valued vector fields  \eqref{xiA}, whose Lie bracket satisfies the $\dbms$ algebra,  mapping real asymptotically flat spacetimes to complex ones (with the appearance of conical singularities) due to an imaginary counterpart of supertranslations. Whether the dual BMS algebra can be understood as a bona fide residual diffeomorphism symmetry for a more general class of asymptotically flat spacetimes, by relaxing some fall-off conditions (see e.g. \cite{Satishchandran:2019pyc,Bieri:2020zki}), remains an interesting open question.

In the case of electromagnetism, the analog role of the dual mass is played by magnetic charges. Their inclusion in the asymptotic phase space has been conjectured to lead to a non-zero $U(1)$   Kac--Moody level \cite{Strominger:2013lka,He:2015zea}. We have shown that, upon a proper split between radiative and zero modes of the EM symplectic potential and a less restrictive boundary condition at time-like infinity, allowing for the presence of magnetic monopoles, the electric and magnetic charge bracket gives rise to a non-zero central charge \eqref{cc}.

This study of the mixed-helicity bracket of higher-spin charges at linear order can be used to investigate the mixed-helicity celestial OPE through the  OPE/bracket correspondence revealed in \cite{Freidel:2021ytz, Freidel:2023gue,Pranzetti:2025flv}. In particular, the issue of associativity of the
celestial OPE
\cite{Mago:2021wje,Ren:2022sws,Ball:2022bgg,Ball:2023sdz} could be addressed by completing the task to verify the validity of the Jacobi identities \eqref{2bq1q} and \eqref{sjYM2}, respectively for the gravitational \eqref{bqq}  and the gauge theory \eqref{scbYM} mixed-helicity brackets.

Moreover, the relevance of studying the mixed-helicity charge bracket can go beyond more theoretical interests. For instance, in \cite{Blanchet:2020ngx,Compere:2022zdz,Blanchet:2023pce}, a
program to holographically reconstruct the radiation signal from binary compact mergers, by matching the gravitational signal multipole moments with the higher-spin charges, was initiated. For this purpose, a dictionary involving real-valued celestial charges and their algebra is crucial in order to have physically realistic applications. We expect the results derived here to provide useful applications for this program.

\section*{Acknowledgments}

We would like to thank Laurent Freidel for helpful discussions and insights. 

\appendix

\section{Dirac bracket for the radiative phase space in gravity}
\subsection{Continuous basis}\la{a:DB} 
Let us start with the asymptotic gravitational  pre-symplectic potential
\be\la{rpp:full}
\theta_{G}=\f1\k\int_{-\infty}^{\infty}du'\left(N\delta C+\bar N \delta\bar C\right),
\ee
where all fields $N$, $C$, $\bar N$, $\bar C$ are independent in absence of further assumptions.
From this, we can read off the  Poisson brackets, here denoted as $\{,\}_{PB}$. The only non-trivial ones are the following
\be
\{N(u,z),C(u',z')\}_{PB}=&\k\delta(u-u')\delta^2(z,z'), \\
\{\bar N(u,z),\bar C(u',z')\}_{PB}=&\k\delta(u-u')\delta^2(z,z').
\ee
 Notice that these brackets are \emph{twice} the ones commonly used in the literature \cite{He:2014laa}. This is because, by treating all fields as independent, we are effectively counting the degrees of freedom of the theory \emph{twice}.

In the following, we show how the construction of the Dirac bracket, once the constraints
\begin{equation}
N = \partial_u \bar C, \qquad \bar N = \partial_u C
\end{equation}
are imposed, corrects this overcounting and yields the expected brackets.

Our approach differs from some works in the literature (see, e.g.\cite{Freidel:2022skz}), where authors start from an already reduced phase space described by
\be
\theta = \frac{2}{\k} \int_{-\infty}^{\infty} du \, N \, \delta C,
\ee
with canonical variables $C$ and $N := \partial_u \bar C$. In that case, the correct number of degrees of freedom is accounted for from the start, and no additional constraints need to be imposed; the brackets can be read off directly from the presymplectic potential. In our case, we decided to start with the real presymplectic potential \eqref{rpp:full}, as it coincides with the original Ashtekar--Streubel \cite{Ashtekar:1978zz,Ashtekar:1981bq} expression and treats both helicities of the shear, $C$ and $\bar C$, on the same footing.

Now we impose the following constraints on the fields
\be
 \chi_1=&N-\p_u\bar C=0; \\
 \chi_2=&\bar N-\p_u C=0.
\ee
These two constraints are \emph{second class}, since their Poisson bracket is non-vanishing, namely
\be\label{chi12}
\{\chi_1(u,z),\chi_2(u',z')\}_{PB}=&-\p_{u'}\{N(u,z),C(u',z')\}_{PB}-\p_u \{\bar C(u,z),\bar N(u',z')\}_{PB}
\cr&=2\k\p_u \delta(u-u')\delta^2(z,z').
\ee
We define the matrix $M_{ij}(u,u',z,z'),\,i,j=1,2$ as follows
\be
M_{ij}(u,u',z,z'):=\{\chi_i(u,z),\chi_j(u',z')\}_{PB}.
\ee
The only non-vanishing components will be $M_{12}(u,u',z,z')$, given by \eqref{chi12}  and $M_{21}(u,u',z,z')=M_{12}(u,u',z,z')$.

The Dirac bracket $\{,\}_{DB}$ (in the rest of the paper simply denoted as $\{,\}$)  of two functions $\phi$, $\psi$ on phase space will be given by
\be
&\{\phi(u,z),\psi(u',z')\}_{DB}=\{\phi(u,z),\psi(u',z')\}_{PB}
\cr&-\sum_{i,j}\int_{-\infty}^\infty du_1\int_{-\infty}^\infty du_2\int_{S_1}\int_{S_2}\{\phi(u,z),\chi_i(u_1,z_1)\}_{PB}M^{-1}_{ij}(u_1,u_2,z_1,z_2)\{\chi_j(u_2,z_2),\psi(u',z')\}_{PB},
\la{DB}
\ee
where $M^{-1}$ is the inverse matrix of $M$, such that
\be
\sum_{j}\int_{-\infty}^\infty du_1\int_{S_1}M_{ij}(u,u_1,z,z_1)M^{-1}_{jk}(u_1,u',z_1,z')=\delta_i^k\delta(u-u')\delta^2(z,z').
\ee
If we pick $i=1$, we obtain
\be
\int_{-\infty}^\infty du_1\int_{S_1}M_{12}(u,u_1,z,z_1)M^{-1}_{2k}(u_1,u',z_1,z')=\delta_2^k\delta(u-u')\delta^2(z,z')
\cr
\iff 2\k\p_u\int_{-\infty}^\infty du_1\int_{S_1}\delta(u-u_1)\delta^2(z,z')M^{-1}_{2k}(u_1,u',z_1,z')=\delta_2^k\delta(u-u')\delta^2(z,z').
\ee
We make an ansatz for $M^{-1}_{2k}$ such that
\be
M^{-1}_{2k}(u_1,u',z_1,z')=\mathcal{M}_{2k}(u_1,u')\delta^2(z_1,z'),
\ee
implying
\be
 2\k\p_u\int_{-\infty}^\infty du_1\int_{S_1}\delta(u-u_1)\delta^2(z,z_1)M^{-1}_{2k}(u_1,u',z_1,z')=\delta_2^k\delta(u-u')\delta^2(z,z')
 \cr
 \implies
 2\k\p_u\mathcal{M}_{2k}(u,u')\delta^2(z,z')=\delta_2^k\delta(u-u')\delta^2(z,z')
 .
\ee
Consequently, the matrix $\cM$ has to solve the differential equation
\be
\p_u\mathcal{M}_{21}(u,u')=\f{1}{2\k}\delta(u-u'),
\ee
whose solution is
\be
\mathcal{M}_{21}(u,u')=\f{1}{2\k}\theta(u-u')+\alpha,
\ee
where $\theta(x)$ in the Heaviside theta function, and $\alpha$ is a constant.
The value of $\alpha$ can be fixed by a boundary/symmetry constraint. We know that each component $M_{ij}(u,u',z,z')$ is anti-symmetric in the exchange $u\leftrightarrow u'$, as appears from \eqref{chi12}, meaning that also the inverse matrix $M_{ij}^{-1}(u,u',z,z')$ should obey this property.
Thus, we require
\be
\mathcal{M}_{21}(u',u)=\f{1}{2\k}\theta(u'-u)+\alpha=-\f{1}{2\k}\theta(u-u')-\alpha=\f{1}{2\k}\theta(u'-u)-\f{1}{2\k}-\alpha
\cr
\implies
\alpha=-\frac{1}{4\k}.
\ee
Consequently, we have
\be
M^{-1}_{21}(u,u',z,z')=\f{1}{4\k}\sgn(u-u')\delta^2(z,z').
\ee
Since $M_{ij}$ is symmetric under the exchange $i\leftrightarrow j$, the same holds for its inverse, so $M_{12}^{-1}=M_{21}^{-1}$. All the other matrix coefficients vanish.
At this point, we are able to compute the following brackets:
\be\label{bcc-app}
&\{\bar C(u,z), C(u',z')\}_{DB}=
\cr&-\int_{-\infty}^\infty du_1\int_{-\infty}^\infty du_2\int_{S_1}\int_{S_2}\{\bar C(u,z),\chi_2(u_1,z_1)\}_{PB}M^{-1}_{21}(u_1,u_2,z_1,z_2)\{\chi_1(u_2,z_2),C(u',z')\}_{PB}
\cr&=\frac{\k}{4}\int_{-\infty}^\infty du_1\int_{-\infty}^\infty du_2\int_{S_1}\int_{S_2}\delta(u-u_1)\delta^2(z,z_1)\sgn(u_1-u_2)\delta^2(z_1,z_2)\delta(u_2-u')\delta^2(z_2,z')
\cr&=\frac{\k}{4}\sgn(u-u')\delta^2(z,z'),
\ee
\be\la{nc}
&\{N(u,z), C(u',z')\}_{DB}=\{N(u,z), C(u',z')\}_{PB}
\cr&-\int_{-\infty}^\infty du_1\int_{-\infty}^\infty du_2\int_{S_1}\int_{S_2}\{N(u,z),\chi_2(u_1,z_1)\}_{PB}M^{-1}_{21}(u_1,u_2,z_1,z_2)\{\chi_1(u_2,z_2),C(u',z')\}_{PB}
\cr&=\k\delta(u-u')\delta^2(z,z')
\cr&-\f\k4\int_{-\infty}^\infty du_1\int_{-\infty}^\infty du_2\int_{S_1}\int_{S_2}\p_{u_1}\delta(u-u_1)\delta^2(z,z_1)\sgn(u_1-u_2)\delta(z_1,z_2)\delta(u_2-u')\delta^2(z_2,z')
\cr&=\frac{\k}{2}\delta(u-u')\delta^2(z,z'),
\ee
where the latter bracket could equivalently be obtained by using $N=\p_u \bar C$ on the bracket \eqref{bcc-app}.

\subsection{Discrete basis} \la{a:DDB}
 Let us start with the presymplectic potential, written in the discrete basis.
\be
\theta_{G}=\f1{2\pi i\k }\sum_{\sigma=\pm}\sum_{n=0}^{\infty}\left(\mathscr{M}_\sigma(n,z)\delta \mathscr{S}^*_{\sigma}(n,z)-\mathscr{M}^*_\sigma(n,z)\delta \mathscr{S}_{\sigma}(n,z)\right),
\ee
where the canonical variables are independent.
From this, we can read off the following non-trivial Poisson brackets
\be
\{\mathscr{M}_{\pm}(n,z),\mathscr{S}^*_{\pm}(m,z')\}_{PB}=&2\pi i\k\delta_{nm}\delta^2(z,z'), \\
\{\mathscr{M}^*_{\pm}(n,z),\mathscr{S}_{\pm}(m,z')\}_{PB}=&-2\pi i\k\delta_{nm}\delta^2(z,z').
\ee
The other possible brackets involving fundamental discrete fields vanish.

Given the expected relations \eqref{eq6} among the memory and the Goldstone variables, we impose the following constraints
\be
 \chi_1=&\mathscr{S}^*_{\pm}(n)-\sum_{m}R_{nm}\mathscr{M}^*_{\pm}(m)=0; \\
 \chi_2=&\mathscr{S}_{\pm}(n)-\sum_{m}R_{nm}\mathscr{M}_{\pm}(m)=0.
\ee
{These relations assume $R_{nm}=R^*_{mn}$, which is motivated by the regularized expression \eqref{eq15}. 

These two constraints are \emph{second class}, and the matrix of the Poisson brackets among them 
\be
M_{ij}(n,m,z,z'):=\{\chi_i(n,z),\chi_j(m,z')\}_{PB}, \quad i,j=1,2
\ee
is such that its only non-vanishing components are
\be\label{chi12-2}
M_{12}(n,m,z,z')=-M_{21}(n,m,z,z')=4\pi i\k R_{nm} \delta^2(z,z')
.
\ee
The Dirac bracket $\{,\}_{DB}$ of two functions $\phi$, $\psi$ of the phase space will be given by
\be
&\{\phi(n,z),\psi(m,z')\}_{DB}=\{\phi(n,z),\psi(m,z')\}_{PB}
\cr&-\sum_{i,j}\sum_{n',m'}\int_{S_1}\int_{S_2}\{\phi(n,z),\chi_i(n',z_1)\}_{PB}M^{-1}_{ij}(n',m',z_1,z_2)\{\chi_j(m',z_2),\psi(m,z')\}_{PB},
\ee
 where $M^{-1}_{ij}$ is such that
\be
\sum_{j}\sum_{n'}\int_{S_1}M_{ij}(n,n',z,z_1)M^{-1}_{jk}(n',m,z_1,z')=\delta_i^k\delta_{nm}\delta^2(z,z').
\ee
We will now derive its explicit expression.

If we pick $i=1$, we obtain
\be
\sum_{n'}\int_{S_1}M_{12}(n,n',z,z_1)M^{-1}_{2k}(n',m,z_1,z')=\delta_2^k\delta_{nm}\delta^2(z,z')
\cr
\iff 4\pi i\k\sum_{n'}\int_{S_1}R_{nn'}\delta^2(z,z')M^{-1}_{2k}(n',m,z_1,z')=\delta_2^k\delta(u-u')\delta^2(z,z').
\ee
We make the following ansatz for $M^{-1}_{2k}$
\be
 M^{-1}_{2k}(n,m,z_1,z')=(4\pi i\k)^{-1}\mathcal{M}_{2k}(n,m)\delta^2(z_1,z'),
\ee
implying
\be
 \sum_{n'}R_{nm}\mathcal{M}_{2k}(n',m)\delta^2(z,z')=\delta_2^k\delta_{nm}\delta^2(z,z')
 .
\ee
So, our matrix has to solve the equation
\be
\sum_{n'}R_{nn'}\mathcal{M}_{21}(n',m)=\delta_{nm},
\ee
which corresponds to \eqref{eq7}. This means that we have $\mathcal{M}_{21}(n,m)=\tilde R_{nm}$, where $\tilde R$ is the inverse matrix of $R$.
Consequently, we have
\be
M^{-1}_{21}(n,m,z,z')=\f{1}{4\pi i\k}\tilde R_{nm}\delta^2(z,z').
\ee
Moreover, since $M_{ij}$ is anti-symmetric under the exchange $i\leftrightarrow j$, the same holds for its inverse, so $M_{12}^{-1}=-M_{21}^{-1}$. All the other matrix coefficients vanish.
At this point, we are able to compute the following brackets:
\be\label{bmm}
&\{\mathscr{M}^*_{\pm}(n,z),\mathscr{M}_{\pm}(m,z')\}_{DB}=
\cr&-\sum_{n',m'}\int_{S_1}\int_{S_2}\{\mathscr{M}^*_{\pm}(n,z),\chi_2(n',z_1)\}_{PB}M^{-1}_{21}(n',m',z_1,z_2)\{\chi_1(m',z_2),\mathscr{M}_{\pm}(m,z')\}_{PB}
\cr&=-\frac{1}{4\pi i\k}(-2\pi i\k)^2\sum_{n',m'}\int_{S_1}\int_{S_2}\delta_{nn'}\delta^2(z,z_1)\tilde R_{n'm'}\delta^2(z_1,z_2)\delta_{m'm}\delta^2(z_2,z')
\cr&=-i\pi\k\tilde R_{nm}\delta^2(z,z'),
\ee

\be\label{bss}
&\{\mathscr{S}^*_{\pm}(n,z),\mathscr{S}_{\pm}(m,z')\}_{DB}=
\cr&-\sum_{n',m'}\int_{S_1}\int_{S_2}\{\mathscr{S}^*_{\pm}(n,z),\chi_2(n',z_1)\}_{PB}M^{-1}_{21}(n',m',z_1,z_2)\{\chi_1(m',z_2),\mathscr{S}_{\pm}(m,z')\}_{PB}
\cr&=-\frac{1}{4\pi i\k}(2\pi i\k)^2\sum_{n',m'}\int_{S_1}\int_{S_2}R_{nn'}\delta^2(z,z_1)\tilde R_{n'm'}\delta^2(z_1,z_2)R_{m'm}\delta^2(z_2,z')
\cr&=-i\pi\k R_{nm}\delta^2(z,z'),
\ee

\be\label{bms}
&\{\mathscr{M}^*_{\pm}(n,z),\mathscr{S}_{\pm}(m,z')\}_{DB}=-2\pi i\k\delta_{nm}\delta^2(z,z')
\cr&-\sum_{n',m'}\int_{S_1}\int_{S_2}\{\mathscr{M}^*_{\pm}(n,z),\chi_2(n',z_1)\}_{PB}M^{-1}_{21}(n',m',z_1,z_2)\{\chi_1(m',z_2),\mathscr{S}_{\pm}(m,z')\}_{PB}
\cr&=-2\pi i\k\delta_{nm}\delta^2(z,z')
-\frac{1}{4\pi i\k}(2\pi \k)^2\sum_{n',m'}\int_{S_1}\int_{S_2}\delta_{nn'}\delta^2(z,z_1)\tilde R_{n'm'}\delta^2(z_1,z_2)R_{m'm}\delta^2(z_2,z')
\cr&=-i\pi\k \delta_{nm}\delta^2(z,z'),
\ee
which is exactly what we expected.

\section{Discrete basis hard charge expression}\la{A:dbcharge}
We want to focus on the positive-energy mode of the hard charge aspect \eqref{q2i}, using the decompositions \eqref{sheardec} and \eqref{newsdec} of the shear and the news, respectively.  We have
\be\la{discreteq2}
{q}_{s+}^2(z)&=\frac14\int_{-\infty}^\infty du\left[\sum_{\ell=0}^s (-)^{l}(\ell+1) 
  D^{s-\ell}\left[\p_{u,0}^{-\ell+1}\left( \frac{(-u)^{s-\ell}}{(s-\ell)!}C_+\right)  D^{\ell}   N^*_+  \right](u,z) \right]
  \cr&=\frac1{8\pi}\sum_{n=0}^\infty \f{i^{n+1}}{n!}\int_{-\infty}^\infty du\left[\sum_{\ell=0}^s (-)^{l}\f{(\ell+1)}{(s-l)!} 
  D^{s-\ell}\left[(s+n-l)!\p_{u,0}^{-\ell+1}\left( \frac{(-u)^{s+n-\ell}}{(s+n-\ell)!}\right)\mS_+(n)  D^{\ell}   N^*_+(u,z) \right] \right],
\ee
where we used \eqref{c-s}. Now, we use that
\be\la{bp-id}
\p_{u,0}^{-\ell+1}\left(\f{u^{s+n-\ell}}{(s+n-\ell)!}\right)=\f{u^{s+n-1}}{(s+n-1)!}.
\ee
 As a consequence, \eqref{discreteq2} becomes
\be\la{discreteq2bis}
{q}_{s+}^2(z)&=i^{-s}\frac1{8\pi}\sum_{n=0}^\infty \f{1}{n!}\int_{-\infty}^\infty du\left[\sum_{\ell=0}^s \f{(\ell+1)}{(s-l)!} 
  D^{s-\ell}\left[(s+n-l)!\left( \frac{(-i)^{s+n-1}u^{s+n-1}}{(s+n-1)!}\right)\mS_+(n)  D^{\ell}   N^*_+(u,z) \right] \right]
\cr&=i^{-s}\frac1{8\pi}\sum_{n=0}^\infty \left[\sum_{\ell=0}^s (\ell+1)\binom{s+n-l}{n} 
  D^{s-\ell}\left[\mS_+(n)  D^{\ell}   \mathscr{M}^*_+(s+n-1) \right] \right],
\ee
where we used \eqref{n-m}.
Then the smeared charge results to be
\be
{Q}_{s+}^2(\tau_s)&=i^{-s}\frac1{8\pi}\sum_{n=0}^\infty \sum_{\ell=0}^s(-)^{s+\ell} (\ell+1)\binom{s+n-l}{n}\int_S\left[D^{s-\ell}\tau_s(z) 
  \mS_+(n)  D^{\ell}   \mathscr{M}^*_+(s+n-1) \right].
\ee
As a further observation, one can notice that the corner charge aspect can be equivalently written as\footnote{ {This can be deduced for instance from  (68) of \cite{Freidel:2021ytz}, and from the fact that the only fundamental field whose bracket with $\bar C$ is non-trivial is $\bar N$.}}
\be
q_s^2(z)=\f14\int_{-\infty}^\infty du \sum_{\ell=0}^s (-)^{s-\ell}\f{(\ell+1)}{(s-\ell)!}D^{s-\ell}\left((\Delta_u-2)_{s-\ell}\p_{u,0}^{1-s}D^\ell\bar C \bar N\right)(u,z)
\ee
where $\Delta_u:=u\p_u+1$. Through analogous steps as above, from this expression one can derive \eqref{eq31}.
}

\section{Mixed-helicity charge bracket in Gravity}

\subsection{Linear order}\la{A.1}
To compute the mixed-helicity bracket of charges, we use the discrete basis and conveniently express the quadratic charge as \eqref{eq31}, to avoid commutation relations where dealing with matrices $R$ and $\tilde R$ is needed. We obtain
\be
\{\bar Q_{s_1+}^1(\bar \tau_{s_1}), Q_{s_2+}^2(\tau_{s_2})\}&=-i^{-(s_1+s_2)}\f2{\kappa^4\pi} \sum_{n=1}^\infty\sum_{l=0}^{s_2} \int_S \int_{S'} \Bigg[(-)^{s_2}(l+1)\f{(3-n)_{s_2-l}}{(s_2-l)!}\times
\cr
&\times
\bar D_z^{s_1+2} \bar \tau_{s_1}(z)D_{z'}^{s_2-l}\tau_{s_2}(z')\{\mathscr{M}_+  (s_1,z),\mathscr{S}^*_+  (n,z')\}\Bigg]D^l_{z'}\mathscr{M}_+  (s_2+n-1,z')
\cr&=-i^{-(s_1+s_2-1)}\f2{\kappa^4\pi}\sum_{l=0}^{s_2} \int_S  \Bigg[(-)^{s_2-l}(l+1)\f{(3-s_1)_{s_2-l}}{(s_2-l)!}\times
\cr
&\times
D^l\left(\bar D^{s_1+2} \bar \tau_{s_1}(z)D^{s_2-l}\tau_{s_2}(z)\right)\Bigg]\mathscr{M}_+  (s_2+s_1-1,z)
\cr&=-i^{-(s_1+s_2-1)}\f2{\kappa^4\pi}\sum_{l=0}^{s_2} \int_S  \Bigg[(l+1)\f{(s_1+s_2-4-l)_{s_2-l}}{(s_2-l)!}\times
\cr
&\times
D^l\left(\bar D^{s_1+2} \bar \tau_{s_1}(z)D^{s_2-l}\tau_{s_2}(z)\right)\Bigg]\mathscr{M}_+  (s_2+s_1-1,z)
\cr
&=-i^{-(s_1+s_2-1)}\f2{\kappa^2} \sum_{l=0}^{s_2} \int_S \Bigg[(l+1)\f{(s_1+s_2-2)_{s_2-l}}{(s_2-l)!}\times
\cr
&\times
D^l\bar D^{s_1+2} \bar \tau_{s_1}(z)D^{s_2-l}\tau_{s_2}(z)\mathscr{M}_+  (s_1+s_2-1,z)\Bigg]. \la{exp2mixed}
\ee
Here, in the second-to-last  step we used the property of the falling factorial $(-)^n(x)_n=(n-x-1)_n$.

The last step follows from
\be
&\sum_{n=0}^{s_2}\Bigg[(n+1)\f{(s_1+s_2-4-n)_{s_2-n}}{(s_2-n)!} D^n(\bar D^{s_1+2} \bar \tau_{s_1}D^{s_2-n}\tau_{s_2})\Bigg]=
\cr
&\sum_{l=0}^{s_2}\f{1}{l!}\sum_{n=l}^{s_2}\Bigg[\f{(n+1)!}{(n-l)!}\f{(s_1+s_2-4-n)_{s_2-n}}{(s_2-n)!} D^l\bar D^{s_1+2} \bar \tau_{s_1}D^{s_2-l}\tau_{s_2}\Bigg]=
\cr
&\sum_{l=0}^{s_2}\Bigg[(l+1)\f{(s_1+s_2-2)_{s_2-l}}{(s_2-l)!} D^l\bar D^{s_1+2} \bar \tau_{s_1}D^{s_2-l}\tau_{s_2}\Bigg],
\ee
where in the last row, we used the relation \cite{Freidel:2022skz}
\be
\sum_{n=l}^{s}\f{(n+1)!}{(n-l)!}\f{(x-n)_{s-n}}{(s-n)!}=\f{(l+1)!}{(s-l)!}{(x+2)_{s-l}}.
\ee
By adding the negative-energy part of the bracket, we get \eqref{eq36}, i.e.
\be
\{\bar Q_{s_1}^1(\bar \tau_{s_1}), Q_{s_2}^2(\tau_{s_2})\}
&=-\f8{\kappa^2} \sum_{l=0}^{s_2} \int_S (-)^{s_1+s_2} \Bigg[(l+1)\f{(s_1+s_2-2)_{s_2-l}}{(s_2-l)!}
D^l\bar D^{s_1+2} \bar \tau_{s_1}(z)D^{s_2-l}\tau_{s_2}(z)\bar N_{s_1+s_2-1}(z)\Bigg],
\la{eq36b}
\ee
or, equivalently, from the second-to-last  passage of \eqref{exp2mixed}, we obtain 
\be
\{\bar Q_{s_1}^1(\bar \tau_{s_1}), Q_{s_2}^2(\tau_{s_2})\}
&=-\f8{\kappa^2} \sum_{n=0}^{s_2} \int_S \Bigg[(-)^{s_1+s_2}(n+1)\f{(s_1+s_2-4-n)_{s_2-n}}{(s_2-n)!}
D^n(\bar D^{s_1+2} \bar \tau_{s_1}D^{s_2-n}\tau_{s_2})\bar{N}_{s_1+s_2-1}\Bigg]
\la{geillergen},
\ee
which perfectly matches the general expression (J.4a) of \cite{Geiller:2024bgf}, up to some notational differences, summarized in the following table.
\begin{center}
    \begin{tabular}{c|c}
       This work   & Reference \cite{Geiller:2024bgf}  \\[2mm]
    \hline\\  
         $ N_s$  & $\bar N_s$ \\[2mm]
          $ D$  & $\eth$ \\[2mm]
        $\k$ & $8$ \\
    \end{tabular}
\end{center}

\subsection{Quadratic order}\la{A.2}
We want to study the quadratic order of the bracket, for spins $s_1,s_2\le1$. In this case, 
the only contribution to $\{\bar Q_{s_1}, Q_{s_2}\}^2_+$ is $\{\bar Q_{s_1}^2, Q_{s_2}^2\}_+$, because $Q_s^3=0$ for $s=0,1$. 

\subsubsection{$\{\bar Q_0,Q_1\}^2$}
We obtain
\be
\{\bar Q_0^2(\bar \tau_0), Q_1^2(\tau_1)\}_+&=-\f{i}{\pi^2\kappa^4}\sum_{n=0}^\infty\sum_{m=0}^\infty \int_{S\times S'}\bigg[(-1-m)\bar \tau_0(z) D_{z'}\tau_1(z')
\bigg(\mathscr{M}_+^*(n,z)\mathscr{S}_+(m,z')\times
\cr
&\times\{\mathscr{S}_+(n+1,z),\mathscr{M}_+^*(m,z')\}+\{\mathscr{M}_+^*(n,z),\mathscr{S}_+(m,z')\}\times
\cr
&\times\mathscr{M}_+^*(m,z')\mathscr{S}_+(n+1,z)\bigg)+
2\bar \tau_0(z)\tau_1(z')\bigg(\mathscr{M}_+^*(n,z)\mathscr{S}_+(m,z')\times
\cr
&\times D_{z'}\{\mathscr{S}_+(n+1,z),\mathscr{M}_+^*(m,z')\}+\{\mathscr{M}_+^*(n,z),\mathscr{S}_+(m,z')\}\times
\cr
&\times D_{z'}\mathscr{M}_+^*(m,z')\mathscr{S}_+(n+1,z) \bigg)\bigg]
\cr&=\f{1}{\pi\kappa^2}\sum_{n=0}^\infty\int_{S}\bigg[\bar \tau_0(z) D\tau_1(z)
\bigg((-2-n)\mathscr{M}_+^*(n,z)\mathscr{S}_+(n+1,z)
\cr
&+(1+n)\mathscr{M}_+^*(n,z)\mathscr{S}_+(n+1,z)\bigg)-
2\bar \tau_0(z)D(\tau_1(z)\mathscr{S}_+(n+1,z))\mathscr{M}_+^*(n,z)
\cr&
- 2\bar \tau_0(z)\tau_1(z)D\mathscr{M}_+^*(n,z)\mathscr{S}_+(n+1,z) \bigg]
\cr&=\f{1}{\pi\kappa^2}\sum_{n=0}^\infty\int_{S}\bigg[-\bar \tau_0(z) D\tau_1(z)
+
2D\bar \tau_0(z)\tau_1(z)\bigg]\mathscr{S}_+(n+1,z)\mathscr{M}_+^*(n,z)
  \cr
&= \bar{Q}_{0+}^2(-\bar \tau_0 D\tau_1+
2\tau_1D\bar \tau_0)
\ee

\subsubsection{$\{\bar Q_1, Q_1\}^2$}

Let us consider

\be
\{\bar Q_{1+}(\bar\tau_1), Q_{1+}(\tau_1)\}^2=\{\bar Q^2_{1+}(\bar\tau_1), Q^2_{1+}(\tau_1)\}.
\ee

The final result is given by the sum of four contributions, which we will manipulate one by one in the following.

\paragraph{Contribution 1}
\be
&-\f{4}{k^4}\sum_{n=0}^\infty\sum_{m=0}^\infty\int_{S\times S'}\bigg[ \f{1}{4\pi^2}(n-3)(m+1) \bar D \bar \tau_1(z) D' \tau_1(z')\big(\{\mathscr{M}_+^*(n,z),\,\mathscr{S}_+(m,z')\}\mathscr{M}_+^*(m,z')\mathscr{S}_+(n,z)+
\cr
&+\mathscr{M}_+^*(n,z)\mathscr{S}_+(m,z')\{\mathscr{M}_+^*(m,z'),\,\mathscr{S}_+(n,z)\}\big)\bigg]
\cr
&=\f{4}{k^2}i\sum_{n=0}^\infty\int_{S}\bigg[ \f{1}{4\pi}(n-3)(n+1) \bar D \bar \tau_1(z) D \tau_1(z)\big(\mathscr{M}_+^*(n,z)\mathscr{S}_+(n,z)-\mathscr{M}_+^*(n,z)\mathscr{S}_+(n,z)\big)\bigg]=0
\ee

\paragraph{Contribution 2}
\be
&\f{8}{k^4}\sum_{n=0}^\infty\sum_{m=0}^\infty\int_{S\times S'}\bigg[ \f{1}{4\pi^2}(n-3) \bar D \bar \tau_1(z)  \tau_1(z')\big(\{\mathscr{M}_+^*(n,z),\,\mathscr{S}_+(m,z')\}\mathscr{S}_+(n,z)D'\mathscr{M}_+^*(m,z')+
\cr
&+\mathscr{M}_+^*(n,z)\mathscr{S}_+(m,z')\{\mathscr{S}_+(n,z),\,D'\mathscr{M}_+^*(m,z')\}\big)\bigg]
\cr
&=-\f{8}{k^2}i\sum_{n=0}^\infty\int_{S}\bigg[ \f{1}{4\pi}(3-n) 
\tau_1(z)D\bar D \bar \tau_1(z)  \mathscr{M}_+^*(n,z)\mathscr{S}_+(n,z)\bigg]
\cr
&=-\f{8}{k^2}i\sum_{n=0}^\infty\int_{S}\bigg[ \f{1}{4\pi}(3-n) 
\bigg(\tau_1(z)\bar D  D \bar\tau_1(z)  \mathscr{M}_+^*(n,z)\mathscr{S}_+(n,z)-\tau_1(z) \bar \tau_1(z)  \mathscr{M}_+^*(n,z)\mathscr{S}_+(n,z)\bigg)\bigg]
\ee

\paragraph{Contribution 3}
\be
&\f{8}{k^4}\sum_{n=0}^\infty\sum_{m=0}^\infty\int_{S\times S'}\bigg[ \f{1}{4\pi^2}(m+1)  \bar \tau_1(z) D' \tau_1(z')\big(\bar D\{\mathscr{M}_+^*(n,z),\,\mathscr{S}_+(m,z')\}\mathscr{S}_+(n,z)\mathscr{M}_+^*(m,z')+
\cr
&+\bar D\mathscr{M}_+^*(n,z)\mathscr{S}_+(m,z')\{\mathscr{S}_+(n,z),\,\mathscr{M}_+^*(m,z')\}\big)\bigg]
\cr
&=-\f{8}{k^2}i\sum_{n=0}^\infty\int_{S}\bigg[ \f{1}{4\pi}(n+1)  \bar D D \tau_1(z)\bar \tau_1(z)\mathscr{S}_+(n,z)\mathscr{M}_+^*(n,z)
\bigg]
\cr
&=-\f{8}{k^2}i\sum_{n=0}^\infty\int_{S}\bigg[ \f{1}{4\pi}(n+1)  \bigg( D \bar D\tau_1(z)\bar \tau_1(z)\mathscr{S}_+(n,z)\mathscr{M}_+^*(n,z)-\tau_1(z)\bar \tau_1(z)\mathscr{S}_+(n,z)\mathscr{M}_+^*(n,z)
\bigg)\bigg]
\ee

\paragraph{Contribution 4}
\be
&-\f{16}{k^4}\sum_{n=0}^\infty\sum_{m=0}^\infty\int_{S\times S'}\bigg[ \f{1}{4\pi^2}  \bar \tau_1(z) \tau_1(z')\big(\bar D\{\mathscr{M}_+^*(n,z),\,\mathscr{S}_+(m,z')\}\mathscr{S}_+(n,z)D'\mathscr{M}_+^*(m,z')+
\cr
&+\bar D\mathscr{M}_+^*(n,z)\mathscr{S}_+(m,z')D'\{\mathscr{S}_+(n,z),\,\mathscr{M}_+^*(m,z')\}\big)\bigg]
\cr&=\f{16}{k^2}i\sum_{n=0}^\infty\int_{S}\bigg[ \f{1}{4\pi}  \big( \bar \tau_1(z)\bar D\tau_1(z)D\mathscr{M}_+^*(n,z)\mathscr{S}_+(n,z)
+\bar \tau_1(z)\tau_1(z)[\bar D,D]\mathscr{M}_+^*(n,z)\mathscr{S}_+(n,z)
\cr&-\tau_1(z) D\bar \tau_1(z)\bar D\mathscr{M}_+^*(n,z)\mathscr{S}_+(n,z)\big)\bigg]
\cr&=\f{16}{k^2}i\sum_{n=0}^\infty\int_{S}\bigg[ \f{1}{4\pi}  \big( \bar \tau_1(z)\bar D\tau_1(z)D\mathscr{M}_+^*(n,z)\mathscr{S}_+(n,z)
-2\bar \tau_1(z)\tau_1(z)\mathscr{M}_+^*(n,z)\mathscr{S}_+(n,z)
\cr&-\tau_1(z) D\bar \tau_1(z)\bar D\mathscr{M}_+^*(n,z)\mathscr{S}_+(n,z)\big)\bigg]
.
\ee

Here we used the commutation relation of covariant derivatives $[\bar D, D]O_s=sO_s$ holding for operators $O_s$ with helicity $s$, when the angular metric is taken to be the round-sphere metric \cite{Freidel:2021ytz}. When the metric is flattened to a plane, the situation is even easier to deal with, since $D$ and $\bar D$ become partial derivatives.

The sum of the four contributions yields
\be
\{\bar Q_{1+}(\bar\tau_1), Q_{1+}(\tau'_1)\}^2&=-\f{4}{k^2}i\sum_{n=0}^\infty\int_{S} \f{1}{2\pi}\bigg[(3-n)\tau'_1(z) \bar D  D \bar \tau_1(z) \mathscr{S}_+(n,z)\mathscr{M}^*_+(n,z)
\cr&+(n+1)\bar \tau_1(z)  D  \bar D \tau'_1(z) \mathscr{S}_+(n,z)\mathscr{M}_+^*(n,z)
\cr
&+2D\bar \tau_1(z)\tau'_1(z)\mathscr{S}_+(n,z)\bar D\mathscr{M}_+^*(n,z)
-2
\bar \tau_1(z)\bar D \tau'_1(z)\mathscr{S}_+(n,z)D\mathscr{M}_+^*(n,z)
\bigg]
\\
&=\f{4}{k^2}i\sum_{n=0}^\infty\int_{S} \f{1}{2\pi}\bigg[\tau'_1(z)   D \bar \tau_1(z) \left[(3-n) \bar D\left(\mathscr{S}_+(n,z)\mathscr{M}^*_+(n,z)\right)
-2\mathscr{S}_+(n,z)\bar D\mathscr{M}_+^*(n,z)
\right]
\cr
&
+\bar \tau_1(z)    \bar D \tau'_1(z) \left[(n+1) D\left(\mathscr{S}_+(n,z)\mathscr{M}_+^*(n,z)
\right)
+2\mathscr{S}_+(n,z)D\mathscr{M}_+^*(n,z)
\right]
\cr
&+4\bar D\tau'_1(z)   D \bar \tau_1(z) \mathscr{S}_+(n,z)\mathscr{M}^*_+(n,z)
\bigg],
\ee
which can be written as
\eqref{quadmixed11}.

\subsection{$0th$-order violation of the Jacobi identities} \la{a:jac0obstr}
Here, we prove the obstruction to the Jacobi identities \eqref{jac0obstr} for the subset of mixed-helicity charges with $s=0,1$ with the restriction $D\bar\tau_1=0=\bar D\tau_1$. We have 

\be
\{\bar Q_1^1(\bar\tau_1),\{\bar Q_0(\bar\tau_0), Q_1(\tau_1)\}^1\}_+&=\left\{\bar Q_{1+}^1(\bar\tau_1),\bar{Q}_{0+}^{1}(-\bar \tau_0 D\tau'_1+
2\tau'_1D\bar \tau_0)+\f2{\kappa^2} \int_S \bar\tau_0D^3\tau'_1 \mathscr{M}^*_+  (0,z)\right\}
\cr&=\left\{\bar Q_{1+}^1(\bar\tau_1),\f2{\kappa^2} \int_S \bar\tau_0D^3\tau'_1 \mathscr{M}^*_+  (0,z)\right\}
\cr&=-\f{4\pi}{\kappa^2}\int_{S}\bar D^{3}\bar\tau_1\bar\tau_0D^3\tau'_1\tilde R_{10},\\
\{Q_1^1(\tau_1),\{\bar Q_1(\bar\tau_1),\bar Q_0(\bar\tau_0) \}^1\}_+&=\{Q_{1+}^1(\tau_1),\bar{Q}_{0+}^1( \bar\tau_0 \bar D\bar\tau_1-
2\bar\tau_1\bar D \bar\tau_0)\}=0,\\
\{\bar Q_0^1(\bar\tau_0),\{Q_1(\tau_1),\bar Q_1(\bar\tau_1) \}^1\}_+&=0,
\ee
where, in the first two brackets, we used $\{\bar Q^1_{s_1}(\bar\tau_{s_1}),\bar Q^1_{s_2}(\bar \tau_{s_2})\}=0\,\forall s_1,s_2$; in the third one, we used the fact that under the aforementioned restrictions we have $\{Q_1(\tau_1),\bar Q_1(\bar \tau_1)\}^1=0$.

\section{Action of charges on non-radiative corner aspects}\la{VectBra}
In this appendix, we compute the canonical action of real charges \eqref{QR} on charge aspects of spin $0,1$ in the non-radiative phase space $\Gamma_{\text{NR}}$.
We use the complex charge aspects \eqref{aspectinf} defined at spatial infinity, where the non-radiative conditions \eqref{NRcond} hold,
expressed in terms of the discrete basis.  
For the soft and hard parts, we respectively have
\be
q_{s+}^1(z)&=-\f{i^{-s}}{4}D^{s+2}\mathscr{M}_+^*(s,z),\\\la{qsquad}
q_{s+}^2(z)&=\f{i^{-s}}{8\pi}\sum_{n=0}^\infty\sum_{l=0}^s\left[(l+1)\f{(s+n-l)_{s-l}}{(s-l)!}D^{s-l}(\mathscr{S}_+(n,z)\mathscr{M}^*_+(s+n-1,z))\right]
\cr
&=\f{i^{s}}{8\pi}\sum_{n=0}^\infty\sum_{l=0}^s\left[(-)^{s-l}(l+1)\f{(3-n)_{s-l}}{(s-l)!}D^{s-l}(\mathscr{S}^*_+(n,z)\mathscr{M}_+(s+n-1,z))\right].
\ee
Here, the hard part has two possible expressions, as already mentioned for the case of smeared charges through equations \eqref{eq30}, \eqref{eq31}. 
In computing Poisson brackets, we always choose the most convenient expression for the hard charge (namely, first or second row of \eqref{qsquad}), case by case, in order to avoid the presence of commutators of the kind $\{\mathscr{S}^*_+(n,z),\mathscr{S}_+(m,z')\}$ or $\{\mathscr{M}^*_+(n,z),\mathscr{M}_+(m,z')\}$  which would require the manipulation of matrices $R_{nm}$ and $\tilde R_{nm}$. 

The action of $\Re Q_0$ on $q_0(z)$ is trivial, since at the linear order, the brackets of both $Q_0$ and $\bar Q_0$ with $q_0$ vanish. For all other cases, we present the explicit calculations below.
{
\subsection{$\{\Re Q_0,j_A\}$}
To derive the action of the real smeared charge $\Re[Q_0(\tau_0)]$ on $q_1$ we start by computing the action of $Q_0(\tau_0)$. We have
\be
\{Q_0^1(\tau_0),q^2_1(z')\}_+
&=\f14\bigg[3D^2\tau_0D\mathscr{M}_+^*(0,z')+D^3\tau_0\mathscr{M}_+^*(0,z')\bigg],
\ee
and
\be
\{Q_0^2(\tau_0),q_1^1(z')\}_+
&=-\f{1}{4} D^3(\tau_0(z')\mathscr{M}^*_+(0,z)).
\ee
Summing up the two pieces, we obtain
\be\la{Qp}
\{Q_0(\tau_0),q_1(z)\}_+^1
&=3D\tau_0q^1_{0+}(0,z)+\tau_0Dq^1_{0+}(0,z).
\ee
The complex conjugate $\bar Q_0(\bar \tau_0)$ gives
\be
\{\bar Q_0^2(\bar\tau_0),q_1^1(z')\}_+
&=-\f{1}{4} D^3(\bar\tau_0(z')\mathscr{M}^*_+(0,z)),
\ee
and
\be
\{\bar Q_0^1(\bar\tau_0),q^2_1(z')\}_+
&=\f14\bigg[-\bar D^2\bar \tau_0 D\mathscr{M}_+(0,z')-3D\bar D^2\bar \tau_0\mathscr{M}_+(0,z')\bigg].
\ee
Thus we have
\be\la{bQp}
\{\bar Q_0(\tau_0),q_1(z)\}_+^1&=-\f{1}{4}\bigg[D^3(\bar\tau_0\mathscr{M}^*_+(0,z))+\bar D^2\bar\tau_0D\mathscr{M}_+(0,z)+3D\bar D^2\bar\tau_0\mathscr{M}_+(0,z)\bigg]
\cr&=3D\bar\tau_0q^1_{0+}(z)+\bar\tau_0Dq^1_{0+}(z)
&
\cr&-\f{1}{4}\bigg[D^3\bar\tau_0\mathscr{M}^*_+(0,z)+3D^2\bar\tau_0D\mathscr{M}^*_+(0,z)+\bar D^2\bar\tau_0D\mathscr{M}_+(0,z)+3D\bar D^2\bar\tau_0\mathscr{M}_+(0,z)\bigg].
\ee
From \eqref{Qp} and \eqref{bQp} we extract the action of $\Re Q_0$ on $q_1$, which is
\be
\{\Re Q_0(\tau_0),q_1(z)\}_+^1&=3DTq^1_{0+}(z)+TDq^1_{0+}(z)
&
\cr&+\f12\bigg[D^3\bar\tau_0N_0(z)+3D^2\bar\tau_0DN_0(z)+\bar D^2\bar\tau_0D\bar N_0(z)+3D\bar D^2\bar\tau_0\bar N_0(z)\bigg]_+
\cr&=3DTq^1_{0+}(z)+TDq^1_{0+}(z)
\cr&+\f12\bigg[D^3\bar\tau_0N_0(z)+3D^2\bar\tau_0DN_0(z)+\bar D(\bar D\bar\tau_0D\bar N_0)-\bar D(\bar \tau_0\bar D D\bar N_0)+\bar\tau_0D\bar D^2 \bar N_0+3\bar\tau_0\bar D\bar N_0
\cr&+3\bar D(\bar DD\bar\tau_0\bar N_0(z))-3\bar D( D\bar\tau_0\bar D\bar N_0)+3D\bar\tau_0\bar D^2\bar N_0+3\bar D\bar \tau_0\bar N_0\bigg]_+
\cr&=3DTm^1(z)-3D\tilde T\tilde m^1(z)+TDm^1(z)-\tilde TD\tilde m^1(z)
&
\cr&\f12\bigg[D^3(\bar\tau_0N_0)+\bar D(\bar D\bar\tau_0D\bar N_0)-\bar D(\bar \tau_0\bar D D\bar N_0)
\cr&+3\bar D(\bar DD\bar\tau_0\bar N_0(z))-3\bar D( D\bar\tau_0\bar D\bar N_0)+3\bar D(\bar \tau_0\bar N_0)\bigg]_+
\la{totalD}
\ee
It is apparent that, when we integrate this expression on the celestial sphere $S$ against the smearing function $\tau_1$, terms on the second and third line of the last step integrate to zero under the hypotheses of globality and holomorphicity of $\tau_1$.
\newline
Finally, we can translate this result in tensor notation, using \eqref{q1ja}.
\eqref{totalD} implies
\be\la{RQpA}
\{\Re Q_0(\tau_0),j_A(z)\}^1
&=\Re(\{\Re Q_0(\tau_0), q_1(z)\}^1\bar m_A)
\cr&=\f32D_ATm^1(z)-\f32D_A\tilde T\tilde m^1(z)+\f12TD_Am^1(z)-\f12\tilde TD_A\tilde m^1(z)
\cr
&+ D_{\langle A}D_{B}D_{C\rangle}
\alpha^{BC}(T,\tilde T)
+D^C\beta_{AC}(T,\tilde T),
\ee
{
where  
\be
\alpha^{BC}=&\f3{16} (TC^{BC}+\tilde T\tilde C^{BC}) \\
\beta_{AC}=&\f18 \bigg[(D_BT+\tilde D_B\tilde T)D_{\langle A}{C_{C\rangle}}^{B}
-(TD_B+\tilde T\tilde D_B)D_{\langle A}{C_{C\rangle}}^{B}
\cr&+3D_B(D_{\langle A}T-\tilde D_{\langle A}\tilde T){C_{C\rangle}}^B
-3(D_{\langle A}T-\tilde D_{\langle A}\tilde T)D_B{C_{C\rangle}}^B\bigg].
\ee
}

In tensor notation, the conditions of wedge and holomorphicity on $\tau_1=\f12(Y^A+i\tilde Y^A)\bar m_A$ translate into \eqref{cY1} and \eqref{cY2}.
When one integrates \eqref{RQpA} contracted with the parameter $Y^A$ obeying conditions \eqref{cY1}, \eqref{cY2} on the celestial sphere $S$, only the first row survives, thus yielding
\be\la{RQ0RQ1}
\{\Re Q_0(\tau_0),\Re Q_1(\tau_1)\}^1=\int_S Y^A\bigg[\f32D_ATm^1-\f32D_A\tilde T\tilde m^1+\f12TD_Am^1-\f12\tilde TD_A\tilde m^1\bigg],
\ee}
as expected.
{\subsection{$\{\Re Q_1,m\}$ and $\{\Re Q_1,\tilde m\}$}
Let us start with the action of $Q_1(\tau_1)$ on $q_0$.
We have 
\be
\{Q_1(\tau_1),q_0(z')\}^1_+&=D\tau_1q_{0+}^1(z')+2D(\tau_1q_{0+}^1(z'))
\cr
&=2\tau_1Dq_{0+}^1(z')+3D\tau_1q_{0+}^1(z').
\ee
For the mixed-helicity bracket, assuming $\tau_1$ global-holomorphic, we have
\be
\{\bar Q_1(\bar\tau_1),q_0(z')\}^1&=\{\bar Q_1^2(\bar\tau_1),q_0^1(z')\}_+
\cr
&=-\f{1}{4}[D^2(\bar\tau_12\bar D\mathscr{M}^*_+(0,z'))+3D^2(\bar D\bar\tau_1\mathscr{M}^*_+(0,z'))]
\cr&=2\bar\tau_1 \bar Dq^1_{0+}(z')+3\bar D\bar\tau_1q^1_{0+}(z').
\ee
This implies
\be\la{RQ1q0}
\{\Re Q_1(\tau_1), q_0(z')\}^1_+&=(\tau_1D+\bar\tau_1\bar D)q_{0+}^1(z')+\f{3}{2}(D\tau_1+\bar D\bar\tau_1)q_{0+}^1(z')
\cr
&=Y^AD_Aq_{0+}^1(z')+\f{3}{2}D_AY^Aq_{0+}^1(z'),
\ee
where the last row is the result translated into tensor notation.
It is straightforward to obtain the following actions of $\Re Q_1$ the covariant and the dual mass from \eqref{RQ1q0}, through \eqref{q0m}
\be
\{\Re Q_1(\tau_1), m(z')\}^1&=\Re\{\Re Q_1(\tau_1), q_0(z')\}^1
\cr
&=Y^AD_Am^1(z')+\f{3}{2}D_AY^Am^1(z'), \\
\{\Re Q_1(\tau_1), \tilde m(z')\}^1&=\Im\{\Re Q_1(\tau_1), q_0(z')\}^1
\cr
&=Y^AD_A\tilde m^1(z')+\f{3}{2}D_AY^A\tilde m^1(z').
\ee
Starting from these two objects, we can build the following bracket
\be
\{\Re Q_0(\tau_0),\Re Q_1(\tau_1)\}&=\int_S\{Tm- \tilde T\tilde m,\Re Q_1(\tau_1)\}
\cr&=-\int_S\left( TY^AD_Am^1+\f32TD_AY^Am-\tilde TY^AD_A\tilde m^1-\f32\tilde TD_AY^A\tilde m\right)
\cr&=\int_S\left(\f12 TY^AD_Am^1+\f32D_ATY^Am-\f12\tilde TY^AD_A\tilde m^1-\f32D_A\tilde TY^A\tilde m\right)
\ee
which matches \eqref{RQ0RQ1} as expected.
\subsection{$\{\Re Q_1,j_A\}$}
We start from the same-helicity piece
\be
\{Q_1(\tau_1),q_1(z')\}^1_+&=2D\tau_1q_{1+}^1(z')+2D(\tau_1q_{1+}^1(z')).
\ee
For the mixed-helicity piece, assuming the wedge and holomorphicity conditions on $\tau_1$, we have
\be
\{\bar Q_1(\bar\tau_1),q_1(z')\}^1_+&=\{\bar Q_1^2(\bar\tau_1),q_1^1(z')\}_+
\cr&=\f{i}{4}[D^3(2\bar\tau_1\bar D\mathscr{M}^*_+(1,z'))+2D^3(\bar D\bar\tau_1\mathscr{M}^*_+(1,z'))]
\cr&=2\bar\tau_1\bar Dq^1_{1+}(z'))+2\bar D\bar\tau_1q^1_{1+}(z').
\ee
Consequently, we have
\be
\{\Re Q_1(\tau_1),q_1(z')\}^1_+&=(D\tau_1+\bar D\bar\tau_1)q_{1+}^1(z')+D\tau_1q_{1+}^1(z')+(\tau_1D+\bar \tau_1\bar D)q_{1+}^1(z').
\ee
In tensor notation,  through \eqref{q1ja}, this implies
\be
\{\Re Q_1(\tau_1),j_A(z')\}^1&=\Re\{\Re Q_1(\tau_1),q_1(z')\bar m_A\}^1
\cr&= D_BY^Bj_A+\f12(D_BY^Bj_A+D_BY_Aj^B-D_AY_Bj^B)+Y^BD_Bj_A
\cr&= D_BY^Bj_A+D_BY_Aj^B+Y^BD_Bj_A,
\ee
where in the last step we used the restriction \eqref{cY1}.
}

\section{Shadow charge bracket in gravity}
\subsection{Propagators}\la{Prop}
Let us prove the properties of the propagator $G_n(z;z')$ we defined in section \ref{shadsec}. Let us consider the identity
\be
\int_{S'}G_{n}(z';z'')D^n_{z'}G_n(z;z')=G_n(z;z''),
\ee
from which we have
\be
(-)^n\int_{S'}D^n_{z'}G_{n}(z';z'')G_n(z;z')=G_n(z;z'')\implies (-)^n D^n_{z'}G_{n}(z';z'')=\delta^2(z',z'')=D^n_{z'}G_n(z'';z').
\ee
Using this property, we obtain
\be
G_n(z;z'')&=\int_{S'}G_{n}(z';z'')D^n_{z'}G_n(z;z')
\cr&=(-)^n\int_{S'}G_{n}(z';z'')D^n_{z'}G_n(z';z)
\cr&=\int_{S'}D^n_{z'}G_{n}(z';z'')G_n(z';z)=(-)^nG_n(z'';z).
\ee 
Let us now prove that the actions of operators $\bar D^{\,s-2}$ and $D^{-(s+2)}$ on the charge $q_s$ commute for $s\le 2$. The proof for $s\ge3$ is analogous. We have

\be
 \bar D^{s-2}D^{-(s+2)}q^1_s(z)
&=\int_{ S'}\int_{S''}\bar G_{2-s}(z';z)G_{s+2}(z'';z')D_{z''}^{s+2}\bar D_{z''}^{2-s}\tilde q_s^1(z'')
\cr
&=(-)^{2-s}\int_{S''}\bar D^{2-s}_{z''}\bar G_{2-s}(z'';z)\tilde q_s^1(z'')
\cr&=\int_{S'}\int_{S''}G_{s+2}(z';z)D_{z''}^{s+2}\bar D^{2-s}_{z''}\bar G_{2-s}(z'';z')\tilde q_s^1(z'')
\cr&=\int_{S'}\int_{S''}G_{s+2}(z';z)\bar G_{2-s}(z'';z')\bar D^{2-s}_{z''}D_{z''}^{s+2}\tilde q_s^1(z'')
= D^{-(s+2)}\bar D^{s-2}q^1_s(z),
\ee
Here, we used the celestial diamond relations (see Fig.~\ref{fig:CD-GR-2}), which allow us to write
$q_s^1=D^{s+2}\bar D^{2-s}\tilde q_s^1$, and
 the defining property \eqref{defG} of the propagator.

\subsection{Linear order}\la{A.6}
We want to manipulate the expression of the linear-order mixed-helicity bracket \eqref{eq36}, to put it in a simpler form. Our final goal is to prove that, by imposing the wedge condition on the positive-helicity charge, namely $ D^{s+2}\tau_s=0$, the bracket always reduces to the simple expression \eqref{bqq}. In order to do this, we will separately analyze different ranges of the values of spins $s_1$, $s_2$.  
\subsubsection{Case $s_1+s_2-3\ge0$}
In this regime, we can manipulate \eqref{eq36} as follows
\be
\{\bar Q_{s_1}^1(\bar \tau_{s_1}), Q_{s_2}^2(\tau_{s_2})\}
&=-\f8{\kappa^2} \sum_{l=0}^{s_2} \int_S (-)^{s_1+s_2} \Bigg[(s_2+1)\f{(s_1+s_2-2)_{l}}{l!}
D^{s_2-l}\bar D^{s_1+2} \bar \tau_{s_1}(z)D^{l}\tau_{s_2}(z)
\cr&-l\f{(s_1+s_2-2)_{l}}{l!}
D^{s_2-l}\bar D^{s_1+2} \bar \tau_{s_1}(z)D^{l}\tau_{s_2}(z)\Bigg]\bar N_{s_1+s_2-1}(z)
\cr
&=-\f8{\kappa^2}\int_S (-)^{s_1+s_2} \Bigg[\sum_{l=0}^{s_2}  \bigg((s_2+1)\f{(s_1+s_2-2)_{l}}{l!}
D^{s_2-l}\bar D^{s_1+2} \bar \tau_{s_1}(z)D^{l}\tau_{s_2}(z)\bigg)
\cr&
- \sum_{l=1}^{s_2} \bigg((s_1+s_2-2)\f{(s_1+s_2-3)_{l-1}}{(l-1)!}
D^{s_2-l}\bar D^{s_1+2} \bar \tau_{s_1}(z)D^{l}\tau_{s_2}(z)\bigg)\Bigg]\bar N_{s_1+s_2-1}(z)
\cr
&=-\f8{\kappa^2}\int_S (-)^{s_1+s_2} \Bigg[\sum_{l=0}^{s_2}  \bigg((s_2+1)\f{(s_1+s_2-3)_{l}}{l!}
D^{s_2-l}\bar D^{s_1+2} \bar \tau_{s_1}(z)D^{l}\tau_{s_2}(z)\bigg)
\cr&
- \sum_{l=0}^{s_2-1} \bigg((s_1-3)\f{(s_1+s_2-3)_{l}}{l!}
D^{s_2-l-1}\bar D^{s_1+2} \bar \tau_{s_1}(z)D^{l+1}\tau_{s_2}(z)\bigg)\Bigg]\bar N_{s_1+s_2-1}(z) \la{dec}
.
\ee
Here, we used the binomial identities
\be
l\binom{n}{l}=\begin{cases}
n\binom{n-1}{l-1} \quad &\text{for} \quad l>0,\\
0 \quad &\text{for} \quad l=0,    \end{cases}
\ee
and
\be
\binom{n}{l}=\begin{cases}
\binom{n-1}{l}+\binom{n-1}{l-1} \quad &\text{for} \quad l>0,\\
1 \quad &\text{for} \quad l=0.    \end{cases}
\ee

\subsubsection{Case $s_1\ge3$}
The case $s_1\ge3$ implies $s_1+s_2-3\ge s_2$, which includes the case of the previous section. Consequently, we can start with the expression \eqref{dec}, and rewrite it as follows

\be
\{\bar Q_{s_1}^1(\bar \tau_{s_1}), Q_{s_2}^2(\tau_{s_2})\}
&=-\f8{\kappa^2}\int_S (-)^{s_1+s_2} \Bigg[\sum_{l=0}^{s_2}  \bigg((s_2+1)\f{(s_1+s_2-3)_{l}}{l!}
D^{s_2-l}\int_{S'}\delta^2(z',z)\bar D^{s_1+2}_{z'} \bar \tau_{s_1}(z')D^{l}\tau_{s_2}(z)\bigg)
\cr&
- \sum_{l=0}^{s_2-1} \bigg((s_1-3)\f{(s_1+s_2-3)_{l}}{l!}
D^{s_2-l-1}\int_{S'}\delta^2(z',z)\bar D^{s_1+2}_{z'} \bar \tau_{s_1}(z')D^{l+1}\tau_{s_2}(z)\bigg)\Bigg]\bar N_{s_1+s_2-1}(z)
\cr&=-\f8{\kappa^2}\int_S  \Bigg[\sum_{l=0}^{s_2}  \bigg((s_2+1)\f{(s_1+s_2-3)_{l}}{l!}
D^{s_1+s_2-2-l}\int_{S'}G_{s_1-2}(z';z)\bar D^{s_1+2}_{z'} \bar \tau_{s_1}(z')D^{l}\tau_{s_2}(z)\bigg)
\cr&
- \sum_{l=0}^{s_2-1} \bigg((s_1-3)\f{(s_1+s_2-3)_{l}}{l!}
D^{s_1+s_2-3-l}\int_{S'}G_{s_1-2}(z';z)\bar D^{s_1+2}_{z'} \bar \tau_{s_1}(z')D^{l+1}\tau_{s_2}(z)\bigg)\Bigg]
\cr&\times(-)^{s_1+s_2}\bar N_{s_1+s_2-1}(z)
\cr&=-\f8{\kappa^2}\int_S  \Bigg[\sum_{l=0}^{s_2}  \bigg((s_2+1)\f{(s_1+s_2-3)_{l}}{l!}
D^{s_1+s_2-2-l}\bar{ \mS}^{-1} [\bar \tau_{s_1}](z)D^{l}\tau_{s_2}(z)\bigg)
\cr&
- \sum_{k=0}^{s_2-1} \bigg((s_1-3)\f{(s_1+s_2-3)_{k}}{k!}
D^{s_1+s_2-3-k}\bar{ \mS}^{-1} [\bar \tau_{s_1}](z)D^{k+1}\tau_{s_2}(z)\bigg)\Bigg]
(-)^{s_1+s_2}\bar N_{s_1+s_2-1}(z).
\ee

Here, we used the defining property of the propagator \eqref{defG}, namely
\be
D_{z}^{s-2}G_{s-2}(z';z)=\delta^2(z';z),
\ee
and the definition \eqref{shadi} for $\bar{ \mS}^{-1}[\bar \tau_s]$.

If we now impose the wedge condition on positive-helicity charges, i.e. $D^{s+2}\tau_s=0$, summations over $l$ and $k$ can be extended until $k_{max}=l_{max}=s_1+s_2-3$. This can be done without problems because terms with $l\ge s_2+2$ or $k\ge s_1+1$ identically vanish by the wedge condition.  More carefully, one should consider terms with $l=s_1+1$ and $k=s_1$. Nevertheless, these two terms turn out to cancel each other as follows
\be
&(s_2+1)\f{(s_1+s_2-3)_{s_2+1}}{(s_2+1)!}
D^{s_1-3}\bar{ \mS}^{-1} [\bar \tau_{s_1}](z)D^{s_2+1}\tau_{s_2}(z)
-  (s_1-3)\f{(s_1+s_2-3)_{s_2}}{s_2!}
D^{s_1-3}\bar{ \mS}^{-1} [\bar \tau_{s_1}](z)D^{s_2+1}\tau_{s_2}(z)=
\cr&=\f{(s_1+s_2-3)_{s_2+1}}{s_2!}
D^{s_1-3}\bar{ \mS}^{-1} [\bar \tau_{s_1}](z)D^{s_2+1}\tau_{s_2}(z)
-  (s_1-3)\f{(s_1+s_2-3)_{s_2}}{s_2!}
D^{s_1-3}\bar{ \mS}^{-1} [\bar \tau_{s_1}](z)D^{s_2+1}\tau_{s_2}(z)
\cr&=\f{(s_1+s_2-3)_{s_2+1}}{s_2!}
D^{s_1-3}\bar{ \mS}^{-1} [\bar \tau_{s_1}](z)D^{s_2+1}\tau_{s_2}(z)
-  (s_1-3)\f{(s_1+s_2-3)_{s_2}}{s_2!}
D^{s_1-3}\bar{ \mS}^{-1} [\bar \tau_{s_1}](z)D^{s_2+1}\tau_{s_2}(z)=0.
\ee
This is because $(s_1+s_2-3)_{s_2+1}
=(s_1+s_2-3)_{s_2}(s_1-3)$.

Given this, one can write
\be
\{\bar Q_{s_1}^1(\bar \tau_{s_1}), Q_{s_2}^2(\tau_{s_2})\}
&=-\f8{\kappa^2}\int_S  \Bigg[\sum_{l=0}^{s_1+s_2-3}  \bigg((s_2+1)\f{(s_1+s_2-3)_{l}}{l!}
D^{s_1+s_2-2-l}\bar{ \mS}^{-1} [\bar \tau_{s_1}](z)D^{l}\tau_{s_2}(z)\bigg)
\cr&
- \sum_{k=0}^{s_1+s_2-3} \bigg((s_1-3)\f{(s_1+s_2-3)_{k}}{k!}
D^{s_1+s_2-3-l}\bar{ \mS}^{-1} [\bar \tau_{s_1}](z)D^{k+1}\tau_{s_2}(z)\bigg)\Bigg]
(-)^{s_1+s_2}\bar N_{s_1+s_2-1}(z)
\cr
&=-\f8{\kappa^2}\int_S (-)^{s_1+s_2} D^{s_1+s_2-3}\Bigg[(s_2+1)D\bar{ \mS}^{-1} [\bar \tau_{s_1}](z)\tau_{s_2}(z)
\cr
&- (s_1-3)
\bar{ \mS}^{-1} [\bar \tau_{s_1}](z)D\tau_{s_2}(z)\Bigg]
\bar N_{s_1+s_2-1}(z)
\cr
&=\bar\mS[\bar{ Q}^1_{s_1+s_2-1}]\bigg((s_2+1)D\bar{ \mS}^{-1} [\bar \tau_{s_1}]\tau_{s_2}
- (s_1-3)
\bar{ \mS}^{-1} [\bar \tau_{s_1}]D\tau_{s_2}\bigg)
.
\ee
Here, in the last step, we  used the celestial diamond relations (see Fig.~\ref{fig:CD-GR-1}), and the complex conjugate of \eqref{sca}, to write $D^{s_1+s_2-3}\bar N_{s_1+s_2-1}=\bar S[\bar q_{s_1+s_1-1}^1]=\bar \mS[\bar q_{s_1+s_1-1}^1]$ 
\subsubsection{Case $s_1+s_2-3\ge0$, $s_1\le2$}

Let us assume that $s_1\le 2$, while keeping $s_1+s_2-3\ge0$. In this case $s_1+s_2-3\le s_2-1$, so summations in \eqref{dec} truncate to $l_{max}=s_1-s_2-3$, because $(s_1+s_2-3)_l=0$ $\forall l>l_{max}$. We thus have
\be
\{\bar Q_{s_1}^1(\bar \tau_{s_1}), Q_{s_2}^2(\tau_{s_2})\}
&=-\f8{\kappa^2} \int_S (-)^{s_1+s_2} D^{s_1+s_2-3}\Bigg[(s_2+1)
D^{3-s_1}\bar D^{s_1+2} \bar \tau_{s_1}(z)\tau_{s_2}(z)
\cr&-(s_1-3)D^{2-s_1}\bar D^{s_1+2} \bar \tau_{s_1}(z)D\tau_{s_2}(z)\Bigg]\bar N_{s_1+s_2-1}(z)
\cr&=\bar\mS[\bar{ Q}^1_{s_1+s_2-1}]\bigg((s_2+1)D\bar{ \mS}^{-1} [\bar \tau_{s_1}]\tau_{s_2}
- (s_1-3)
\bar{ \mS}^{-1} [\bar \tau_{s_1}]D\tau_{s_2}\bigg).
\ee
Here, we used the complex conjugate of \eqref{sca}, defining the action of the map $\bar \mS$ on the charge aspects $\bar q_s$, and the definition \eqref{shadi} for $\bar{ \mS}^{-1}[\bar \tau_s]$.

\subsubsection{Case $s_1+s_2-3=-1$}
Let us consider the case $s_1+s_2=2$. In this case we rewrite \eqref{eq36} as
\be
\{\bar Q_{s_1}^1(\bar \tau_{s_1}), Q_{s_2}^2(\tau_{s_2})\}
&=-\f8{\kappa^2} \int_S  (s_2+1)
D^{s_2}\bar D^{s_1+2} \bar \tau_{s_1}(z)\tau_{s_2}(z) D\bar\mS[\bar{ q}^1_{1}](z)
\cr&=\bar\mS[\bar{ Q}^1_{1}]\bigg((s_2+1)D\bar{ \mS}^{-1} [\bar \tau_{s_1}]\tau_{s_2}
- (s_1-3)
\bar{ \mS}^{-1} [\bar \tau_{s_1}]D\tau_{s_2}\bigg).
\ee
Here, we used the celestial diamond relations for $s=1$ (see Fig.~\ref{fig:CD-GR-2}) and the complex conjugate of \eqref{sca}, to define the action of $\bar \mS$ on $\bar q_1^1$, and the definition \eqref{shadi} for $\bar{ \mS}^{-1}[\bar \tau_s]$.

\subsubsection{Case $s_1+s_2-3=-2$}

Let us take $s_1+s_2=1$. In this case we rewrite \eqref{eq36} as
\be
\{\bar Q_{s_1}^1(\bar \tau_{s_1}), Q_{s_2}^2(\tau_{s_2})\}
&=\f8{\kappa^2} \int_S  \Bigg[(s_2+1)
D^{s_2}\bar D^{s_1+2} \bar \tau_{s_1}(z)\tau_{s_2}(z)
-s_2
D^{s_2-1}\bar D^{s_1+2} \bar \tau_{s_1}(z)D\tau_{s_2}(z)\Bigg]D^2\bar\mS[\bar{q}^1_{0}](z)
\cr
&=\bar\mS[\bar{ Q}^1_{0}]  \bigg((s_2+1)
D^{3-s_1}\bar D^{s_1+2} \bar \tau_{s_1}\tau_{s_2}
-(s_1-3)
D^{2-s_1}\bar D^{s_1+2} \bar \tau_{s_1}D\tau_{s_2}
\cr&+s_1
D^{s_2}\bar D^{s_1+2} \bar \tau_{s_1}(z)D^2\tau_{s_2}
-s_2
D^{s_2-1}\bar D^{s_1+2} \bar \tau_{s_1}D^3\tau_{s_2}\bigg)
\cr&=\bar\mS[\bar{ Q}^1_{0}]\bigg((s_2+1)D\bar{ \mS}^{-1} [\bar \tau_{s_1}]\tau_{s_2}
- (s_1-3)
\bar{ \mS}^{-1} [\bar \tau_{s_1}]D\tau_{s_2}
\cr&+(-)^{s_2}\bar D^{s_1+2}\bar \tau_{s_1}D^{s_2+2}\tau_{s_2}\bigg)
.
\ee
Here, in the second row, we used the celestial diamond relations for $s=0$ (see Fig.~\ref{fig:CD-GR-2}) and the complex conjugate of \eqref{sca}, for the action of $\bar \mS$ on $\bar q_0^1$;
in the second-to-last row, we used that $s_2=1-s_1$. In the last step we used the definition \eqref{shadi} for $\bar{ \mS}^{-1}[\bar \tau_s]$. We observe that the term on the last row drops out if one imposes $D^{s+2}\tau_s=0$, thus giving the expected formula.
\bigskip

We conclude that for general spins $s_1,s_2$, assuming $D^{s+2}\tau_s=0$, the mixed-helicity commutator assumes the form
\be\{\bar Q_{s_1}(\bar \tau_{s_1}), Q_{s_2}(\tau_{s_2})\}^1=\{\bar Q_{s_1}^1(\bar \tau_{s_1}), Q_{s_2}^2(\tau_{s_2})\}=\bar\mS[\bar{ Q}^1_{s_1+s_2-1}]\bigg((s_2+1)D\bar{ \mS}^{-1} [\bar \tau_{s_1}]\tau_{s_2}- (s_1-3)\bar{ \mS}^{-1} [\bar \tau_{s_1}]D\tau_{s_2}\bigg).\ee

Using the identities \eqref{shadsmearb}, it can be rewritten as

\be
\{\bar Q_{s_1}(\bar \tau_{s_1}), Q_{s_2}(\tau_{s_2})\}^1=\bar{ Q}^1_{s_1+s_2-1}\bigg(\bar \mS\big[(s_2+1)D\bar{ \mS}^{-1} [\bar \tau_{s_1}]\tau_{s_2}
- (s_1-3)
\bar{ \mS}^{-1} [\bar \tau_{s_1}]D\tau_{s_2}\big]\bigg).
\ee

\subsection{Quadratic order} \la{Shad2}
In this appendix, we examine the second-order bracket for spins $s_1,s_2\le1$ and determine under which conditions the structure \eqref{bqq} survives. The wedge restriction $D^{s+2}\tau_s=0$ is assumed throughout, but we will find that additional constraints on the smearing parameters are required.

\subsubsection{$\{\bar Q_0, Q_1\}^2$} 
We have

\be
\{\bar{Q}_{0}(\bar{\tau}_0), Q_1(\tau_1)\}^2_+
&= \bar{Q}_{0+}^2(- {\bar\tau_0} D\tau_1+
2\tau_1D{\bar\tau_0})
\cr&= \bar\mS[\bar{Q}_{0+}^{2}]\bar{ \mS}^{-1}(- {\bar\tau_0} D\tau_1+
2\tau_1D{\bar\tau_0})
\cr&= \bar\mS[\bar{Q}_{0+}^{2}] (3 D^2\bar D^2{\bar\tau_0} D\tau_1+
2\tau_1D^3\bar D^2{\bar\tau_0}
\cr&
+3 D^2{\bar\tau_0} \bar DD\bar D\tau_1+6 \bar DD^2{\bar\tau_0} D\bar D\tau_1
+2\bar D^2\tau_1D^3{\bar\tau_0}
+4\bar D\tau_1D^3\bar D{\bar\tau_0})
\cr&= \bar\mS[\bar{Q}_{0+}^{2}] (3 \bar \mS^{-1}[\bar \tau_0] D\tau_1+
2\tau_1D\bar \mS^{-1}[\bar \tau_0]
\cr&
+3 D^2{\bar\tau_0} \bar DD\bar D\tau_1+6 \bar DD^2{\bar\tau_0} D\bar D\tau_1
+2\bar D^2\tau_1D^3{\bar\tau_0}
+4\bar D\tau_1D^3\bar D{\bar\tau_0}).
\la{bQ0Q1-app}
\ee
We observe that the structure is the same as the linear order if we impose the additional restriction $\bar D\tau_1=0$, in which case the last line in \eqref{bQ0Q1-app} vanishes.

\subsubsection{$\{\bar Q_1, Q_0\}^2$}
In this case, we have
\be
\{ \bar{ Q}_1(\bar{\tau}_1),Q_0( \tau_0)\}^2_+
&= \bar{Q}_{0+}^2( \tau_0 \bar D{\bar\tau_1}-
2{\bar\tau_1}\bar D \tau_0)
\cr&= \bar\mS[\bar{Q}_{0+}^{2}]\bar{ \mS}^{-1}( \tau_0 \bar D{\bar\tau_1}-
2{\bar\tau_1}\bar D \tau_0)
\cr&= \bar\mS[\bar{Q}_{0+}^{2}](2D\tau_0 D\bar D^3{\bar\tau_1}+\tau_0 D^2\bar D^3{\bar\tau_1}
\cr& -3D^2\bar D^2\tau_0 \bar D{\bar\tau_1}-6D\bar D^2\tau_0 D\bar D{\bar\tau_1}-3\bar D^2\tau_0 D^2\bar D{\bar\tau_1}
\cr&-
2D^2{\bar\tau_1}\bar D^3 \tau_0-
4D{\bar\tau_1} D\bar D^3 \tau_0-
2{\bar\tau_1} D^2\bar D^3 \tau_0)
\cr&= \bar\mS[\bar{Q}_{0+}^{2}](2D\tau_0 \bar \mS^{-1}[\bar\tau_1]+\tau_0 D\bar \mS^{-1}[\bar\tau_1]
\cr& -3D^2\bar D^2\tau_0 \bar D{\bar\tau_1}-6D\bar D^2\tau_0 D\bar D{\bar\tau_1}-3\bar D^2\tau_0 D^2\bar D{\bar\tau_1}
\cr&-
2D^2{\bar\tau_1}\bar D^3 \tau_0-
4D{\bar\tau_1} D\bar D^3 \tau_0-
2{\bar\tau_1} D^2\bar D^3 \tau_0).
\la{bQ1Q0-app}
\ee
Again, the structure is the same as the linear order if we impose the additional restriction $\bar D \tau_0=0$, with the last two lines of \eqref{bQ1Q0-app} vanishing.

\subsubsection{$\{\bar Q_1, Q_1\}^2$}

Here, we see that, under the restriction $\bar D \tau_1'=0$, the bracket \eqref{quadmixed11} reduces to
\be
\{\bar Q_{1}(\bar\tau_1), Q_{1}(\tau'_1)\}^2_+
&=\bar Q^2_{1+}(2\tau_1'D\bar\tau_1)
\cr&=2\bar Q^2_{1+}(\bar \mS[D\bar D^3(\tau_1'D\bar\tau_1)])
\cr&=2\bar Q^2_{1+}(\bar \mS[D\tau_1'\bar D^3D\bar\tau_1'+\tau_1'D\bar D^3D\bar\tau_1'])
\cr&=\bar Q^2_{1+}(\bar \mS[2D\tau_1'\bar \mS^{-1}[\bar\tau_1']+2\tau_1'D\bar \mS^{-1}[\bar\tau_1']),
\ee
as expected.

\subsection{Jacobi identities}\la{JI}

Let us consider the Jacobi identity \eqref{2q1bq}. First of all, we observe that its first term reduces to $\{\bar Q_{s_1}^1(\bar \tau_1),\{Q_{s_2}(\tau_{s_3}),Q_{s_3}(\tau_{s_3})\}^2\}$ due to the globality assumption. If we use \eqref{bqq}, together with the fact that the  structure of the bracket \eqref{winf} survives at the quadratic order, we can rephrase \eqref{2q1bq} as
\be
\bar Q^1_{s_1+s_2+s_3-2}(\{\bar\tau_{s_1},\{\tau_{s_2},\tau_{s_3}\}\}+\{\tau_{s_2},\{\tau_{s_3},\bar\tau_{s_1}\}\}+\{\tau_{s_3},\{\bar\tau_{s_1},\tau_{s_2}\}\}
)=0,\la{jacbq}
\ee
where
\be
\{\tau_{s_1},\tau_{s_2}\}&=(s_2+1)\tau_{s_2}D\tau_{s_1}-(s_1+1)\tau_{s_1}D\tau_{s_2},\\
\{\bar\tau_{s_1},\tau_{s_2}\}&=\bar \mS[(s_2+1)\tau_{s_2}D\bar{ \mS}^{-1}[\bar \tau_{s_1}]-(s_1-3)\bar{ \mS}^{-1}[\bar \tau_{s_1}]D\tau_{s_2}].
\ee

We want to prove \eqref{jacbq}.
We compute its three pieces separately. We will mark with the same color terms cancelling out in the summation. We have
\be
\{\tau_{s_3},\{\bar\tau_{s_1},\tau_{s_2}\}\}&=(s_2+1)\{\tau_{s_3},\bar \mS[\tau_{s_2}D\bar{ \mS}^{-1}[\bar \tau_{s_1}]]\}-(s_1-3)\{\tau_{s_3},\bar \mS[\bar{ \mS}^{-1}[\bar \tau_{s_1}]D\tau_{s_2}]\}
\cr&=(s_2+1)\bar \mS[(s_3+1)\tau_{s_3}D(\tau_{s_2}D\bar{ \mS}^{-1}[\bar \tau_{s_1}])-(s_1+s_2-4)\tau_{s_2}D\bar{ \mS}^{-1}[\bar \tau_{s_1}]D\tau_{s_3}]
\cr&-(s_1-3)\bar \mS[(s_3+1)\tau_{s_3}D(\bar{ \mS}^{-1}[\bar \tau_{s_1}]D\tau_{s_2})-(s_1+s_2-4)\bar{ \mS}^{-1}[\bar \tau_{s_1}]D\tau_{s_2}D\tau_{s_3}]
\cr&=\bar \mS[-{\color{purple}({s_2}{s_3}+{s_2}+{s_3})\tau_{s_3}D\tau_{s_2}D\bar{ \mS}^{-1}[\bar \tau_{s_1}]}
\cr&-{\color{red}({s_2}{s_3}+{s_2}+{s_3}+1)\tau_{s_3}\tau_{s_2}D^2\bar{ \mS}^{-1}[\bar \tau_{s_1}]}
\cr&+({\color{green}s_2s_1-3s_2+s_1-4}+{\color{brown}s_2^2})\tau_{s_2}D\bar{ \mS}^{-1}[\bar \tau_{s_1}]D\tau_{s_3}
\cr&+{\color{blue}(s_1s_3-3s_3+s_1-4)\tau_{s_3}D\bar{ \mS}^{-1}[\bar \tau_{s_1}]D\tau_{s_2}}
\cr&+{\color{cyan}(s_1s_3-3s_3+s_1-3)\tau_{s_3}\bar{ \mS}^{-1}[\bar \tau_{s_1}]D^2\tau_{s_2}}
\cr&-({\color{orange}s_1^2-7s_1+12}+{\color{lightgray}s_1s_2-3s_2})\bar{ \mS}^{-1}[\bar \tau_{s_1}]D\tau_{s_2}D\tau_{s_3}],
\ee
\be
\{\tau_{s_2},\{\tau_{s_3},\bar\tau_{s_1}\}\}
&=\bar \mS[{\color{lime}({s_3}{s_2}+{s_3}+{s_2})\tau_{s_2}D\tau_{s_3}D\bar{ \mS}^{-1}[\bar \tau_{s_1}]}
\cr&+{\color{red}({s_3}{s_2}+{s_3}+{s_2}+1)\tau_{s_2}\tau_{s_3}D^2\bar{ \mS}^{-1}[\bar \tau_{s_1}]}
\cr&-({\color{blue}s_3s_1-3s_3+s_1-4}+{\color{olive}s_3^2})\tau_{s_3}D\bar{ \mS}^{-1}[\bar \tau_{s_1}]D\tau_{s_2}
\cr&-{\color{green}(s_1s_2-3s_2+s_1-4)\tau_{s_2}D\bar{ \mS}^{-1}[\bar \tau_{s_1}]D\tau_{s_3}}
\cr&-{\color{magenta}(s_1s_2-3s_2+s_1-3)\tau_{s_2}\bar{ \mS}^{-1}[\bar \tau_{s_1}]D^2\tau_{s_3}}
\cr&+({\color{orange}s_1^2-7s_1+12}+{\color{teal}s_1s_3-3s_3})\bar{ \mS}^{-1}[\bar \tau_{s_1}]D\tau_{s_3}D\tau_{s_2}],
\ee
\be
\{\bar \tau_{s_1},\{\tau_{s_2},\tau_{s_3}\}\}&=-\bar \mS[(s_2+s_3)\{\tau_{s_2},\tau_{s_3}\}D\bar{ \mS}^{-1}[\bar \tau_{s_1}]-(s_1-3)\bar{ \mS}^{-1}[\bar\tau_{s_1}]D\{\tau_{s_2},\tau_{s_3}\}]
\cr&=\bar \mS[-({\color{brown}s_2^2}+{\color{lime}s_3s_2+s_2+s_3})\tau_{s_2}D\tau_{s_3}D\bar{ \mS}^{-1}[\bar \tau_{s_1}]
\cr&+({\color{purple}s_3s_2+s_2+s_3}+{\color{olive}s_3^2})\tau_{s_3}D\tau_{s_2}D\bar{ \mS}^{-1}[\bar\tau_{s_1}]
\cr&+{\color{lightgray}(s_2s_1-3s_2)\bar{ \mS}^{-1}[\bar\tau_{s_1}]D\tau_{s_2}D\tau_{s_3}}
\cr&+{\color{magenta}(s_2s_1-3s_2+s_1-3)\bar{ \mS}^{-1}[\bar\tau_{s_1}]\tau_{s_2}D^2\tau_{s_3}}
\cr&-{\color{teal}(s_3s_1-3s_3)\bar{ \mS}^{-1}[\bar\tau_{s_1}]D\tau_{s_3}D\tau_{s_2}}
\cr&-{\color{cyan}(s_3s_1-3s_3+s_1-3)\bar{ \mS}^{-1}[\bar\tau_{s_1}]\tau_{s_3}D^2\tau_{s_2}}]\,.
\ee

As we see, all terms cancel out in the summation, and the identity \eqref{jacbq} is satisfied.

\section{Mixed-helicity charge bracket in Yang--Mills}\la{C}
\subsection{Linear order}

We conveniently express the hard charge aspect $r_{s_1+}^2$ using the second row of \eqref{r2+}, to compute the action
 \be
 \{  r^{2a}_{s_1+}(z_1),\mathscr{F}^b_+(s_2,z_2)\}
 &=
 g^2_{YM}i^{s_1}f^{ab}{}_c\sum_{n=0}^{s_1} 
(-)^{n}
\f{(s_1 +s_2-1)_{s_1-n}}{(s_1-n)!} D_1^{s_1-n}   \left(\mathscr{F}^c_+(s_1+s_2,z_1)  D_1^{n} \d^2(z_1,z_2)\right)
\cr
&=g^2_{YM}i^{s_1}f^{ab}{}_c
\sum_{\ell=0}^{s_1}
\left(\sum_{n=\ell}^{s_1}
(-)^{s_1+n}
\f{(s_1 +s_2-1)_{n}}{n!} 
\f{(n)_{\ell}}{\ell!} 
\right)
D_1^{\ell}\mathscr{F}^c_+(s_1+s_2,z_1) D_1^{s_1-\ell} \d^2(z_1,z_2)
\cr
&= g^2_{YM} i^{s_1}f^{ab}{}_c
\sum_{\ell=0}^{s_1}
\f1{(s_1+s_2-\ell-1)}\f{(s_1 +s_2-1)!}{\ell!(s_1-\ell)!(s_2-2)!} 
D_1^{\ell}\mathscr{F}^c_+(s_1+s_2,z_1) D_1^{s_1-\ell} \d^2(z_1,z_2)
\cr
&= g^2_{YM}i^{-s_1}f^{ab}{}_c
\sum_{p=0}^{s_1}
\left(\sum_{\ell=0}^{s_1-p}
\f{(-)^{\ell+p}}{(s_1+s_2-\ell-1)}\f{(s_1 +s_2-1)!}{\ell!p!(s_1-\ell-p)!(s_2-2)!} 
\right)
\cr
&\times
D_2^{s_1-p}\mathscr{F}^c_+(s_1+s_2,z_2) D_1^{p} \d^2(z_1,z_2)
\cr
&= g^2_{YM} i^{s_1}f^{ab}{}_c
\sum_{p=0}^{s_1}
\f{(s_1+s_2-p-2)_{s_2-2}}{(s_2-2)!}
D_2^{p}\mathscr{F}^c_+(s_1+s_2,z_2) D_1^{s_1-p} \d^2(z_1,z_2)\,,
\la{r2F}
 \ee
 where we used
 \be
 \sum_{n=\ell}^{s_1}
(-)^{s_1+n}
\f{(s_1 +s_2-1)_{n}}{n!} 
\f{(n)_{\ell}}{\ell!} &=
\f1{(s_1+s_2-\ell-1)}\f{(s_1 +s_2-1)!}{\ell!(s_1-\ell)!(s_2-2)!}\,,
 \\
 \sum_{\ell=0}^{s_1-p}
\f{(-)^{\ell+p}}{(s_1+s_2-\ell-1)}\f{(s_1 +s_2-1)!}{\ell!p!(s_1-\ell-p)!(s_2-2)!} 
&=(-)^{s_1}\f{(s_2+p-2)_p}{p! }\,.
 \ee

 The action \eqref{r2F}, can be used to compute the bracket
 \be
\{r^{2a}_{s_1+}(z_1),\bar r^{1b}_{s_2+}(z_2)\}
&=-i^{s_1+s_2}g^2_{YM}f^{ab}{}_{c}\sum_{n=0}^{s_1}
\f{(s_1+s_2-n-2)_{s_1-n}}{(s_1-n)!}
\bar D^{s_2+1}_{2}(D^{s_1-n}_1\delta^2(z_1,z_2)D_{2}^{n}\mathscr{F}^c_+(s_1+s_2,z_2)).
\la{posrcomm-app}
\ee
The negative-energy counterpart of \eqref{posrcomm-app} can be obtained by performing the  substitutions $r_{s+}\rightarrow\bar r_{s-}$, $D\rightarrow\bar D$, $\mathscr{F}_+(s)\rightarrow\mathscr{F}_-(s)$.
Putting the two sectors together, one obtains
\be
\{r^{2a}_{s_1}(z_1),\bar r^{1b}_{s_2}(z_2)\}
&=g^2_{YM}f^{ab}{}_{c}\sum_{n=0}^{s_1}
\f{(s_1+s_2-n-2)_{s_1-n}}{(s_1-n)!}
\bar D^{s_2+1}_{2}(D^{s_1-n}_1\delta^2(z_1,z_2)D_{2}^{n} \bar F^c_{s_1+s_2}(z_2)),
\la{ymba-app}
\ee
where one uses that $F_s(z)=
-i^{-s}\mathscr{F}^*_+(s,z)-i^s\mathscr{F}_-(s,z)$.

We obtain the charge bracket by integrating the aspects against the smearing functions on the celestial sphere, as follows 
\be
\{R^{2}_{s_1}(\tau_{s_1}),\bar R^{1}_{s_2}(\bar\tau_{s_2})\}
&=g^2_{YM}\sum_{n=0}^{s_1}\int_{S}(-)^{s_1+s_2+1}D^{n}(D^{s_1-n}\tau^a_{s_1}\bar D^{s_2+1}\bar\tau^b_{s_2})f^{ab}{}_{c}
\f{(s_1+s_2-n-2)_{s_1-n}}{(s_1-n)!}
 \bar F^c_{s_1+s_2}(z)
\cr&=g^2_{YM}\sum_{k=0}^{s_1}\sum_{n=k}^{s_1}\int_{S}(-)^{s_1+s_2+1}D^{s_1-k}\tau^a_{s_1}D^{k}\bar D^{s_2+1}\bar\tau^b_{s_2}f^{ab}{}_{c}
\f{(n)_k}{k!}\f{(s_1+s_2-n-2)_{s_1-n}}{(s_1-n)!}
 \bar F^c_{s_1+s_2}(z)
\cr&=g^2_{YM}\sum_{k=0}^{s_1}\int_{S}(-)^{s_1+s_2+1}D^{s_1-k}\tau^a_{s_1}D^{k}\bar D^{s_2+1}\bar\tau^b_{s_2}f^{ab}{}_{c}
\f{(s_1+s_2-1)_{s_1-k}}{(s_1-k)!}
 \bar F^c_{s_1+s_2}(z)
\cr&=-ig^2_{YM}\sum_{k=0}^{s_1}\int_{S}(-)^{s_1+s_2}\binom{s_1+s_2-1}{k}\tr([D^{s_1-k}\bar D^{s_2+1}\bar\tau_{s_2},D^{k}\tau_{s_1}]
 \bar F_{s_1+s_2}(z)),
\ee
where we used
\be
\sum_{n=k}^{s_1}\f{(n)_k}{k!}\f{(s_1+s_2-n-2)_{s_1-n}}{(s_1-n)!}=\f{(s_1+s_2-1)_{s_1-k}}{(s_1-k)!},
\ee
and
\be
\tr([T_a,T_b]_\Ag T_c)=if_{abc}.
\ee
If we apply the same manipulations to the two pieces of the bracket \eqref{r1br2}, we obtain \eqref{smearedmixedfull}.

\subsection{Quadratic order}
Here, we want to compute the quadratic-order contribution to the bracket, for spin $s_1,s_2=0$.
We have that the only contribution to $\{ R_0,\bar R_0\}_+^2$ is $\{ R^2_0,\bar R^2_0\}_+$, since $R^3_0=0$.
\subsubsection{ $\{ R_0,\bar R_0\}^2$}
 We will use
\be
\{R^2_{0+}(\tau_0),\mathscr{A}^{* c}_{+}(n,z)\}=-g^2_{YM}{f_{ab}}^c\tau^a_0(z)\mathscr{A}^{* b}_+(n,z)
\ee
and
\be
\{R^2_{0+}(\tau_0),\mathscr{F}^{ c}_{+}(n,z)\}=-g^2_{YM}{f_{ab}}^c\tau^a_0(z)\mathscr{F}_+^{ b}(n,z).
\ee

Using these two relations, we obtain

\be
\{R^2_{0+}(\tau_0),\bar R^2_{0+}(\bar \tau'_0)\}&=\bigg\{R^2_{0+}(\tau_0),-\f{i}{2\pi}f_{cde}\sum_{l=0}^\infty\int_S\bar \tau'^c_0(z)\mathscr{A}^{* e}_+(l,z)\mathscr{F}^d_+(l,z)\bigg\}
\cr
&=-\f{i}{2\pi}f_{cde}\sum_{l=0}^\infty\int_S\bar \tau'^c_0(z)(\{R^2_{0+}(\tau_0(z)),\mathscr{A}^{* e}_+(l,z)\}\mathscr{F}^d_+(l,z)+\mathscr{A}^{* e}_+(l,z)\{R^2_{0+}(\tau_0),\mathscr{F}^d_+(l,z)\})
\cr
&=-\f{i}{2\pi}g^2_{YM}[f_{ced}{f_{ab}}^e+f_{aed}{f_{bc}}^e]\sum_{l=0}^\infty\int_S\bar \tau'^c_0(z)\tau^a_0(z)\mathscr{A}^{* b}_+(l,z)\mathscr{F}^d_+(l,z)
\cr
&=\f{1}{2\pi}g^2_{YM}f_{bed}\sum_{l=0}^\infty\int_S[\bar \tau'_0,\tau_0]^e_\Ag(z)\mathscr{A}^{* b}_+(l,z)\mathscr{F}^d_+(l,z)
\cr
&=i{g^2_{YM}}\bar R^2_{0+}([\bar \tau'_0,\tau_0]_\Ag).
\ee
Here, in the second-last step, we used the Jacobi identity for the structure constants
\be
f_{ced}{f_{ab}}^e+f_{aed}{f_{bc}}^e+f_{ced}{f_{ab}}^e+f_{bed}{f_{ca}}^e=0.
\ee 

With the usual substitutions, one obtains an analogous relation for the negative-energy sector $\{ R_0,\bar R_0\}_-$, and putting the two pieces together, one finds \eqref{quadYM}.

\section{Shadow charge bracket in Yang-Mills} 
\subsection{Linear order}\la{D.3}
In this appendix, we prove that the bracket \eqref{smearedmixedfull}, under the wedge restriction $D^{s+1}\tau_s=0$, can always be written in the simple form \eqref{scbYM}. The proof is organized by considering the different ranges of $s_1,s_2$ separately.

\subsubsection{Case ${s_2}\ge1$}
If ${s_2}\ge1$, we can extend the summations in \eqref{smearedmixedfull} up to $m_{max}={s_1}+{s_2}-1$, because all terms with $m>{s_1}$ vanish by the wedge restriction $D^{s+1}\tau_s=0$.
Here we have that \eqref{smearedmixedfull} reduces to
 \be
\{R_{{s_1}}(\tau_{s_1}),\bar R_{{s_2}}(\bar \tau_{{s_2}})\}^1
&=-ig^2_{YM}\int_S\biggl(\sum_{m=0}^{{s_1}+{s_2}-1}(-)^{{s_1}+{s_2}}\binom{{s_1}+{s_2}-1}{m}\tr([D^{{s_1}-m}\bar D^{{s_2}+1}\bar \tau_{{s_2}},D^m\tau_{s_1}]_\Ag \bar{F}_{{s_1}+{s_2}}(z))
\biggr)
\cr&=-ig^2_{YM}\int_S\biggl(\sum_{m=0}^{{s_1}+{s_2}-1}(-)^{{s_1}+{s_2}}\binom{{s_1}+{s_2}-1}{m}\cr&\times\tr([D^{{s_1}+{s_2}-1-m}\int_{S'}G_{{s_2}-1}(z';z)\bar D_{z'}^{{s_2}+1}\bar \tau_{{s_2}}(z'),D^m\tau_{s_1}(z)]_\Ag \bar{F}_{{s_1}+{s_2}}(z))
\biggr)
\biggr)
\cr&=-ig^2_{YM}\int_S\biggl(\sum_{m=0}^{{s_1}+{s_2}-1}(-)^{{s_1}+{s_2}}\binom{{s_1}+{s_2}-1}{m}\tr([D^{{s_1}+{s_2}-1-m}\bar{ \mS}^{-1}[\bar \tau_{{s_2}}],D^m\tau_{s_1}]_\Ag \bar{F}_{{s_1}+{s_2}}(z))
\biggr)
\cr&=-ig^2_{YM}\int_S\biggl((-)^{{s_1}+{s_2}}\tr(D^{{s_1}+{s_2}-1}[\bar{ \mS}^{-1}[\bar \tau_{{s_2}}],\tau_{s_1}]_\Ag \bar{F}_{{s_1}+{s_2}}(z))
\biggr)
\cr&=-ig^2_{YM}\bar\mS[\bar { R}^1_{{s_1}+{s_2}}]([\tau_{s_1},\bar{ \mS}^{-1}[\bar \tau_{{s_2}}]]_\Ag).
 \ee
Here, in the third step we used the definition \eqref{shadYib} of $\bar\mS^{-1}[\bar \tau_{s_2}]$;
in the last step, we used the celestial diamond relations for $s\ge1$ (see Fig. \ref{fig:CD-YM-1}) and the complex conjugate of \eqref{scaY}, to write $D^{s_1+s_2-1}\bar F_{s_1+s_2}=\bar S[r_{s_1+s_2}]=\bar \mS[r_{s_1+s_2}]$.

\subsubsection{Case ${s_1}+{s_2}-1\ge0$, ${s_2}\le 1$}
In this case, summations in \eqref{smearedmixedfull} truncate to $m_{max}={s_1}+{s_2}-1$, and we have
 \be
\{R_{{s_1}}(\tau_{s_1}),\bar R_{{s_2}}(\bar \tau_{{s_2}})\}^1
&=-ig^2_{YM}\int_S\biggl(\sum_{m=0}^{{s_1}+{s_2}-1}(-)^{{s_1}+{s_2}}\binom{{s_1}+{s_2}-1}{m}\tr([D^{{s_1}-m}\bar D^{{s_2}+1}\bar \tau_{{s_2}},D^m\tau_{s_1}]_\Ag \bar{F}_{{s_1}+{s_2}}(z))
\biggr)
\cr&=-ig^2_{YM}\int_S\biggl(\sum_{m=0}^{{s_1}+{s_2}-1}(-)^{{s_1}+{s_2}}\binom{{s_1}+{s_2}-1}{m}\cr&\times\tr([D^{{s_1}+{s_2}-1-m}D^{1-{s_2}}\bar D^{{s_2}+1}\bar \tau_{{s_2}},D^m\tau_{s_1}]_\Ag \bar{F}_{{s_1}+{s_2}}(z))
\biggr)
\cr&=-ig^2_{YM}\int_S\biggl(\sum_{m=0}^{{s_1}+{s_2}-1}(-)^{{s_1}+{s_2}}\binom{{s_1}+{s_2}-1}{m}\tr([D^{{s_1}+{s_2}-1-m}\bar{ \mS}^{-1}[\bar \tau_{{s_2}}],D^m\tau_{s_1}]_\Ag \bar{F}_{{s_1}+{s_2}}(z))
\biggr)
\cr&=-ig^2_{YM}\int_S\biggl((-)^{{s_1}+{s_2}}\tr(D^{{s_1}+{s_2}-1}[\bar{ \mS}^{-1}[\bar \tau_{{s_2}}],\tau_{s_1}]_\Ag \bar{F}_{{s_1}+{s_2}}(z))
\biggr)
\cr&=-ig^2_{YM}\bar\mS[\bar { R}^1_{{s_1}+{s_2}}]([\tau_{s_1},\bar{ \mS}^{-1}[\bar \tau_{{s_2}}]]_\Ag).
 \ee
 Here, once again,  we used \eqref{shadYib}, the celestial diamond relations for spin $s\ge1$ and the complex conjugate of \eqref{scaY} for the definition of the maps $\bar \mS$ and $\bar\mS^{-1}$.

 \subsubsection{Case ${s_1}+{s_2}=0$}
In this case
 \be
\{R_{0}(\tau_0),\bar R_{0}(\bar \tau_{0})\}^1
&=-ig^2_{YM}\int_S\biggl(\f{(-1)_0}{0!}\tr([\bar D\bar \tau_{0},\tau_0]_\Ag \bar{F}_{0}(z))
\biggr)
\cr&=-ig^2_{YM}\int_S\tr([\bar D\bar \tau_{0},\tau_0]_\Ag \bar{F}_{0}(z))
\cr&=ig^2_{YM}\int_S\tr([D\bar D\bar \tau_{0},\tau_0]_\Ag \bar\mS[\bar{ r}_{0}](z))
\cr&=-ig^2_{YM}\bar\mS[\bar { R}^1_{0}]([\tau_0,\bar{ \mS}^{-1}[ \bar \tau_{0}]]_\Ag )
.
 \ee
Here we used the celestial diamond relations for spin $s=0$ (see Fig. \ref{fig:CD-YM-2}) to write $\bar F_0=D\bar\mS[\bar r_0]$.

 \subsection{Jacobi identities}\la{jacYM}
 We compute the three pieces of the Jacobi identity \eqref{sjYM1} separately. We find
 \be\{R_{s_1}(\tau_{s_1}),\{R_{s_2}(\tau_{s_2}),R_{s_3}(\bar\tau_{s_3}\}\}^1&=-ig^2_{YM}\{R_{s_1}(\tau_{s_1}),\bar R_{s_2+s_3}(\bar \mS[\tau_{s_2},\bar \mS^{-1}[\bar\tau_{s_3}]]_\Ag)\}^1
 \cr&=-g^4_{YM}\bar R_{s_1+s_2+s_3}(\bar\mS[\tau_{s_1},[\tau_{s_2},\bar \mS^{-1}[\bar\tau_{s_3}]]_\Ag]_\Ag),
 \ee
  \be\{R_{s_2}(\tau_{s_2}),\{\bar R_{s_3}(\bar\tau_{s_3}),R_{s_1}(\tau_{s_1})\}\}^1&=ig^2_{YM}\{R_{s_2}(\tau_{s_2}),\bar R_{s_1+s_3}(\bar \mS[\tau_{s_1},\bar \mS^{-1}[\bar\tau_{s_3}]]_\Ag)\}^1
 \cr&=g^4_{YM}\bar R_{s_1+s_2+s_3}(\bar\mS[\tau_{s_2},[\tau_{s_1},\bar \mS^{-1}[\bar\tau_{s_3}]]_\Ag]_\Ag),
 \ee
  \be\{\bar R_{s_3}(\bar\tau_{s_3}),\{R_{s_1}(\tau_{s_1}),R_{s_2}(\tau_{s_2})\}\}^1&=-ig^2_{YM}\{\bar R_{s_3}(\tau_{s_3}), R_{s_1+s_3}([\tau_{s_1},\tau_{s_2}]_\Ag)\}^1
 \cr&=g^4_{YM}\bar R_{s_1+s_2+s_3}(\bar\mS[[\tau_{s_1},\tau_{s_2}]_\Ag,\bar \mS^{-1}[\bar\tau_{s_3}]]_\Ag).
 \ee
 We see that the sum of the three terms is zero, thus the identity is verified.

\bibliographystyle{bib-style2.bst}
\bibliography{biblio-w.bib}

@article{Simmons-Duffin:2012juh,
    author = "Simmons-Duffin, David",
    title = "{Projectors, Shadows, and Conformal Blocks}",
    eprint = "1204.3894",
    archivePrefix = "arXiv",
    primaryClass = "hep-th",
    doi = "10.1007/JHEP04(2014)146",
    journal = "JHEP",
    volume = "04",
    pages = "146",
    year = "2014"
}

@article{Emond:2020lwi,
    author = "Emond, William T. and Huang, Yu-Tin and Kol, Uri and Moynihan, Nathan and O'Connell, Donal",
    title = "{Amplitudes from Coulomb to Kerr-Taub-NUT}",
    eprint = "2010.07861",
    archivePrefix = "arXiv",
    primaryClass = "hep-th",
    doi = "10.1007/JHEP05(2022)055",
    journal = "JHEP",
    volume = "05",
    pages = "055",
    year = "2022"
}

@article{Kol:2020zth,
    author = "Kol, Uri",
    title = "{Dual Komar Mass, Torsion and Riemann-Cartan Manifolds}",
    eprint = "2010.07887",
    archivePrefix = "arXiv",
    primaryClass = "hep-th",
    month = "10",
    year = "2020"
}

@article{Kol:2019nkc,
    author = "Kol, Uri and Porrati, Massimo",
    title = "{Properties of Dual Supertranslation Charges in Asymptotically Flat Spacetimes}",
    eprint = "1907.00990",
    archivePrefix = "arXiv",
    primaryClass = "hep-th",
    reportNumber = "CERN-TH-2019-069",
    doi = "10.1103/PhysRevD.100.046019",
    journal = "Phys. Rev. D",
    volume = "100",
    number = "4",
    pages = "046019",
    year = "2019"
}

@article{Huang:2019cja,
    author = "Huang, Yu-Tin and Kol, Uri and O'Connell, Donal",
    title = "{Double copy of electric-magnetic duality}",
    eprint = "1911.06318",
    archivePrefix = "arXiv",
    primaryClass = "hep-th",
    reportNumber = "NCTS-TH/1908",
    doi = "10.1103/PhysRevD.102.046005",
    journal = "Phys. Rev. D",
    volume = "102",
    number = "4",
    pages = "046005",
    year = "2020"
}

@article{Freidel:2021fxf,
    author = "Freidel, Laurent and Oliveri, Roberto and Pranzetti, Daniele and Speziale, Simone",
    title = "{The Weyl BMS group and Einstein\textquoteright{}s equations}",
    eprint = "2104.05793",
    archivePrefix = "arXiv",
    primaryClass = "hep-th",
    doi = "10.1007/JHEP07(2021)170",
    journal = "JHEP",
    volume = "07",
    pages = "170",
    year = "2021"
}

@article{Freidel:2024jyf,
    author = "Freidel, Laurent and Moosavian, Seyed Faroogh and Pranzetti, Daniele",
    title = "{On the definition of the spin charge in asymptotically-flat spacetimes}",
    eprint = "2403.19547",
    archivePrefix = "arXiv",
    primaryClass = "hep-th",
    month = "3",
    year = "2024"
}

@article{Freidel:2022skz,
    author = "Freidel, Laurent and Pranzetti, Daniele and Raclariu, Ana-Maria",
    title = "{A discrete basis for celestial holography}",
    eprint = "2212.12469",
    archivePrefix = "arXiv",
    primaryClass = "hep-th",
    month = "12",
    year = "2022"
}

@article{White:2014qia,
    author = "White, C. D.",
    title = "{Diagrammatic insights into next-to-soft corrections}",
    eprint = "1406.7184",
    archivePrefix = "arXiv",
    primaryClass = "hep-th",
    doi = "10.1016/j.physletb.2014.08.041",
    journal = "Phys. Lett. B",
    volume = "737",
    pages = "216--222",
    year = "2014"
}

@article{Strominger:2021lvk,
    author = "Strominger, Andrew",
    title = "{w(1+infinity) and the Celestial Sphere}",
    eprint = "2105.14346",
    archivePrefix = "arXiv",
    primaryClass = "hep-th",
    month = "5",
    year = "2021"
}

@article{Freidel:2021qpz,
    author = "Freidel, Laurent and Pranzetti, Daniele",
    title = "{Gravity from symmetry: Duality and impulsive waves}",
    eprint = "2109.06342",
    archivePrefix = "arXiv",
    primaryClass = "hep-th",
    month = "9",
    year = "2021"
}

@article{Pasterski:2021fjn,
	archiveprefix = {arXiv},
	author = {Pasterski, Sabrina and Puhm, Andrea and Trevisani, Emilio},
	eprint = {2105.03516},
	month = {5},
	primaryclass = {hep-th},
	reportnumber = {CPHT-RR015.032021},
	title = {{Celestial Diamonds: Conformal Multiplets in Celestial CFT}},
	year = {2021}}

@article{Pasterski:2021dqe,
	archiveprefix = {arXiv},
	author = {Pasterski, Sabrina and Puhm, Andrea and Trevisani, Emilio},
	eprint = {2105.09792},
	month = {5},
	primaryclass = {hep-th},
	reportnumber = {CPHT-RR037.052021},
	title = {{Revisiting the Conformally Soft Sector with Celestial Diamonds}},
	year = {2021}}

@article{Donnay:2018neh,
	archiveprefix = {arXiv},
	author = {Donnay, Laura and Puhm, Andrea and Strominger, Andrew},
	doi = {10.1007/JHEP01(2019)184},
	eprint = {1810.05219},
	journal = {JHEP},
	pages = {184},
	primaryclass = {hep-th},
	title = {{Conformally Soft Photons and Gravitons}},
	volume = {01},
	year = {2019}}

@article{Kapec:2016jld,
	archiveprefix = {arXiv},
	author = {Kapec, Daniel and Mitra, Prahar and Raclariu, Ana-Maria and Strominger, Andrew},
	doi = {10.1103/PhysRevLett.119.121601},
	eprint = {1609.00282},
	journal = {Phys. Rev. Lett.},
	number = {12},
	pages = {121601},
	primaryclass = {hep-th},
	title = {{2D Stress Tensor for 4D Gravity}},
	volume = {119},
	year = {2017}}

@article{Donnay:2020guq,
	archiveprefix = {arXiv},
	author = {Donnay, Laura and Pasterski, Sabrina and Puhm, Andrea},
	doi = {10.1007/JHEP09(2020)176},
	eprint = {2005.08990},
	journal = {JHEP},
	pages = {176},
	primaryclass = {hep-th},
	reportnumber = {CPHT-RR022.042020},
	title = {{Asymptotic Symmetries and Celestial CFT}},
	volume = {09},
	year = {2020}}

@article{Campiglia:2015yka,
	archiveprefix = {arXiv},
	author = {Campiglia, Miguel and Laddha, Alok},
	doi = {10.1007/JHEP04(2015)076},
	eprint = {1502.02318},
	journal = {JHEP},
	pages = {076},
	primaryclass = {hep-th},
	title = {{New symmetries for the Gravitational S-matrix}},
	volume = {04},
	year = {2015}}

@article{Ashtekar:1981sf,
	author = {Ashtekar, A.},
	doi = {10.1103/PhysRevLett.46.573},
	journal = {Phys. Rev. Lett.},
	pages = {573--576},
	title = {{Asymptotic Quantization of the Gravitational Field}},
	volume = {46},
	year = {1981}}

@article{Geiller:2021gdk,
    author = "Geiller, Marc and Jai-akson, Puttarak and Osumanu, Abdulmajid and Pranzetti, Daniele",
    title = "{Electromagnetic duality and central charge from first order formulation}",
    eprint = "2107.05443",
    archivePrefix = "arXiv",
    primaryClass = "hep-th",
    journal = "Grav. Cosmol.",
    volume = "1",
    pages = "4",
    year = "2026"
}

@article{Cachazo:2014fwa,
	archiveprefix = {arXiv},
	author = {Cachazo, Freddy and Strominger, Andrew},
	eprint = {1404.4091},
	month = {4},
	primaryclass = {hep-th},
	title = {{Evidence for a New Soft Graviton Theorem}},
	year = {2014}}

@article{Pasterski:2016qvg,
	archiveprefix = {arXiv},
	author = {Pasterski, Sabrina and Shao, Shu-Heng and Strominger, Andrew},
	doi = {10.1103/PhysRevD.96.065026},
	eprint = {1701.00049},
	journal = {Phys. Rev. D},
	number = {6},
	pages = {065026},
	primaryclass = {hep-th},
	title = {{Flat Space Amplitudes and Conformal Symmetry of the Celestial Sphere}},
	volume = {96},
	year = {2017}}

@article{Kapec:2014opa,
	archiveprefix = {arXiv},
	author = {Kapec, Daniel and Lysov, Vyacheslav and Pasterski, Sabrina and Strominger, Andrew},
	doi = {10.1007/JHEP08(2014)058},
	eprint = {1406.3312},
	journal = {JHEP},
	pages = {058},
	primaryclass = {hep-th},
	title = {{Semiclassical Virasoro symmetry of the quantum gravity $ \mathcal{S}$-matrix}},
	volume = {08},
	year = {2014}}

@article{Strominger:2013jfa,
	archiveprefix = {arXiv},
	author = {Strominger, Andrew},
	doi = {10.1007/JHEP07(2014)152},
	eprint = {1312.2229},
	journal = {JHEP},
	pages = {152},
	primaryclass = {hep-th},
	title = {{On BMS Invariance of Gravitational Scattering}},
	volume = {07},
	year = {2014}}

@article{Strominger:2017zoo,
	archiveprefix = {arXiv},
	author = {Strominger, Andrew},
	eprint = {1703.05448},
	month = {3},
	primaryclass = {hep-th},
	title = {{Lectures on the Infrared Structure of Gravity and Gauge Theory}},
	year = {2017}}

@article{Barnich:2016lyg,
	archiveprefix = {arXiv},
	author = {Barnich, Glenn and Troessaert, C\'edric},
	doi = {10.1007/JHEP03(2016)167},
	eprint = {1601.04090},
	journal = {JHEP},
	pages = {167},
	primaryclass = {gr-qc},
	title = {{Finite BMS transformations}},
	volume = {03},
	year = {2016}}

@article{Blanchet:2020ngx,
	archiveprefix = {arXiv},
	author = {Blanchet, Luc and Comp\`ere, Geoffrey and Faye, Guillaume and Oliveri, Roberto and Seraj, Ali},
	doi = {10.1007/JHEP02(2021)029},
	eprint = {2011.10000},
	journal = {JHEP},
	pages = {029},
	primaryclass = {gr-qc},
	title = {{Multipole expansion of gravitational waves: from harmonic to Bondi coordinates}},
	volume = {02},
	year = {2021}}

@article{Barnich:2013axa,
	archiveprefix = {arXiv},
	author = {Barnich, Glenn and Troessaert, C\'edric},
	doi = {10.1007/JHEP11(2013)003},
	eprint = {1309.0794},
	journal = {JHEP},
	pages = {003},
	primaryclass = {hep-th},
	title = {{Comments on holographic current algebras and asymptotically flat four dimensional spacetimes at null infinity}},
	volume = {11},
	year = {2013}}

@article{Compere:2019gft,
	archiveprefix = {arXiv},
	author = {Comp\`ere, Geoffrey and Oliveri, Roberto and Seraj, Ali},
	doi = {10.1007/JHEP10(2020)116},
	eprint = {1912.03164},
	journal = {JHEP},
	pages = {116},
	primaryclass = {gr-qc},
	title = {{The Poincar\'e and BMS flux-balance laws with application to binary systems}},
	volume = {10},
	year = {2020}}

@article{Ashtekar:1981bq,
	author = {Ashtekar, A. and Streubel, M.},
	doi = {10.1098/rspa.1981.0109},
	journal = {Proc. Roy. Soc. Lond. A},
	pages = {585--607},
	title = {{Symplectic Geometry of Radiative Modes and Conserved Quantities at Null Infinity}},
	volume = {376},
	year = {1981}}

@article{Campiglia:2014yka,
	archiveprefix = {arXiv},
	author = {Campiglia, Miguel and Laddha, Alok},
	doi = {10.1103/PhysRevD.90.124028},
	eprint = {1408.2228},
	journal = {Phys. Rev. D},
	number = {12},
	pages = {124028},
	primaryclass = {hep-th},
	title = {{Asymptotic symmetries and subleading soft graviton theorem}},
	volume = {90},
	year = {2014}}

@article{Barnich:2009se,
	archiveprefix = {arXiv},
	author = {Barnich, Glenn and Troessaert, Cedric},
	doi = {10.1103/PhysRevLett.105.111103},
	eprint = {0909.2617},
	journal = {Phys. Rev. Lett.},
	pages = {111103},
	primaryclass = {gr-qc},
	reportnumber = {ULB-TH-09-24},
	title = {{Symmetries of asymptotically flat 4 dimensional spacetimes at null infinity revisited}},
	volume = {105},
	year = {2010}}

@article{Mason:2025pbz,
    author = "Mason, Lionel and Sharma, Atul",
    title = "{Chiral higher-spin theories from twistor space}",
    eprint = "2505.09419",
    archivePrefix = "arXiv",
    primaryClass = "hep-th",
    month = "5",
    year = "2025"
}

@article{Campiglia:2020qvc,
    author = "Campiglia, Miguel and Peraza, Javier",
    title = "{Generalized BMS charge algebra}",
    eprint = "2002.06691",
    archivePrefix = "arXiv",
    primaryClass = "gr-qc",
    doi = "10.1103/PhysRevD.101.104039",
    journal = "Phys. Rev. D",
    volume = "101",
    number = "10",
    pages = "104039",
    year = "2020"
}

@article{Campiglia:2024uqq,
    author = "Campiglia, Miguel and Sudhakar, Adarsh",
    title = "{Gravitational Poisson brackets at null infinity compatible with smooth superrotations}",
    eprint = "2408.13067",
    archivePrefix = "arXiv",
    primaryClass = "gr-qc",
    doi = "10.1007/JHEP12(2024)170",
    journal = "JHEP",
    volume = "12",
    pages = "170",
    year = "2024"
}

@article{Miller:2025wpq,
    author = "Miller, Noah",
    title = "{Spacetime $Lw_{1+\infty}$ Symmetry and Self-Dual Gravity in Plebanski Gauge}",
    eprint = "2504.07176",
    archivePrefix = "arXiv",
    primaryClass = "hep-th",
    month = "4",
    year = "2025"
}

@article{Bu:2022iak,
    author = "Bu, Wei and Heuveline, Simon and Skinner, David",
    title = "{Moyal deformations, W$_{1+\infty}$ and celestial holography}",
    eprint = "2208.13750",
    archivePrefix = "arXiv",
    primaryClass = "hep-th",
    doi = "10.1007/JHEP12(2022)011",
    journal = "JHEP",
    volume = "12",
    pages = "011",
    year = "2022"
}

@article{Guevara:2025tsm,
    author = "Guevara, Alfredo and Himwich, Elizabeth and Miller, Noah",
    title = "{Generating Hodges' Graviton MHV Formula with an $Lw_{1+\infty}$ Ward Identity}",
    eprint = "2506.05460",
    archivePrefix = "arXiv",
    primaryClass = "hep-th",
    month = "6",
    year = "2025"
}

@article{Banerjee:2023bni,
    author = "Banerjee, Shamik and Mandal, Raju and Misra, Sagnik and Panda, Sudhakar and Paul, Partha",
    title = "{All OPEs invariant under the infinite symmetry algebra for gluons on the celestial sphere}",
    eprint = "2311.16796",
    archivePrefix = "arXiv",
    primaryClass = "hep-th",
    doi = "10.1103/PhysRevD.110.026020",
    journal = "Phys. Rev. D",
    volume = "110",
    number = "2",
    pages = "026020",
    year = "2024"
}

@article{Banerjee:2023jne,
    author = "Banerjee, Shamik and Kulkarni, Harshal and Paul, Partha",
    title = "{Celestial OPE in self-dual gravity}",
    eprint = "2311.06485",
    archivePrefix = "arXiv",
    primaryClass = "hep-th",
    doi = "10.1103/PhysRevD.109.086017",
    journal = "Phys. Rev. D",
    volume = "109",
    number = "8",
    pages = "086017",
    year = "2024"
}

@article{Ball:2022bgg,
    author = "Ball, Adam",
    title = "{Celestial locality and the Jacobi identity}",
    eprint = "2211.09151",
    archivePrefix = "arXiv",
    primaryClass = "hep-th",
    doi = "10.1007/JHEP01(2023)146",
    journal = "JHEP",
    volume = "01",
    pages = "146",
    year = "2023"
}

@article{Ball:2024oqa,
    author = "Ball, Adam",
    title = "{Currents in celestial CFT}",
    eprint = "2407.13558",
    archivePrefix = "arXiv",
    primaryClass = "hep-th",
    doi = "10.1142/S0217732324300076",
    journal = "Mod. Phys. Lett. A",
    volume = "39",
    number = "29n30",
    pages = "2430007",
    year = "2024"
}

@article{Pranzetti:2025flv,
    author = "Pranzetti, Daniele and Salluce, Domenico Giuseppe",
    title = "{Double-soft limit and celestial shadow OPE from charge bracket}",
    eprint = "2510.26520",
    archivePrefix = "arXiv",
    primaryClass = "hep-th",
    month = "10",
    year = "2025"
}

@article{Costello:2022wso,
    author = "Costello, Kevin and Paquette, Natalie M.",
    title = "{Celestial holography meets twisted holography: 4d amplitudes from chiral correlators}",
    eprint = "2201.02595",
    archivePrefix = "arXiv",
    primaryClass = "hep-th",
    doi = "10.1007/JHEP10(2022)193",
    journal = "JHEP",
    volume = "10",
    pages = "193",
    year = "2022"
}

@article{Kmec:2024nmu,
    author = "Kmec, Adam and Mason, Lionel and Ruzziconi, Romain and Yelleshpur Srikant, Akshay",
    title = "{Celestial $Lw_{1+\infty}$ charges from a twistor action}",
    eprint = "2407.04028",
    archivePrefix = "arXiv",
    primaryClass = "hep-th",
    doi = "10.1007/JHEP10(2024)250",
    journal = "JHEP",
    volume = "10",
    pages = "250",
    year = "2024"
}

@article{Hu:2023geb,
    author = "Hu, Yangrui and Pasterski, Sabrina",
    title = "{Detector operators for celestial symmetries}",
    eprint = "2307.16801",
    archivePrefix = "arXiv",
    primaryClass = "hep-th",
    doi = "10.1007/JHEP12(2023)035",
    journal = "JHEP",
    volume = "12",
    pages = "035",
    year = "2023"
}

@article{Agrawal:2024sju,
    author = "Agrawal, Shreyansh and Charalambous, Panagiotis and Donnay, Laura",
    title = "{Celestial sw$_{1+\infty}$ algebra in Einstein-Yang-Mills theory}",
    eprint = "2412.01647",
    archivePrefix = "arXiv",
    primaryClass = "hep-th",
    doi = "10.1007/JHEP03(2025)208",
    journal = "JHEP",
    volume = "03",
    pages = "208",
    year = "2025"
}

@article{Cresto:2025ubl,
    author = "Cresto, Nicolas and Freidel, Laurent",
    title = "{Asymptotic higher spin symmetries. Part IV. Einstein-Yang-Mills theory}",
    eprint = "2505.04327",
    archivePrefix = "arXiv",
    primaryClass = "hep-th",
    doi = "10.1007/JHEP12(2025)097",
    journal = "JHEP",
    volume = "12",
    pages = "097",
    year = "2025"
}

@article{Cresto:2025bfo,
    author = "Cresto, Nicolas",
    title = "{Asymptotic higher spin symmetries III: Noether realization in Yang{\textendash}Mills theory}",
    eprint = "2501.08856",
    archivePrefix = "arXiv",
    primaryClass = "hep-th",
    doi = "10.1007/s11005-025-02027-7",
    journal = "Lett. Math. Phys.",
    volume = "115",
    number = "6",
    pages = "133",
    year = "2025"
}

@article{Kmec:2025ftx,
    author = "Kmec, Adam and Mason, Lionel and Ruzziconi, Romain and Sharma, Atul",
    title = "{S-algebra in gauge theory: twistor, spacetime and holographic perspectives}",
    eprint = "2506.01888",
    archivePrefix = "arXiv",
    primaryClass = "hep-th",
    doi = "10.1088/1361-6382/ae0673",
    journal = "Class. Quant. Grav.",
    volume = "42",
    number = "19",
    pages = "195008",
    year = "2025"
}

@article{Godazgar:2019dkh,
	archiveprefix = {arXiv},
	author = {Godazgar, Hadi and Godazgar, Mahdi and Pope, C.N.},
	doi = {10.1007/JHEP10(2019)123},
	eprint = {1908.01164},
	journal = {JHEP},
	pages = {123},
	primaryclass = {hep-th},
	title = {{Dual gravitational charges and soft theorems}},
	volume = {10},
	year = {2019}}

@article{Godazgar:2018dvh,
	archiveprefix = {arXiv},
	author = {Godazgar, Hadi and Godazgar, Mahdi and Pope, C.N.},
	doi = {10.1007/JHEP03(2019)057},
	eprint = {1812.06935},
	journal = {JHEP},
	pages = {057},
	primaryclass = {hep-th},
	title = {{Tower of subleading dual BMS charges}},
	volume = {03},
	year = {2019}}

@article{Hosseinzadeh:2018dkh,
	archiveprefix = {arXiv},
	author = {Hosseinzadeh, V. and Seraj, A. and Sheikh-Jabbari, M.M.},
	doi = {10.1007/JHEP08(2018)102},
	eprint = {1806.01901},
	journal = {JHEP},
	pages = {102},
	primaryclass = {hep-th},
	reportnumber = {IPM-P-2018-030},
	title = {{Soft Charges and Electric-Magnetic Duality}},
	volume = {08},
	year = {2018}}

@article{Godazgar:2018qpq,
	archiveprefix = {arXiv},
	author = {Godazgar, Hadi and Godazgar, Mahdi and Pope, C.N.},
	doi = {10.1103/PhysRevD.99.024013},
	eprint = {1812.01641},
	journal = {Phys. Rev. D},
	number = {2},
	pages = {024013},
	primaryclass = {hep-th},
	reportnumber = {MI-TH-1812},
	title = {{New dual gravitational charges}},
	volume = {99},
	year = {2019}}

@article{Compere:2018ylh,
	archiveprefix = {arXiv},
	author = {Comp\`{e}re, Geoffrey and Fiorucci, Adrien and Ruzziconi, Romain},
	doi = {10.1007/JHEP11(2018)200},
	eprint = {1810.00377},
	journal = {JHEP},
	pages = {200},
	primaryclass = {hep-th},
	slaccitation = {%%CITATION = ARXIV:1810.00377;%%},
	title = {{Superboost transitions, refraction memory and super-Lorentz charge algebra}},
	volume = {11},
	year = {2018}}

@article{Barnich:2011mi,
	archiveprefix = {arXiv},
	author = {Barnich, Glenn and Troessaert, Cedric},
	doi = {10.1007/JHEP12(2011)105},
	eprint = {1106.0213},
	journal = {JHEP},
	pages = {105},
	primaryclass = {hep-th},
	reportnumber = {ULB-TH-11-10},
	slaccitation = {%%CITATION = ARXIV:1106.0213;%%},
	title = {{BMS charge algebra}},
	volume = {12},
	year = {2011}}

@article{Bondi:1960jsa,
	author = {Bondi, H.},
	doi = {10.1038/186535a0},
	journal = {Nature},
	number = {4724},
	pages = {535-535},
	slaccitation = {%%CITATION = NATUA,186,535;%%},
	title = {{Gravitational Waves in General Relativity}},
	volume = {186},
	year = {1960}}

@article{BMS,
	author = {Bondi, H. and van der Burg, M. G. J. and Metzner, A. W. K.},
	doi = {10.1098/rspa.1962.0161},
	journal = {Proc. Roy. Soc. Lond.},
	pages = {21-52},
	slaccitation = {%%CITATION = PRSLA,A269,21;%%},
	title = {{Gravitational waves in general relativity. 7. Waves from axisymmetric isolated systems}},
	volume = {A269},
	year = {1962}}

@article{Sachs62,
	author = {Sachs, R.K.},
	journal = {J.Math.Phys.},
	pages = {908-914},
	slaccitation = {%%CITATION = JMAPA,3,908;%%},
	title = {{On the characteristic initial value problem in gravitational theory}},
	volume = {3},
	year = {1962}}

@book{HenneauxBook,
	author = {Henneaux, M. and Teitelboim, C.},
	publisher = {Princeton},
	slaccitation = {%%CITATION = INSPIRE-345963;%%},
	title = {{Quantization of gauge systems}},
	year = {1992}}

@article{Pate:2019lpp,
    author = "Pate, Monica and Raclariu, Ana-Maria and Strominger, Andrew and Yuan, Ellis Ye",
    title = "{Celestial operator products of gluons and gravitons}",
    eprint = "1910.07424",
    archivePrefix = "arXiv",
    primaryClass = "hep-th",
    doi = "10.1142/S0129055X21400031",
    journal = "Rev. Math. Phys.",
    volume = "33",
    number = "09",
    pages = "2140003",
    year = "2021"
}

@article{Guevara:2021abz,
    author = "Guevara, Alfredo and Himwich, Elizabeth and Pate, Monica and Strominger, Andrew",
    title = "{Holographic symmetry algebras for gauge theory and gravity}",
    eprint = "2103.03961",
    archivePrefix = "arXiv",
    primaryClass = "hep-th",
    doi = "10.1007/JHEP11(2021)152",
    journal = "JHEP",
    volume = "11",
    pages = "152",
    year = "2021"
}

@article{He:2014laa,
    author = "He, Temple and Lysov, Vyacheslav and Mitra, Prahar and Strominger, Andrew",
    title = "{BMS supertranslations and Weinberg\textquoteright{}s soft graviton theorem}",
    eprint = "1401.7026",
    archivePrefix = "arXiv",
    primaryClass = "hep-th",
    doi = "10.1007/JHEP05(2015)151",
    journal = "JHEP",
    volume = "05",
    pages = "151",
    year = "2015"
}

@article{Stieberger:2018onx,
    author = "Stieberger, Stephan and Taylor, Tomasz R.",
    title = "{Symmetries of Celestial Amplitudes}",
    eprint = "1812.01080",
    archivePrefix = "arXiv",
    primaryClass = "hep-th",
    reportNumber = "MPP-2018-292",
    doi = "10.1016/j.physletb.2019.03.063",
    journal = "Phys. Lett. B",
    volume = "793",
    pages = "141--143",
    year = "2019"
}

@article{Adamo:2014baa,
    author = "Adamo, Tim and Newman, E. T.",
    title = "{The Kerr-Newman metric: A Review}",
    eprint = "1410.6626",
    archivePrefix = "arXiv",
    primaryClass = "gr-qc",
    doi = "10.4249/scholarpedia.31791",
    journal = "Scholarpedia",
    volume = "9",
    pages = "31791",
    year = "2014"
}

@article{Erbin:2016lzq,
    author = "Erbin, Harold",
    title = "{Janis-Newman algorithm: generating rotating and NUT charged black holes}",
    eprint = "1701.00037",
    archivePrefix = "arXiv",
    primaryClass = "gr-qc",
    reportNumber = "LPTENS-17-02",
    doi = "10.3390/universe3010019",
    journal = "Universe",
    volume = "3",
    number = "1",
    pages = "19",
    year = "2017"
}

@article{Talbot:1969bpa,
    author = "Talbot, C. J.",
    title = "{Newman-Penrose approach to twisting degenerate metrics}",
    doi = "10.1007/BF01645269",
    journal = "Commun. Math. Phys.",
    volume = "13",
    number = "1",
    pages = "45--61",
    year = "1969"
}

@article{Ashtekar:1978zz,
    author = "Ashtekar, A. and Hansen, R. O.",
    title = "{A unified treatment of null and spatial infinity in general relativity. I - Universal structure, asymptotic symmetries, and conserved quantities at spatial infinity}",
    doi = "10.1063/1.523863",
    journal = "J. Math. Phys.",
    volume = "19",
    pages = "1542--1566",
    year = "1978"
}

@article{Pasterski:2017kqt,
    author = "Pasterski, Sabrina and Shao, Shu-Heng",
    title = "{Conformal basis for flat space amplitudes}",
    eprint = "1705.01027",
    archivePrefix = "arXiv",
    primaryClass = "hep-th",
    doi = "10.1103/PhysRevD.96.065022",
    journal = "Phys. Rev. D",
    volume = "96",
    number = "6",
    pages = "065022",
    year = "2017"
}

@article{Cheung:2016iub,
    author = "Cheung, Clifford and de la Fuente, Anton and Sundrum, Raman",
    title = "{4D scattering amplitudes and asymptotic symmetries from 2D CFT}",
    eprint = "1609.00732",
    archivePrefix = "arXiv",
    primaryClass = "hep-th",
    reportNumber = "CALT-TH-2016-024, UMD-PP-017-010",
    doi = "10.1007/JHEP01(2017)112",
    journal = "JHEP",
    volume = "01",
    pages = "112",
    year = "2017"
}

@article{Himwich:2021dau,
    author = "Himwich, Elizabeth and Pate, Monica and Singh, Kyle",
    title = "{Celestial Operator Product Expansions and ${\rm w}_{1+\infty}$ Symmetry for All Spins}",
    eprint = "2108.07763",
    archivePrefix = "arXiv",
    primaryClass = "hep-th",
    month = "8",
    year = "2021"
}

@article{Guevara:2021tvr,
    author = "Guevara, Alfredo",
    title = "{Celestial OPE blocks}",
    eprint = "2108.12706",
    archivePrefix = "arXiv",
    primaryClass = "hep-th",
    month = "8",
    year = "2021"
}

@article{Ferrara:1972uq,
    author = "Ferrara, S. and Grillo, A. F. and Parisi, G. and Gatto, R.",
    title = "{The shadow operator formalism for conformal algebra. Vacuum expectation values and operator products}",
    doi = "10.1007/BF02907130",
    journal = "Lett. Nuovo Cim.",
    volume = "4S2",
    pages = "115--120",
    year = "1972"
}

@article{Freidel:2021dfs,
    author = "Freidel, Laurent and Pranzetti, Daniele and Raclariu, Ana-Maria",
    title = "{Sub-subleading Soft Graviton Theorem from Asymptotic Einstein's Equations}",
    eprint = "2111.15607",
    archivePrefix = "arXiv",
    primaryClass = "hep-th",
    month = "11",
    year = "2021"
}

@article{Fotopoulos:2019vac,
    author = "Fotopoulos, Angelos and Stieberger, Stephan and Taylor, Tomasz R. and Zhu, Bin",
    title = "{Extended BMS Algebra of Celestial CFT}",
    eprint = "1912.10973",
    archivePrefix = "arXiv",
    primaryClass = "hep-th",
    doi = "10.1007/JHEP03(2020)130",
    journal = "JHEP",
    volume = "03",
    pages = "130",
    year = "2020"
}

@article{Fotopoulos:2019tpe,
    author = "Fotopoulos, Angelos and Taylor, Tomasz R.",
    title = "{Primary Fields in Celestial CFT}",
    eprint = "1906.10149",
    archivePrefix = "arXiv",
    primaryClass = "hep-th",
    doi = "10.1007/JHEP10(2019)167",
    journal = "JHEP",
    volume = "10",
    pages = "167",
    year = "2019"
}

@article{Fan:2019emx,
    author = "Fan, Wei and Fotopoulos, Angelos and Taylor, Tomasz R.",
    title = "{Soft Limits of Yang-Mills Amplitudes and Conformal Correlators}",
    eprint = "1903.01676",
    archivePrefix = "arXiv",
    primaryClass = "hep-th",
    doi = "10.1007/JHEP05(2019)121",
    journal = "JHEP",
    volume = "05",
    pages = "121",
    year = "2019"
}

@article{Pasterski:2017ylz,
    author = "Pasterski, Sabrina and Shao, Shu-Heng and Strominger, Andrew",
    title = "{Gluon Amplitudes as 2d Conformal Correlators}",
    eprint = "1706.03917",
    archivePrefix = "arXiv",
    primaryClass = "hep-th",
    doi = "10.1103/PhysRevD.96.085006",
    journal = "Phys. Rev. D",
    volume = "96",
    number = "8",
    pages = "085006",
    year = "2017"
}

@article{Adamo:2021lrv,
    author = "Adamo, Tim and Mason, Lionel and Sharma, Atul",
    title = "{Celestial $w_{1+\infty}$ symmetries from twistor space}",
    eprint = "2110.06066",
    archivePrefix = "arXiv",
    primaryClass = "hep-th",
    month = "10",
    year = "2021"
}

@article{Ball:2021tmb,
    author = "Ball, Adam and Narayanan, Sruthi and Salzer, Jakob and Strominger, Andrew",
    title = "{Perturbatively Exact $w_{1+\infty}$ Asymptotic Symmetry of Quantum Self-Dual Gravity}",
    eprint = "2111.10392",
    archivePrefix = "arXiv",
    primaryClass = "hep-th",
    month = "11",
    year = "2021"
}

@article{Adamo:2021zpw,
    author = "Adamo, Tim and Bu, Wei and Casali, Eduardo and Sharma, Atul",
    title = "{Celestial operator products from the worldsheet}",
    eprint = "2111.02279",
    archivePrefix = "arXiv",
    primaryClass = "hep-th",
    month = "11",
    year = "2021"
}

@article{Adamo:2019ipt,
    author = "Adamo, Tim and Mason, Lionel and Sharma, Atul",
    title = "{Celestial amplitudes and conformal soft theorems}",
    eprint = "1905.09224",
    archivePrefix = "arXiv",
    primaryClass = "hep-th",
    reportNumber = "IMPERIAL-TP-TA-2019-02",
    doi = "10.1088/1361-6382/ab42ce",
    journal = "Class. Quant. Grav.",
    volume = "36",
    number = "20",
    pages = "205018",
    year = "2019"
}

@article{Magnea:2025zut,
    author = "Magnea, Lorenzo and Zunino, Enrico",
    title = "{Non-abelian soft radiation data for a celestial theory}",
    eprint = "2512.22104",
    archivePrefix = "arXiv",
    primaryClass = "hep-th",
    month = "12",
    year = "2025"
}

@article{Ren:2022sws,
    author = "Ren, Lecheng and Spradlin, Marcus and Yelleshpur Srikant, Akshay and Volovich, Anastasia",
    title = "{On effective field theories with celestial duals}",
    eprint = "2206.08322",
    archivePrefix = "arXiv",
    primaryClass = "hep-th",
    doi = "10.1007/JHEP08(2022)251",
    journal = "JHEP",
    volume = "08",
    pages = "251",
    year = "2022"
}

@article{Mago:2021wje,
    author = "Mago, Jorge and Ren, Lecheng and Srikant, Akshay Yelleshpur and Volovich, Anastasia",
    title = "{Deformed w(1+infinity) Algebras in the Celestial CFT}",
    eprint = "2111.11356",
    archivePrefix = "arXiv",
    primaryClass = "hep-th",
    month = "11",
    year = "2021"
}

@article{Geiller:2024bgf,
    author = "Geiller, Marc",
    title = "{Celestial $w_{1+\infty}$ charges and the subleading structure of asymptotically-flat spacetimes}",
    eprint = "2403.05195",
    archivePrefix = "arXiv",
    primaryClass = "hep-th",
    month = "3",
    year = "2024"
}

@article{Freidel:2023gue,
    author = "Freidel, Laurent and Pranzetti, Daniele and Raclariu, Ana-Maria",
    title = "{On infinite symmetry algebras in Yang-Mills theory}",
    eprint = "2306.02373",
    archivePrefix = "arXiv",
    primaryClass = "hep-th",
    doi = "10.1007/JHEP12(2023)009",
    journal = "JHEP",
    volume = "12",
    pages = "009",
    year = "2023"
}

@article{Freidel:2021ytz,
    author = "Freidel, Laurent and Pranzetti, Daniele and Raclariu, Ana-Maria",
    title = "{Higher spin dynamics in gravity and w1+\ensuremath{\infty} celestial symmetries}",
    eprint = "2112.15573",
    archivePrefix = "arXiv",
    primaryClass = "hep-th",
    doi = "10.1103/PhysRevD.106.086013",
    journal = "Phys. Rev. D",
    volume = "106",
    number = "8",
    pages = "086013",
    year = "2022"
}

@article{Compere:2022zdz,
    author = "Comp{\`e}re, Geoffrey and Oliveri, Roberto and Seraj, Ali",
    title = "{Metric reconstruction from celestial multipoles}",
    eprint = "2206.12597",
    archivePrefix = "arXiv",
    primaryClass = "hep-th",
    doi = "10.1007/JHEP11(2022)001",
    journal = "JHEP",
    volume = "11",
    pages = "001",
    year = "2022"
}

@article{Ball:2023sdz,
    author = "Ball, Adam and Hu, Yangrui and Pasterski, Sabrina",
    title = "{Multicollinear singularities in celestial CFT}",
    eprint = "2309.16602",
    archivePrefix = "arXiv",
    primaryClass = "hep-th",
    doi = "10.1007/JHEP02(2024)219",
    journal = "JHEP",
    volume = "02",
    pages = "219",
    year = "2024"
}

@article{Satishchandran:2019pyc,
    author = "Satishchandran, Gautam and Wald, Robert M.",
    title = "{Asymptotic behavior of massless fields and the memory effect}",
    eprint = "1901.05942",
    archivePrefix = "arXiv",
    primaryClass = "gr-qc",
    doi = "10.1103/PhysRevD.99.084007",
    journal = "Phys. Rev. D",
    volume = "99",
    number = "8",
    pages = "084007",
    year = "2019"
}

@article{Bieri:2020zki,
    author = "Bieri, Lydia",
    title = "{New Effects in Gravitational Waves and Memory}",
    eprint = "2010.09207",
    archivePrefix = "arXiv",
    primaryClass = "gr-qc",
    doi = "10.1103/PhysRevD.103.024043",
    journal = "Phys. Rev. D",
    volume = "103",
    number = "2",
    pages = "024043",
    year = "2021"
}

@article{Blanchet:2023pce,
    author = "Blanchet, Luc and Comp{\`e}re, Geoffrey and Faye, Guillaume and Oliveri, Roberto and Seraj, Ali",
    title = "{Multipole expansion of gravitational waves: memory effects and Bondi aspects}",
    eprint = "2303.07732",
    archivePrefix = "arXiv",
    primaryClass = "gr-qc",
    doi = "10.1007/JHEP07(2023)123",
    journal = "JHEP",
    volume = "07",
    pages = "123",
    year = "2023"
}

@article{Campiglia:2021bap,
    author = "Campiglia, Miguel and Laddha, Alok",
    title = "{BMS Algebra, Double Soft Theorems, and All That}",
    eprint = "2106.14717",
    archivePrefix = "arXiv",
    primaryClass = "hep-th",
    month = "6",
    year = "2021"
}

@article{Freidel:2018fsk,
    author = "Freidel, Laurent and Pranzetti, Daniele",
    title = "{Electromagnetic duality and central charge}",
    eprint = "1806.03161",
    archivePrefix = "arXiv",
    primaryClass = "hep-th",
    doi = "10.1103/PhysRevD.98.116008",
    journal = "Phys. Rev. D",
    volume = "98",
    number = "11",
    pages = "116008",
    year = "2018"
}

@article{Strominger:2015bla,
    author = "Strominger, Andrew",
    title = "{Magnetic Corrections to the Soft Photon Theorem}",
    eprint = "1509.00543",
    archivePrefix = "arXiv",
    primaryClass = "hep-th",
    doi = "10.1103/PhysRevLett.116.031602",
    journal = "Phys. Rev. Lett.",
    volume = "116",
    number = "3",
    pages = "031602",
    year = "2016"
}

@article{Strominger:2013lka,
    author = "Strominger, Andrew",
    title = "{Asymptotic Symmetries of Yang-Mills Theory}",
    eprint = "1308.0589",
    archivePrefix = "arXiv",
    primaryClass = "hep-th",
    doi = "10.1007/JHEP07(2014)151",
    journal = "JHEP",
    volume = "07",
    pages = "151",
    year = "2014"
}

@article{He:2015zea,
    author = "He, Temple and Mitra, Prahar and Strominger, Andrew",
    title = "{2D Kac-Moody Symmetry of 4D Yang-Mills Theory}",
    eprint = "1503.02663",
    archivePrefix = "arXiv",
    primaryClass = "hep-th",
    doi = "10.1007/JHEP10(2016)137",
    journal = "JHEP",
    volume = "10",
    pages = "137",
    year = "2016"
}

@article{Cresto:2024fhd,
    author = "Cresto, Nicolas and Freidel, Laurent",
    title = "{Asymptotic higher spin symmetries I: covariant wedge algebra in gravity}",
    eprint = "2409.12178",
    archivePrefix = "arXiv",
    primaryClass = "hep-th",
    doi = "10.1007/s11005-025-01921-4",
    journal = "Lett. Math. Phys.",
    volume = "115",
    number = "2",
    pages = "39",
    year = "2025"
}

@article{Cresto:2024mne,
    author = "Cresto, Nicolas and Freidel, Laurent",
    title = "{Asymptotic Higher Spin Symmetries II: Noether Realization in Gravity}",
    eprint = "2410.15219",
    archivePrefix = "arXiv",
    primaryClass = "hep-th",
    month = "10",
    year = "2024"
}

@article{Pano:2023slc,
    author = "Pano, Yorgo and Puhm, Andrea and Trevisani, Emilio",
    title = "{Symmetries in Celestial CFT$_{d}$}",
    eprint = "2302.10222",
    archivePrefix = "arXiv",
    primaryClass = "hep-th",
    reportNumber = "CPHT-RR071.122022",
    doi = "10.1007/JHEP07(2023)076",
    journal = "JHEP",
    volume = "07",
    pages = "076",
    year = "2023"
}

@article{Strominger:2021mtt,
    author = "Strominger, Andrew",
    title = "{$w_{1+\infty}$ Algebra and the Celestial Sphere: Infinite Towers of Soft Graviton, Photon, and Gluon Symmetries}",
    eprint = "2105.14346",
    archivePrefix = "arXiv",
    primaryClass = "hep-th",
    doi = "10.1103/PhysRevLett.127.221601",
    journal = "Phys. Rev. Lett.",
    volume = "127",
    number = "22",
    pages = "221601",
    year = "2021"
}

\end{document}